\documentclass[letterpaper, 10pt]{article}
\usepackage[utf8]{inputenc}
\usepackage[top=1in, bottom=1in, left=1in, right=1in]{geometry}
\usepackage{float}
\usepackage{color}
\usepackage{tikz}

\usepackage{titling}
\RequirePackage[explicit]{titlesec}
\titlespacing*{\section}{0pt}{0.5em}{0.3pt}
\titlespacing*{\subsection}{0pt}{0.35em}{0pt}
\titlespacing*{\subsubsection}{0pt}{0.25em}{0pt}

\usepackage{caption}
\RequirePackage{fancyhdr}
\usepackage{lastpage}
\usepackage{empheq}

\usepackage{mathpazo}
\usepackage{xcolor}
\definecolor{GT}{RGB}{0, 0, 0}
\definecolor{Conv}{RGB}{247,37,133}
\definecolor{MC}{RGB}{114,9,183}
\definecolor{MVN}{RGB}{67,97,238}
\definecolor{FE}{RGB}{76,201,240}

\newcommand{\GTline}{\raisebox{2pt}{\tikz{\draw[-,GT,solid,line width = 1pt](0,0) -- (3.5mm,0);}}}
\newcommand{\Convline}{\raisebox{2pt}{\tikz{\draw[-,Conv,dashed,line width = 1pt](0,0) -- (3.5mm,0);}}}
\newcommand{\MCline}{\raisebox{2pt}{\tikz{\draw[-,MC,solid,line width = 1pt](0,0) -- (3.5mm,0);}}}
\newcommand{\MVNline}{\raisebox{2pt}{\tikz{\draw[-,MVN,solid,line width = 1pt](0,0) -- (3.5mm,0);}}}
\newcommand{\FEline}{\raisebox{2pt}{\tikz{\draw[-,FE,solid,line width = 1pt](0,0) -- (3.5mm,0);}}}

\usepackage[
  colorlinks=true,
  linkcolor=blue,
  anchorcolor=blue,
  citecolor=blue,
  filecolor=blue,
  menucolor=blue,
  runcolor=blue,
  urlcolor=blue]{hyperref}
\usepackage{amsmath, amsthm, amssymb, amsfonts}
\usepackage{upgreek}
\usepackage{dsfont}

\usepackage{enumerate}
\usepackage{enumitem}

\usepackage{graphicx}
\graphicspath{{}}
\usepackage{subcaption}

\usepackage{multirow}

\usepackage[numbers, sort&compress]{natbib}

\title{Stochastic Particle Advection Velocimetry (SPAV): Theory, Simulations, and Proof-of-Concept Experiments}

\author{
  Ke Zhou$^1$,
  Jiaqi Li$^{2,3}$,
  Jiarong Hong$^{2,3}$, and
  Samuel J. Grauer$^{1,4,5,}$\thanks{Corresponding author: \href{mailto:sgrauer@psu.edu}{sgrauer@psu.edu}}\vspace*{.15em}\\
  {\small $^1$Department of Mechanical Engineering, Pennsylvania State University}\vspace*{-.25em}\\
  {\small $^2$Department of Mechanical Engineering, University of Minnesota}\vspace*{-.25em}\\
  {\small $^3$Saint Anthony Falls Laboratory, University of Minnesota}\vspace*{-.25em}\\
  {\small $^4$Institute for Computational and Data Sciences, Pennsylvania State University}\vspace*{-.25em}\\
  {\small $^5$Erlangen Graduate School in Advanced Optical Technologies, Friedrich-Alexander-Universit{\"a}t Erlangen-N{\"u}rnberg}}
\date{}

\begin{document}

\maketitle
\setcounter{footnote}{5}
\vspace*{-2em}

\begin{abstract}
Particle tracking velocimetry (PTV) is widely used to measure time-resolved, three-dimensional velocity and pressure fields in fluid dynamics research. Inaccurate localization and tracking of particles is a key source of error in PTV, especially for single camera defocusing, plenoptic imaging, and digital in-line holography (DIH) sensors. To address this issue, we developed stochastic particle advection velocimetry (SPAV): a statistical data loss that improves the accuracy of PTV. SPAV is based on an explicit particle advection model that predicts particle positions over time as a function of the estimated velocity field. The model can account for non-ideal effects like drag on inertial particles. A statistical data loss that compares the tracked and advected particle positions, accounting for arbitrary localization and tracking uncertainties, is derived and approximated. We implement our approach using a physics-informed neural network, which simultaneously minimizes the SPAV data loss, a Navier--Stokes physics loss, and a wall boundary loss, where appropriate.  Results are reported for simulated and experimental DIH-PTV measurements of laminar and turbulent flows. Our statistical approach significantly improves the accuracy of PTV reconstructions compared to a conventional data loss, resulting in an average reduction of error close to 50\%. Furthermore, our framework can be readily adapted to work with other data assimilation techniques like state observer, Kalman filter, and adjoint--variational methods.\par\vspace{.5em}

\noindent\textbf{Keywords:} Particle tracking velocimetry, data assimilation, digital in-line holography, physics-informed neural network
\end{abstract}
\vspace{1.5em}

\section{Introduction}
\label{sec:intro}
Measurement techniques that can capture time-resolved, three-dimensional (3D), three-component (3C) velocity fields are essential to the study of fluid dynamics \cite{Scarano2012, Kim2012}. For instance, 3D3C flow diagnostics have been deployed to support research on turbulent boundary layers \cite{Schneiders2017, Schroder2009}, unsteady jets and wakes \cite{Coriton2015, Zhu2017}, turbulent combustion \cite{weinkauff2013}, and an array of biomedical flows \cite{deLima2007, Hegner2015}, among many other targets. Moreover, reliable velocity and pressure data are needed to develop and validate computational fluid dynamics (CFD) models for engineering design, especially in scenarios that feature variable density mixing, high flow speeds, reactions, and non-ideal fluids \cite{Dimotakis2005}. The measurements should be accurate, precise, well-resolved in space and time, and non-intrusive. Optical diagnostics have the potential to meet these requirements, especially particle-based techniques like particle image and tracking velocimetry (PIV and PTV), but there are often significant trade-offs between spatial and temporal resolution and the pressure field is not readily accessible \cite{Schneiders2015}. While PIV can be combined with tomographic reconstruction for 3D3C velocimetry, this approach suffers from reconstruction errors, like ghost particles \cite{deSilva2012}, and spatial filtering \cite{Scarano2012}, so PTV is often preferred. This paper reports a new framework for 3D PTV that can enhance the accuracy of velocity and pressure field estimates, regardless of the imaging modality, and enable the use of PTV with particles that do not perfectly follow the flow.\par

Three-dimensional PTV generally proceeds as follows. Tracer particles are seeded into the flow or arise spontaneously, e.g., atomized droplets, bubbles in a liquid, snowfall, etc. The particles are illuminated, usually with a laser, and images of the measurement domain are recorded with one or more cameras. Next, a localization algorithm is employed to identify the 3D position of each particle in each frame, and the particles are followed across successive frames with a tracking algorithm, resulting in Lagrangian particle trajectories called ``tracks''. These steps may be performed synergistically, as in the Shake-The-Box (STB) method \cite{Schanz2016}. Many applications call for Eulerian velocity and/or pressure fields, which requires post-processing of the particle tracks. For instance, velocity fields are commonly utilized to identify coherent structures in turbulent flow \cite{Schanz2016}, estimate a dissipation rate field to support modeling \cite{Schneiders2017}, infer pressure fields to study aeronautics and surface loading \cite{Jux2020}, and so on. Unfortunately, localization, tracking, and post-processing are all subject to appreciable uncertainties \cite{Bhattacharya2020}, and \textit{regularization} is often required to reduce the effects of noise and produce estimates that are consistent with the governing physics.\par

\begin{figure}[ht]
    \centering
    \subcaptionbox{\label{fig:localization:triangulation}}{
        \includegraphics[height=4cm]{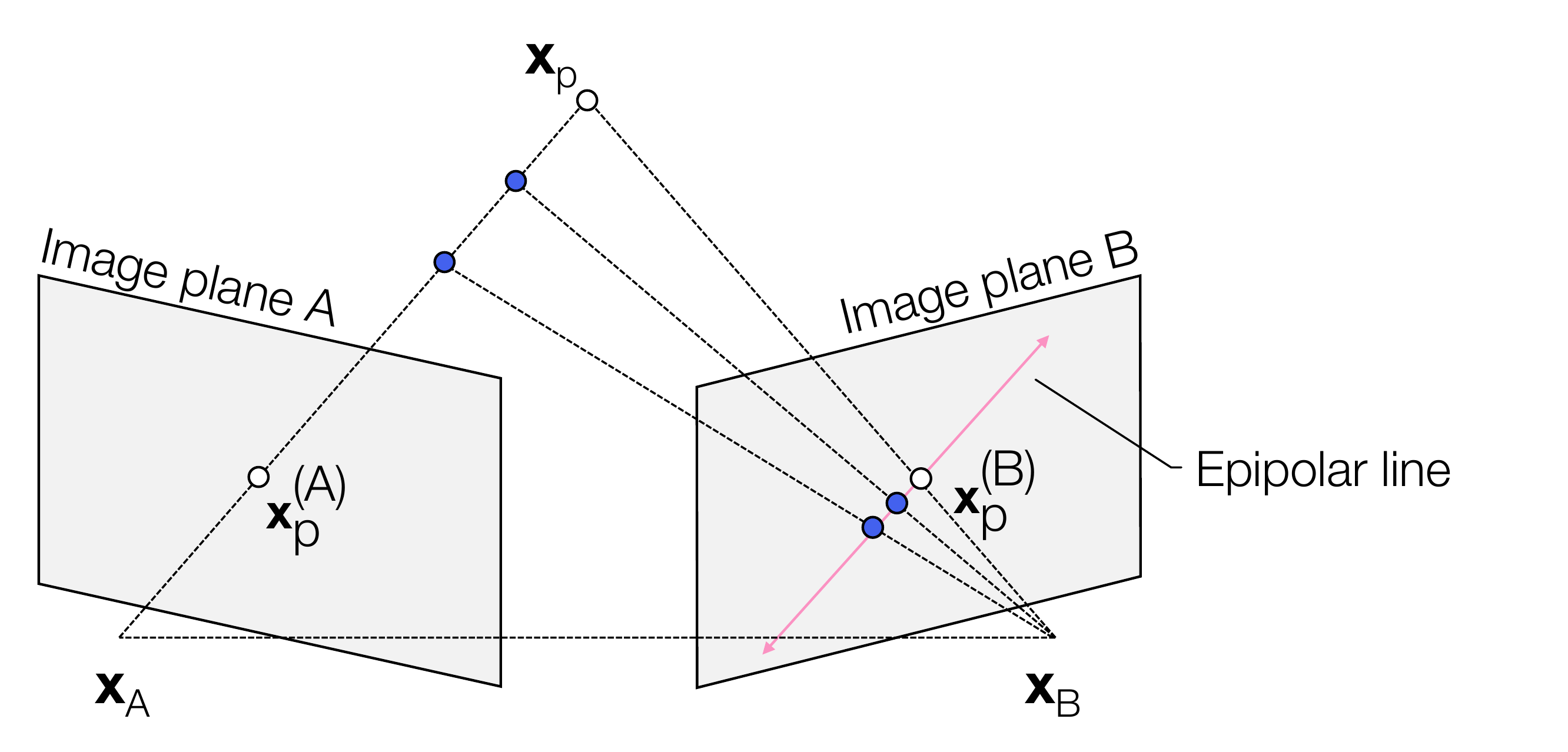}}\quad
    \subcaptionbox{\label{fig:localization:refocus}}{
        \includegraphics[height=4cm]{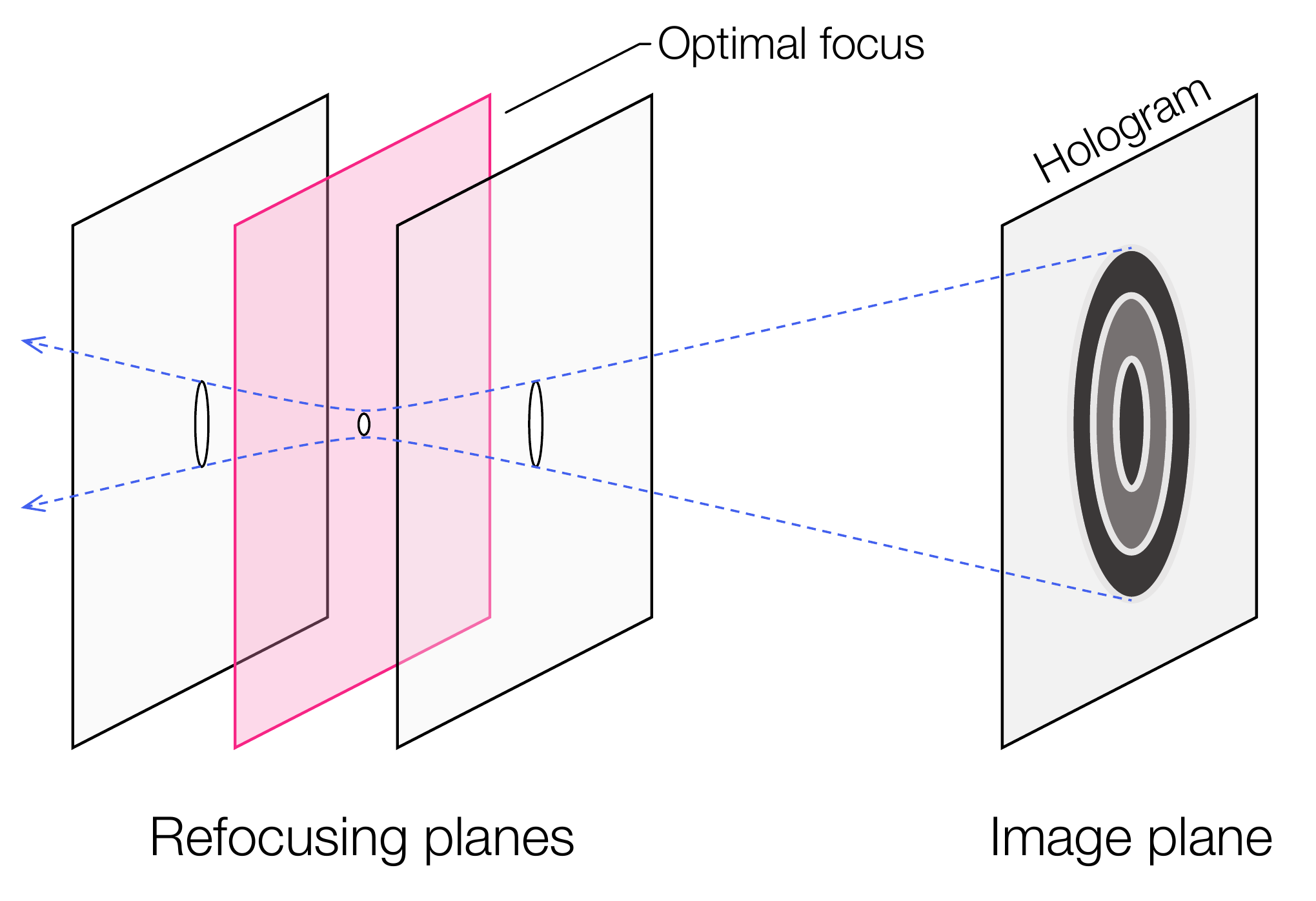}}
    \caption{Particle localization using (a) triangulation and (b) numerical refocusing.}
    \label{fig:localization}
\end{figure}

Particle localization techniques fall into two broad categories: triangulation and numerical refocusing, which generally correspond to distinct imaging modalities. These techniques are sketched in Fig.~\ref{fig:localization}. Triangulation is typically employed to locate particles from multi-camera images of Mie scattering off the particles.\footnote{Other data, e.g., plenoptic images of Mie scattered light from a particle field, can also be processed with a triangulation algorithm to locate particles \cite{Fischer2022}.} This is the most common form of PTV and is usually conducted with three or more cameras. Each particle's location in a single image corresponds to a line-of-sight through physical space, so the lines-of-sight transecting one particle from multiple images can be leveraged to triangulate that particle's position in 3D space \cite{Maas1993}. This procedure requires the cameras to be calibrated in a consistent global coordinate system, and individual particles must be identified in two or more simultaneous images to obtain the requisite lines-of-sight. Particle identification is typically carried out through a probabilistic, sequential search across all views, as set out by Wieneke \cite{Wieneke2008}. Ambiguous particle matches and experimental factors like variable laser illumination, background noise, and calibration drift limit the accuracy of triangulation, especially at high seeding densities \cite{Schanz2016}. These errors are contingent on the number and position of cameras, resulting in a potentially complex spatial distribution of localization uncertainties.\par

Numerical refocusing is a depth sensing procedure used in the context of single-camera and limited-angle systems, such as a plenoptic camera \cite{Hall2017, Nobes2014}, synthetic aperture (SA) array \cite{Bajpayee2014}, or digital in-line holography (DIH) setup \cite{Toloui2017}. The detected light field or hologram is refocused (``backpropagated'') onto a series of planes oriented parallel to the sensor. Particles are assumed to lie in the plane which gives rise to the sharpest image, i.e., where the refocused diameter is at a minimum. However, while it is trivial to determine the 2D location of particles in an image, the accuracy of depth sensing is relatively poor due to the finite depth-of-field (DoF) of a real imaging system \cite{Katz2010}. Consequently, particle position estimates exhibit an anisotropic distribution of uncertainty that is elongated normal to the sensor plane \cite{Gao2013}. Various algorithms have been designed to improve this procedure by incorporating additional information, including the minimum intensity or intensity variance of a particle stream \cite{Tian2010, Wu2013}, maximum gradient of particle edges \cite{Yang2012}, and combinations of these criteria \cite{Gao2013, Huang2021}. Nevertheless, improvements are limited and large uncertainties remain prevalent in PTV experiments that require refocusing. We demonstrate our technique on DIH data that is processed with a refocusing algorithm, but the method generalizes to triangulation methods.\par

Once the tracer particles have been localized, tracking is performed to link them across two or more sequential frames to form a Lagrangian track and compute particle displacements. This is done by algorithms that minimize a cost function designed to be low for tracks that faithfully follow a single particle and high for tracks that are incomplete or contain an erroneous match. The simplest strategy is to link particles with the smallest displacement between frames, called ``nearest neighbor'' tracking. However, this approach can only be used in scenarios where particle displacements are much smaller than particle spacing. In other words, the performance of nearest neighbor tracking is poor at high seeding densities \cite{Malik1993}. To solve this issue, researchers have developed an array of algorithms that exploit known properties of velocity fields. Examples include multi-frame schemes that promote \textit{temporal} smoothness by penalizing improbably large acceleration events \cite{Malik1993, Dracos1996} and \textit{spatial} smoothness by penalizing nearby tracks with divergent shapes \cite{Baek1996}. Recent PTV algorithms have also combined the tracking and localization steps to enhance the accuracy of both, as is the case in the STB method \cite{Schanz2016}. STB is tailor-made for triangulation from Mie scattering images and has been successfully deployed to measure densely-seeded turbulent flows \cite{Schroder2015, Schroder2018}. Another common scheme is Crocker--Grier tracking \cite{Crocker1996}, which is agnostic to the localization technique (unlike STB) and has been utilized to track a wide array of objects including micro-organisms \cite{Mallery2019}, cells \cite{Cheong2009}, and colloidal particles \cite{Roller2018} in fluid dynamics and biology research. Taking the Crocker--Grier approach, particle positions are linked across time but not corrected, meaning that significant localization errors may persist in the tracks.\par

Lagrangian particle velocity data is spatially sparse and PTV practicioners usually use the particle tracks to estimate Eulerian flow fields. Early conversion techniques filled the gaps between tracks with na{\"i}ve interpolation \cite{Agui1987}, which is prone to oversmoothing in some cases and overfitting in others, leading to large errors \cite{Casa2013}. Most PTV algorithms attempt to reduce these errors by incorporating the governing physics into the reconstruction procedure. This approach yields an optimization problem which seeks to satisfy experimental data and known flow physics, often using conventional CFD methods, termed data assimilation (DA). Two popular DA techniques for PTV are FlowFit \cite{Gesemann2016} and Vortex-in-Cell+ (VIC+) \cite{Schneiders2016}, both of which (partially) incorporate the Navier--Stokes equations to improve the accuracy and resolution of velocity field estimates. These 3D3C fields are post-processed to determine the pressure field by (1) direct integration of pressure gradients along one or more paths \cite{Baur1999, Liu2006} or (2) solving a global pressure Poisson equation \cite{Gurka1999, DeKat2012}. DA algorithms that sequentially estimate velocity and pressure normally feature a complex numerical scheme and can only partially assimilate the governing physics. Moreover, the localization and tracking errors from previous steps tend to propagate through the DA algorithm, leading to significant errors in the Eulerian fields.\par

In this work, we introduce a novel framework for PTV DA: \textit{particle advection velocimetry} (PAV). Observed particle tracks are compared to synthetic tracks obtained by advecting particles with the current estimate of the velocity field, which facilitates large timesteps and non-ideal tracers (inertial particles). Tremendous effort has been devoted to uncertainty quantification (UQ) for PTV systems \cite{Mallery2019, Hall2017, Gao2013, Bhattacharya2020}. We thus derive and demonstrate a statistical approach to PAV called stochastic PAV (SPAV), which accounts for arbitrary localization uncertainties. Our method provides a simple and robust framework to take advantage of existing PTV UQ results to improve reconstructions. We utilize the SPAV data loss to simultaneously estimate velocity and pressure fields from error-laden particle tracks. While SPAV may be implemented in conjunction with most PTV DA techniques, we employ a physics-informed neural network (PINN) \cite{Raissi2019} in this work. SPAV can accommodate any PTV modality, and we demonstrate it with simulated and experimental DIH-PTV examples, showing marked improvements over conventional techniques at a reasonable computational cost.\par

\section{Particle advection velocimetry}
\label{sec:method}
\begin{figure}[t]
    \centering
    \subcaptionbox[t]{\label{fig:method:conventional}}{
        \includegraphics[height=1.2in]{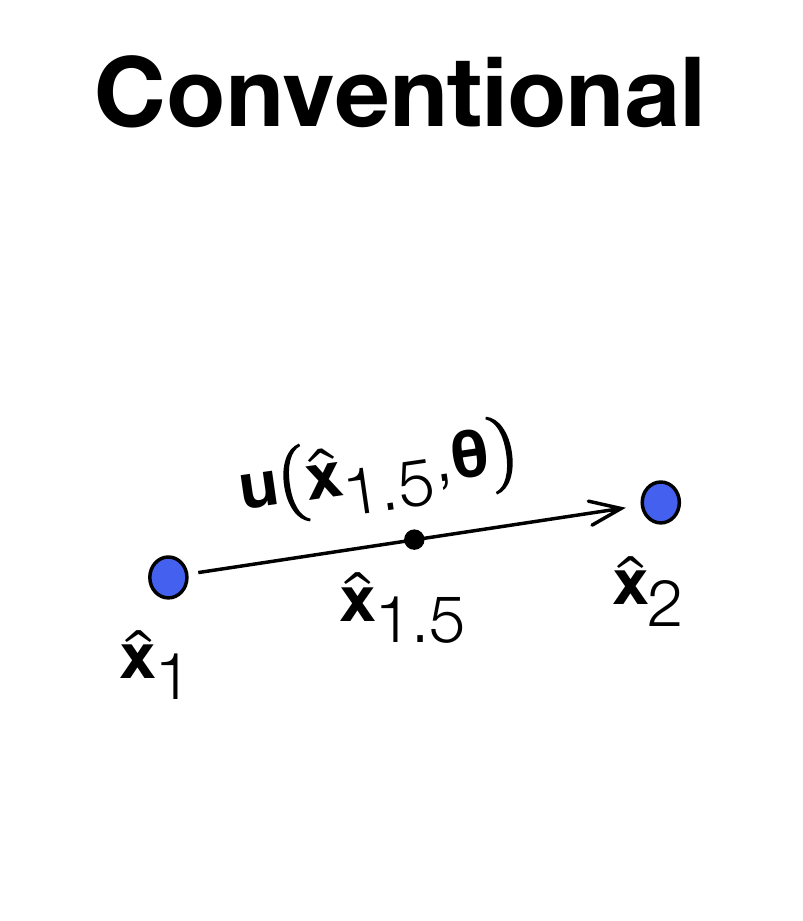}}\quad
    \subcaptionbox{\label{fig:method:PAV}}{
        \includegraphics[height=1.2in]{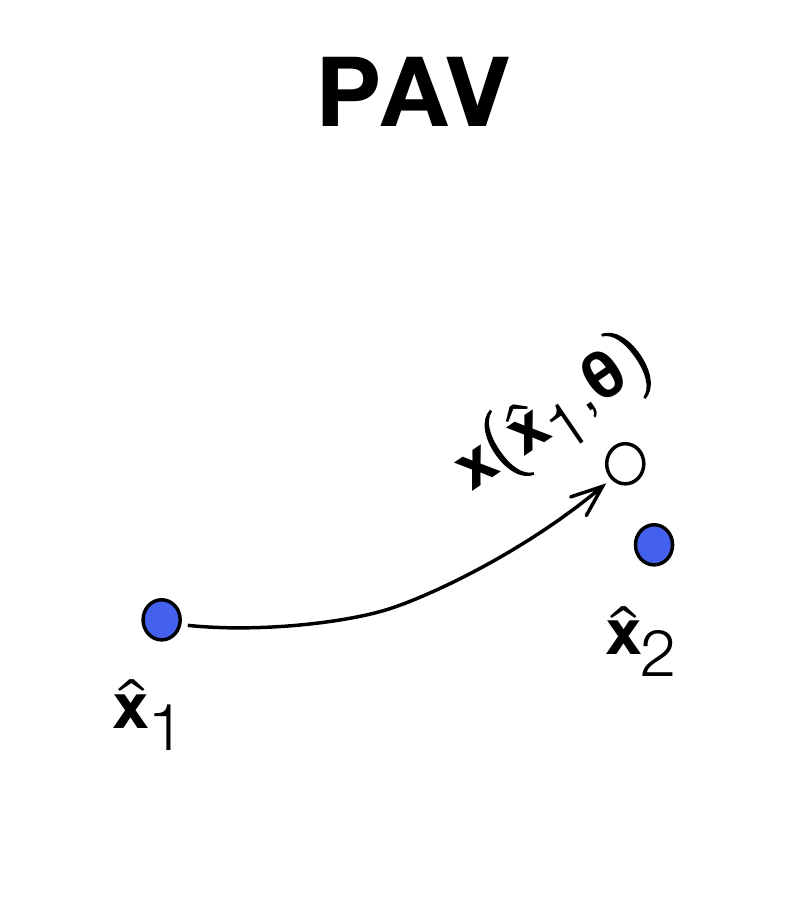}}\quad
    \subcaptionbox{\label{fig:method:SPAV-MC}}{
        \includegraphics[height=1.2in]{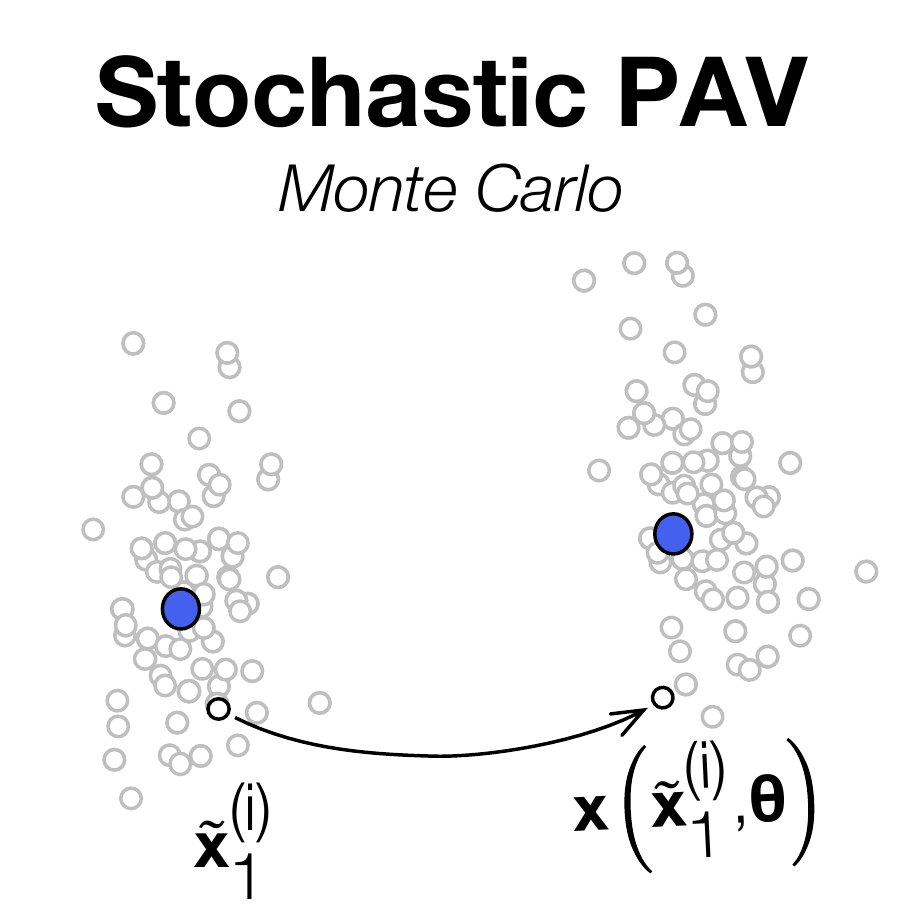}}\quad
    \subcaptionbox{\label{fig:method:SPAV-MVN}}{
        \includegraphics[height=1.2in]{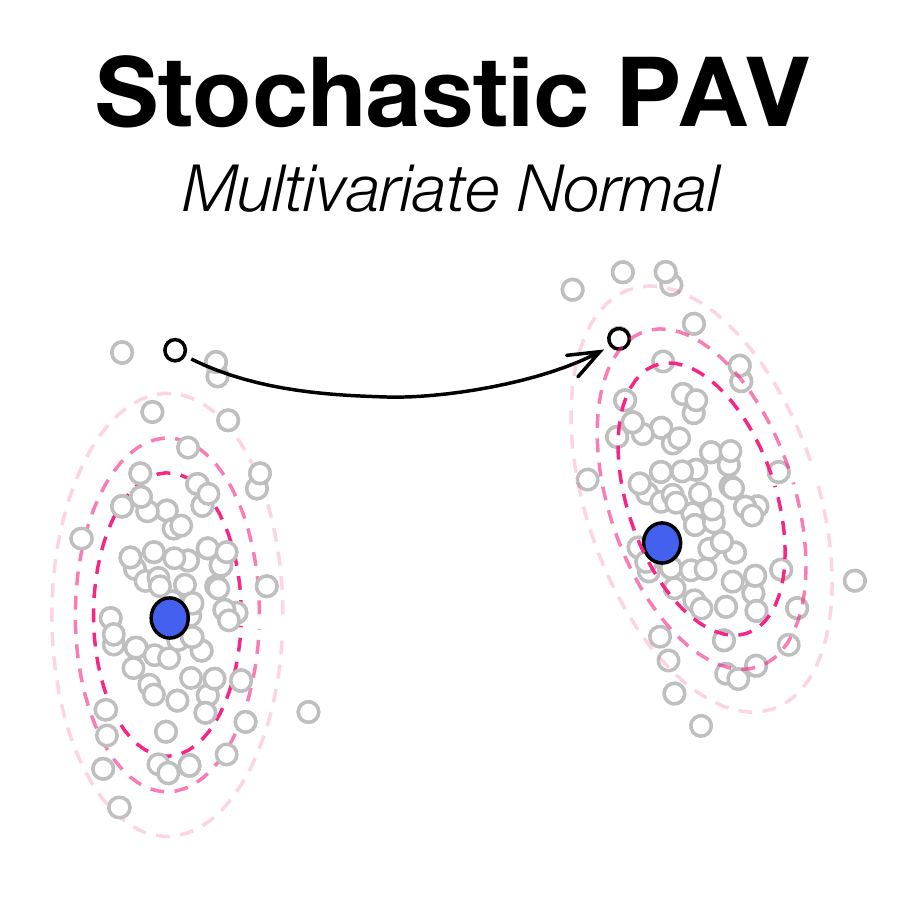}}\quad
    \subcaptionbox{\label{fig:method:SPAV-FE}}{
        \includegraphics[height=1.2in]{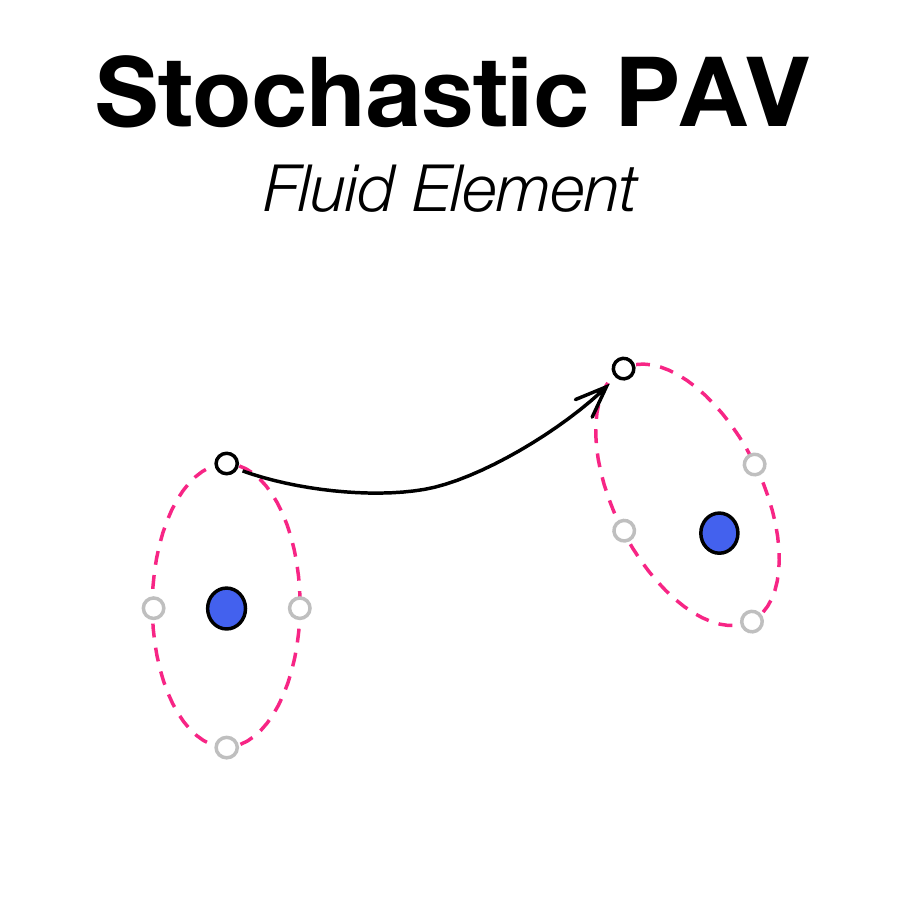}}
    \caption{Data loss terms for PTV: (a) conventional velocimetry, (b) PAV, (c) SPAV using Monte Carlo simulation, (d) SPAV using an MVN fit, (e) SPAV using an advected fluid element. Large blue dots indicate measured positions, small black and grey dots indicate samples from a particle position PDF.}
    \label{fig:method}
\end{figure}

The goal of PTV DA is to reconstruct Eulerian velocity and pressure fields that are consistent with Lagrangian particle tracks extracted from the image data and the equations governing fluid dynamics. Incorporating physical measurements into a numerical simulation requires a model of the measurement procedure. The fidelity of reconstructions is contingent upon the accuracy of this model, and this section describes an improved framework for modeling PTV measurements in a DA algorithm.  We utilize a PINN for PTV DA, but there are many alternative techniques 
like adjoint--variational and observer-based algorithms that can be combined with SPAV. These and other DA methods are discussed in \ref{app:DA}. To begin, we review the conventional objective loss for PTV DA, followed by a presentation of our PAV and SPAV losses. A pictorial diagram of these techniques can be seen in Fig.~\ref{fig:method}. This section concludes with a brief description of PINNs applied to PTV DA.\par

\subsection{Conventional data loss}
\label{sec:method:conv}
The vast majority of DA algorithms for PTV employ a displacement-based velocity estimate in the data loss term, e.g., \cite{Gesemann2016, Schneiders2016, Wang2022a, Han2021, Soto2022, Cai2022}, as illustrated in Fig.~\ref{fig:method:conventional}. The velocity of a particle is calculated from its displacement between two frames,
\begin{equation}
    \hat{\mathbf{u}}_{1.5} \approx \frac{\hat{\mathbf{x}}_2 - \hat{\mathbf{x}}_1}{\Delta t},
    \label{equ:velocity estimate}
\end{equation}
where $\mathbf{x} = [x,y,z]^\mathrm{T}$ and $\mathbf{u} = [u,v,w]^\mathrm{T}$ are position and velocity vectors, $\hat{(\cdot)}$ indicates a measured or estimated quantity, and $\Delta t$ is the interval between frame one and two. Subscripts in Eq.~\eqref{equ:velocity estimate} indicate the timestep, and the velocity is imputed to the particle at the midpoint, $\hat{\mathbf{x}}_{1.5} = (\hat{\mathbf{x}}_{2} - \hat{\mathbf{x}}_{1})/2$. The displacement-based (a.k.a. ``conventional'') data loss for velocimetry compares velocity estimates from Eq.~\eqref{equ:velocity estimate} to the output of a model,
\begin{equation}
    \mathcal{L}_\mathrm{data}^\mathrm{conv} = \frac{1}{N} \sum_{i=1}^N \left\lVert \hat{\mathbf{u}}_{1.5}^{(i)} - \mathbf{u}\mathopen{}\left(\hat{\mathbf{x}}_{1.5}^{(i)}, \boldsymbol\uptheta\right) \right\rVert_2^2,
    \label{equ:data loss:displacement}
\end{equation}
where the index $i$ loops over $N$ particle pairs and $\mathbf{u}(\mathbf{x}, \boldsymbol\uptheta)$ is the velocity field at $\mathbf{x}$ outputted by a numerical model. This model is parameterized by the vector $\boldsymbol\uptheta$, which could signify weights and biases in a PINN, discrete velocity and pressure data corresponding to a CFD mesh, etc. The superscript ``conv'' indicates the conventional data loss for PTV DA.\par

\subsection{Vanilla PAV}
\label{sec:method:PAV}
In ``particle advection velocimetry'', we replace the velocity-wise comparison in Eq.~\eqref{equ:data loss:displacement} with a position-wise comparison. Assuming the particle is an ideal tracer (i.e., it faithfully follows the flow), the advected position may be calculated by integrating the estimated velocity field,
\begin{equation}
    \frac{\mathrm{d}\mathbf{x}}{\mathrm{d}t} = \mathbf{u}\mathopen{} \;\Longleftrightarrow\;
    \mathbf{x}_2\mathopen{}\left(\mathbf{x}_1, \boldsymbol\uptheta\right) = \int_{t_1}^{t_2} \mathbf{u}\mathopen{}\left[\mathbf{x}\mathopen{}\left(t\right), \boldsymbol\uptheta\right] \mathrm{d}t + \mathbf{x}_1,
    \label{equ:particle advection:tracer}
\end{equation}
where $\mathbf{x}(t)$ is the solution to the differential equation on the left side of Eq.~\eqref{equ:particle advection:tracer} initialized at $\mathbf{x}_1$. Our PAV data loss sums the distance between tracked particle positions in frame two, denoted $\hat{\mathbf{x}}_2$, and the positions obtained by advecting particles from the tracked location in frame one, $\hat{\mathbf{x}}_1$, via the current estimate of the velocity field, $\boldsymbol\uptheta$:
\begin{equation}
    \mathcal{L}_\mathrm{data}^\mathrm{PAV} = \frac{1}{N} \sum_{i=1}^N \left\lVert \hat{\mathbf{x}}_2^{(i)} - \mathbf{x}_2\mathopen{}\left(\hat{\mathbf{x}}_1^{(i)}, \boldsymbol\uptheta\right) \right\rVert_2^2.
    \label{equ:data loss:advection}
\end{equation}
This procedure is depicted in Fig.~\ref{fig:method:PAV}. Equation~\eqref{equ:particle advection:tracer} may be computed using a variety of numerical schemes like Runge--Kutta methods. Moreover, while Eq.~\eqref{equ:particle advection:tracer} applies to ideal tracer particles that perfectly follow the flow, non-ideal effects like drag or thermophoresis may be included by explicitly modeling all the forces acting on a particle,
\begin{equation}
    m\frac{\mathrm{d}^2\mathbf{x}}{\mathrm{d}t^2} = \mathbf{f}_\mathrm{p}.
    \label{equ:particle advection:inertial particle}
\end{equation}
In this expression, $\mathbf{f}_\mathrm{p}$ is the net force (comprising drag, buoyancy, etc.) on a particle of mass $m$. Implementing a PAV loss based on Eq.~\eqref{equ:particle advection:inertial particle} can potentially enable the reconstruction of Eulerian fields from PTV tracks in new physical regimes.\par

\subsection{Stochastic PAV}
\label{sec:method:SPAV}
The PAV data loss in Eq.~\eqref{equ:data loss:advection} assumes that the particle positions $\hat{\mathbf{x}}_1$ and $\hat{\mathbf{x}}_2$ have been faithfully recovered. In reality, these quantities are often subject to large, anisotropic errors that depend on the imaging setup and particle tracking algorithms. Localization errors act as noise, limiting the degree to which the physical model can be optimized (via a phyiscs loss) if the positions are treated as a known quantity. Moreover, anisotropic errors may lead to biased velocity fields. These problems also apply to the conventional velocity loss in Eq.~\eqref{equ:data loss:displacement}, since both $\hat{\mathbf{u}}_{1.5}$ and $\hat{\mathbf{x}}_{1.5}$ are adversely affected by localization errors. Therefore, an alternative data loss that accounts for measurement uncertainty is needed.\par

Stochastic PAV is based on the chance of measuring the particle positions $\hat{\mathbf{x}}_1$ and $\hat{\mathbf{x}}_2$ subject to the velocity field given by $\boldsymbol\uptheta$ and the measurement uncertainty. This corresponds to a likelihood probability density function (PDF),
\begin{equation}
    P\mathopen{}\left(\hat{\mathbf{x}}_2| \hat{\mathbf{x}}_1, \boldsymbol\uptheta\right) = \int
    P\mathopen{}\left(\hat{\mathbf{x}}_2| \mathbf{x}_1, \boldsymbol\uptheta\right)
    P\mathopen{}\left(\mathbf{x}_1| \hat{\mathbf{x}}_1\right) \mathrm{d}\mathbf{x}_1,
    \label{equ:likelihood}
\end{equation}
which contains two key PDFs. First, $P(\mathbf{x}_1| \hat{\mathbf{x}}_1)$ is an error model, which describes the probability of a particle being located at $\mathbf{x}_1$ if it was measured at $\hat{\mathbf{x}}_1$. Second, $P(\hat{\mathbf{x}}_2| \mathbf{x}_1, \boldsymbol\uptheta)$ describes the chance of measuring the particle at $\hat{\mathbf{x}}_2$ for a known starting point, $\mathbf{x}_1$, and the velocity field given by $\boldsymbol\uptheta$. The latter PDF contains an error model as well as the advection model from Eq.~\eqref{equ:particle advection:tracer} or \eqref{equ:particle advection:inertial particle}. Equation~\eqref{equ:likelihood} ought to be maximized, so the SPAV data loss, which must be minimized, comprises the aggregate negative log likelihood of the tracked particles,
\begin{equation}
    \mathcal{L}_\mathrm{data}^\mathrm{SPAV} = -\sum_{i=1}^N \log\mathopen{}\left[P\mathopen{}\left(\hat{\mathbf{x}}_2^{(i)}\Big| \hat{\mathbf{x}}_1^{(i)}, \boldsymbol\uptheta\right)\right].
    \label{equ:data loss:stochastic}
\end{equation}
Implementing this loss requires a model of localization errors. It should be noted that $P(\hat{\mathbf{x}}_2| \hat{\mathbf{x}}_1, \boldsymbol\uptheta)$ can be augmented to include the probability of tracking errors, ghost particles, and the like.\par

While any error model can be incorporated into Eq.~\eqref{equ:likelihood}, we illustrate SPAV with the widely-applicable multivariate normal (MVN) model,
\begin{equation}
    P\mathopen{}\left(\mathbf{x}|\hat{\mathbf{x}}\right) = \det\mathopen{}\left(2\pi\boldsymbol\Gamma\right)^{-1/2} \exp\mathopen{}\left[-\frac{1}{2}\left(\mathbf{x} - \hat{\mathbf{x}}\right)^\mathrm{T} \boldsymbol\Gamma^{-1} \left(\mathbf{x} - \hat{\mathbf{x}}\right)\right],
    \label{equ:error model}
\end{equation}
where $\mathbf{x}$ is the (unknown) true particle position, $\hat{\mathbf{x}}$ is the measured position, and $\boldsymbol\Gamma$ is a covariance matrix which describes the magnitude and orientation of measurement errors. Many realistic localization errors can be approximated using an MVN PDF. Conveniently, $P(\mathbf{x}_1| \hat{\mathbf{x}}_1)$ can be directly computed using Eq.~\eqref{equ:error model}. The advected particle PDF is obtained by tracing a particle from $\mathbf{x}_1$ to $\mathbf{x}_2$ with Eq.~\eqref{equ:particle advection:tracer} or \eqref{equ:particle advection:inertial particle}, $P(\mathbf{x}_2) = P[\mathbf{x}_2(\mathbf{x}_1,\boldsymbol\uptheta)]$, such that
\begin{equation}
    P\mathopen{}\left(\hat{\mathbf{x}}_2| \mathbf{x}_1, \boldsymbol\uptheta\right) = P\mathopen{}\left[\hat{\mathbf{x}}_2|
    \mathbf{x}_2\mathopen{}\left(\mathbf{x}_1, \boldsymbol\uptheta\right)\right].
    \label{equ:advected particle PDF}
\end{equation}
For MVN errors, this calculation may be performed using Eq.~\eqref{equ:error model} because the MVN model is symmetric, i.e., $P(\mathbf{x}|\hat{\mathbf{x}}) = P(\hat{\mathbf{x}}|\mathbf{x})$. Otherwise, separate expressions need to be provided for the ``localization PDF'', $P(\mathbf{x}|\hat{\mathbf{x}})$, and the ``measurement PDF'', $P(\hat{\mathbf{x}}|\mathbf{x})$. These functions must be obtained through a numerical or experimental UQ procedure \cite{Mallery2019, Hall2017, Gao2013, Bhattacharya2020}.\par

Unfortunately, the SPAV likelihood is a nonlinear function of $\boldsymbol\uptheta$, even when an MVN distribution is employed to model the localization errors. Therefore, $P(\hat{\mathbf{x}}_2| \hat{\mathbf{x}}_1, \boldsymbol\uptheta)$ must be approximated using a numerical technique. We devised three approximations that are suitable for SPAV.\par

\subsubsection{Monte Carlo sampling}
\label{sec:method:SPAV:MC}
The most precise method for calculating the SPAV likelihood is Monte Carlo simulation \cite{Mooney1997}, as shown in Fig.~\ref{fig:method:SPAV-MC}. This is done by sampling particle locations from the localization PDF at frame one, advecting those samples over the measurement interval, and computing the average measurement loss for the tracked positions at frame two. Formally,
\begin{equation}
    P\mathopen{}\left(\hat{\mathbf{x}}_2| \hat{\mathbf{x}}_1, \boldsymbol\uptheta\right) \approx \frac{1}{M} \sum_{j=1}^M P\mathopen{}\left(\hat{\mathbf{x}}_2\Big| \tilde{\mathbf{x}}_1^{(j)}, \boldsymbol\uptheta\right),
    \label{equ:MC approximation}
\end{equation}
where $M$ samples of $\mathbf{x}_1$, denoted $\tilde{\mathbf{x}}_1$, are drawn from $P(\mathbf{x}_1| \hat{\mathbf{x}}_1)$; $P(\hat{\mathbf{x}}_2| \tilde{\mathbf{x}}_1, \boldsymbol\uptheta)$ is given by the measurement PDF, $P(\hat{\mathbf{x}}_2|\mathbf{x}_2)$, evaluated in terms of the advected position, $\mathbf{x}_2(\tilde{\mathbf{x}}_1, \boldsymbol\uptheta)$. While this produces an accurate likelihood PDF for large values of $M$, it may also be cost-intensive. The high cost per particle limits the batch size used in training, which can slow or altogether prevent the progress of an optimization algorithm.\par

\subsubsection{Multivariate normal approximation}
\label{sec:method:SPAV:MVN}
If an MVN error model is suitable and the measurement interval is short, then the distribution of advected particles can be represented using a rotated and sheared MVN distribution as shown in Fig.~\ref{fig:method:SPAV-MVN}. To do this, samples of $\mathbf{x}_1$ are drawn from $P(\mathbf{x}_1| \hat{\mathbf{x}}_1)$ and advected using Eq.~\eqref{equ:particle advection:tracer} or \eqref{equ:particle advection:inertial particle}. The mean and covariance of the resulting (presumed) MVN distribution are
\begin{subequations}
    \begin{align}
        \hat{\boldsymbol\upmu}_2 &= \frac{1}{M} \sum_{j=1}^{M} \mathbf{x}_2\mathopen{}\left(\tilde{\mathbf{x}}_1^{(j)}, \boldsymbol\uptheta\right) \quad\text{and} \label{equ:MVN estimates:mean}\\
        \hat{\boldsymbol\Gamma}_2 &= \frac{1}{M} \sum_{j=1}^{M}
        \left[\mathbf{x}_2\mathopen{}\left(\tilde{\mathbf{x}}_1^{(j)}, \boldsymbol\uptheta\right) - \hat{\boldsymbol\upmu}_2\right] \left[\mathbf{x}_2\mathopen{}\left(\tilde{\mathbf{x}}_1^{(j)}, \boldsymbol\uptheta\right) - \hat{\boldsymbol\upmu}_2\right]^\mathrm{T}, \label{equ:MVN estimates:covariance}
    \end{align}
    \label{equ:MVN estimates}%
\end{subequations}
such that the advected particle PDF is
\begin{equation}
    P\mathopen{}\left(\mathbf{x}_2| \hat{\mathbf{x}}_1, \boldsymbol\uptheta\right) = \det\mathopen{}\left(2\pi\hat{\boldsymbol\Gamma}_2\right)^{-1/2} \exp\mathopen{}\left[-\frac{1}{2}\left(\mathbf{x}_2 - \hat{\boldsymbol\upmu}_2\right)^\mathrm{T} \hat{\boldsymbol\Gamma}_2^{-1} \left(\mathbf{x}_2 - \hat{\boldsymbol\upmu}_2\right)\right].
    \label{equ:advected MVN distribution}
\end{equation}
This expression can be utilized to calculate the SPAV PDF,
\begin{equation}
    P\mathopen{}\left(\hat{\mathbf{x}}_2| \hat{\mathbf{x}}_1, \boldsymbol\uptheta\right) = \int P\mathopen{}\left(\hat{\mathbf{x}}_2| \mathbf{x}_2\right) P\mathopen{}\left(\mathbf{x}_2| \hat{\mathbf{x}}_1, \boldsymbol\uptheta\right) \mathrm{d}\mathbf{x}_2,
    \label{equ:convolution}
\end{equation}
where $P(\hat{\mathbf{x}}_2| \mathbf{x}_2)$ is given by Eq.~\eqref{equ:error model} because the symmetric MVN error model has been assumed. The convolution of two MVN distributions in Eq.~\eqref{equ:convolution} has an exact expression,
\begin{equation}
    P\mathopen{}\left(\hat{\mathbf{x}}_2| \hat{\mathbf{x}}_1, \boldsymbol\uptheta\right) =
    \det\mathopen{}\left[2\pi\left(\boldsymbol\Gamma + \hat{\boldsymbol\Gamma}_2\right) \right]^{-1/2} \exp\mathopen{}\left[-\frac{1}{2}\left(\hat{\mathbf{x}}_2 - \hat{\boldsymbol\upmu}_2\right)^\mathrm{T} \left(\boldsymbol\Gamma + \hat{\boldsymbol\Gamma}_2\right)^{-1} \left(\hat{\mathbf{x}}_2 - \hat{\boldsymbol\upmu}_2\right)\right],
    \label{equ:MVN approximation}
\end{equation}
where $\boldsymbol\Gamma$ is the generic measurement uncertainty from Eq.~\eqref{equ:error model}. This approximation is more stable than Monte Carlo simulation, but the cost is not significantly lower since many samples are needed for estimates of $\hat{\boldsymbol\upmu}_2$ and $\hat{\boldsymbol\Gamma}_2$ to converge. Moreover, the advected distribution is not necessarily MVN when there is significant shear or rigid body rotation along the particle's path. As a result, there is a complex trade-off between the increased efficiency and reduced accuracy of the MVN approximation.\par

\subsubsection{Fluid element approximation}
\label{sec:method:SPAV:FE}
An even cheaper MVN approximation may be realized by advecting an ellipsoidal fluid element that comprises six points; a four-point, 2D example of this is illustrated in Fig.~\ref{fig:method:SPAV-FE}. The points are located along the principle axes of $\boldsymbol\Gamma$, which may be extracted through an eigenvalue decomposition,
\begin{equation}
    \mathbf{Q}\Lambda\mathbf{Q}^\mathrm{T} = \boldsymbol\Gamma,
    \label{equ:eigenvalue decomposition}
\end{equation}
where $\mathbf{Q} = [\mathbf{q}_1, \mathbf{q}_2, \mathbf{q}_3]$ contains the principle directions of $\boldsymbol\Gamma$ and $\boldsymbol\Lambda = \mathrm{diag}([\lambda_1^2, \lambda_2^2, \lambda_3^2])$ contains the corresponding variances. A natural choice is to place the points at $\hat{\mathbf{x}}_1 \pm \lambda_1\mathbf{q}_1$, $\hat{\mathbf{x}}_1 \pm \lambda_2\mathbf{q}_2$, and $\hat{\mathbf{x}}_1 \pm \lambda_3\mathbf{q}_3$ and advect them to the next measurement interval,
\begin{equation}
    \mathbf{x}_2^{+1} = \mathbf{x}_2\mathopen{}\left(\hat{\mathbf{x}}_1 + \lambda_1\mathbf{q}_1, \boldsymbol\uptheta\right),\;
    \mathbf{x}_2^{-1} = \mathbf{x}_2\mathopen{}\left(\hat{\mathbf{x}}_1 - \lambda_1\mathbf{q}_1, \boldsymbol\uptheta\right),\;
    \mathbf{x}_2^{+2} = \mathbf{x}_2\mathopen{}\left(\hat{\mathbf{x}}_1 + \lambda_2\mathbf{q}_2, \boldsymbol\uptheta\right),\; \dots
    \label{equ:element advection}
\end{equation}
and the advected centroid is taken to be the mean,
\begin{equation}
    \hat{\boldsymbol\upmu}_2 = \frac{1}{6}\left(\mathbf{x}_2^{+1} + \mathbf{x}_2^{-1} + \mathbf{x}_2^{+2} + \mathbf{x}_2^{-2} + \mathbf{x}_2^{+3} + \mathbf{x}_2^{-3}\right).
    \label{equ:element mean}
\end{equation}
Next, we collect the advected vertices into a matrix, center them, and perform a singular value decomposition,
\begin{equation}
    \mathbf{U}\Sigma\mathbf{V}^\mathrm{T} = \left[\mathbf{x}_2^{+1}, \mathbf{x}_2^{-1}, \mathbf{x}_2^{+2}, \mathbf{x}_2^{-2}, \mathbf{x}_2^{+3}, \mathbf{x}_2^{-3}\right] - \hat{\boldsymbol\upmu}_2\mathbf{1}^\mathrm{T},
    \label{equ:SVD}
\end{equation}
where $\mathbf{1}$ is a $6\times1$ vector of ones. Lastly, these elements are used to estimate the advected covariance matrix,
\begin{equation}
    \hat{\boldsymbol\Gamma}_2 = \frac{1}{2}\mathbf{U}\Sigma(\mathbf{U}\Sigma)^\mathrm{T}.
    \label{equ:element covariance}
\end{equation}
Given $\hat{\boldsymbol\upmu}_2$ and $\hat{\boldsymbol\Gamma}_2$ from Eqs.~\eqref{equ:element mean} and \eqref{equ:element covariance}, Eq.~\eqref{equ:MVN approximation} may be used to compute the SPAV likelihood PDF. This procedure is less accurate than the Monte Carlo or sample-based MVN techniques, but it is also hundreds to thousands of times cheaper, dramatically increasing the maximum batch size.\par

It should be noted that the fluid element technique can be conducted using multiple ellipsoids with distinct radii. These ellipsoids could be weighted by distance from the centroid, which may help to stabilize the technique in regions of high shear and/or rotation.\par

\subsection{Velocimetry with a PINN}
\label{sec:method:PINN}
\begin{figure}[ht]
    \centering
    \includegraphics[width=.95\textwidth]{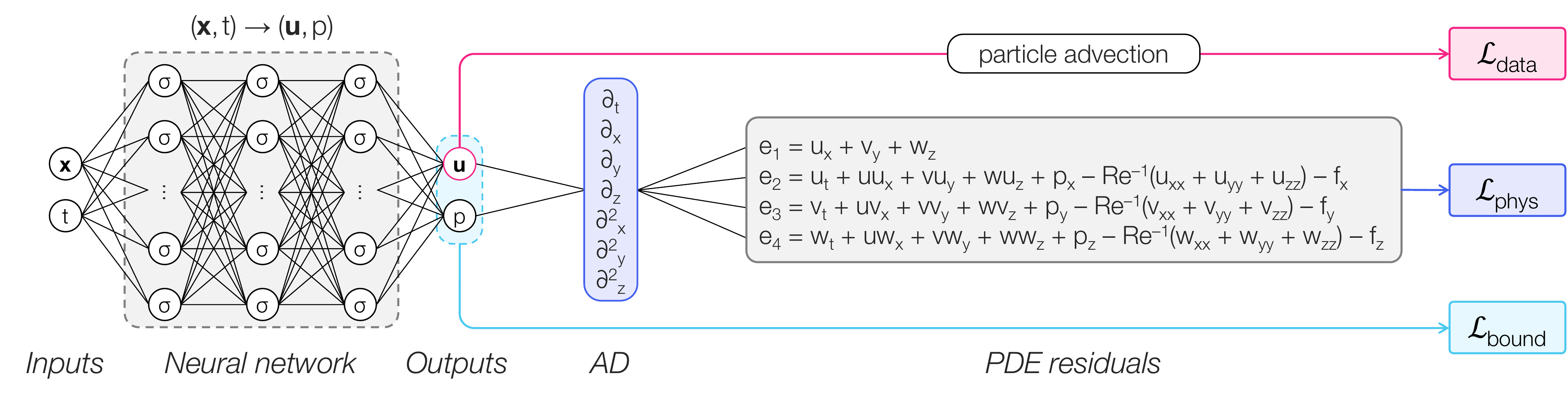}
    \vspace*{2mm}
    \caption{Architecture of a PINN for velocimetry. The neural network constitutes a functional representation of the flow, which can be used to compute data, physics, and boundary condition losses. These losses are added up and minimized via backpropagation to obtain a realistic function.}
    \label{fig:PINN architecture}
\end{figure}

The above conventional, PAV, and SPAV data losses are readily implemented in a PINN-based framework for DA.\footnote{See \ref{app:DA} for an overview of alternative DA techniques, most of which are compatible with SPAV.} Physics-informed neural networks are deep, feedforward neural networks that can approximately solve a forward or inverse problem that is governed by differential equations \cite{Raissi2019}. Figure~\ref{fig:PINN architecture} shows a PINN set up for velocimetry. In effect, the network is a functional representation of the flow, mapping spatio-temporal inputs to velocity and pressure outputs, $(x, y, z, t) \rightarrow (u, v, w, p)$. Additional output fields may be added as necessary. Partial derivatives of the PINN are efficiently computed using automatic differentiation, and these quantities are plugged into the governing equations. Residuals from this procedure are added up in a ``physics loss'', which must be minimized to obtain a physically plausible function. Similarly, synthetic measurements can be computed using the outputted flow fields and a measurement model and then compared to real data, e.g., using any of the $\mathcal{L}_\mathrm{data}$ expressions in Eqs.~\eqref{equ:data loss:displacement}, \eqref{equ:data loss:advection}, and \eqref{equ:data loss:stochastic}, each of which constitutes a ``data loss''. An aggregate loss, physics + data, is minimized by a backpropagation algorithm to produce a network that yields physical flow fields that reproduce the observed particle trajectories.\par

For a simple PTV scenario, a suitable physics loss can be constructed from the incompressible Navier--Stokes equations, although the technology is easily extended to compressible and reactive flows. These equations are written in non-dimensional form and rearranged to obtain physics residuals,
\begin{subequations}
    \begin{align}
        \delta_1 &= u_x + v_y + w_z, \label{equ:physics residuals:continuity}\\
        \delta_2 &= u_t + u\,u_x + v\,u_y + w\,u_z + p_x - Re^{-1} \left(u_{xx} + u_{yy} + u_{zz} \right) - f_\mathrm{x}, \label{equ:physics residuals:x mom}\\
        \delta_3 &= u_t + u\,v_x + v\,v_y + w\,v_z + p_y - Re^{-1} \left(v_{xx} + v_{yy} + v_{zz} \right) - f_\mathrm{y}, \quad\text{and} \label{equ:physics residuals:y mom}\\
        \delta_4 &= u_t + u\,w_x + v\,w_y + w\,w_z + p_z - Re^{-1} \left(w_{xx} + w_{yy} + w_{zz} \right) - f_\mathrm{z}, \label{equ:physics residuals:z mom}
    \end{align}
    \label{equ:physics residuals}%
\end{subequations}
where $(\cdot)_x$, $(\cdot)_y$, $(\cdot)_z$, and $(\cdot)_t$ are partial derivatives of the PINN, $Re$ is the Reynolds number, and $\mathbf{f} = [f_\mathrm{x}, f_\mathrm{y}, f_\mathrm{z}]^\mathrm{T}$ is a density-normalized forcing term that is generally set to zero. A physics loss is constructed by integrating Eq.~\eqref{equ:physics residuals} over the measurement domain,
\begin{equation}
    \mathcal{L}_\mathrm{phys} = \frac{1}{\left|\mathcal{V} \times \mathcal{T}\right|}  \int_\mathcal{T} \iiint_\mathcal{V} \left\lVert \left[\delta_1, \delta_2, \delta_3, \delta_4\right]^\mathrm{T} \right\rVert_2^2 \mathrm{d}\mathcal{V} \,\mathrm{d}t.
    \label{equ:physics loss}
\end{equation}
In this expression, $\mathcal{V}$ and $\mathcal{T}$ are the measurement volume and interval, $\lVert\cdot\rVert_2$ is the Euclidean norm, and the residuals $\delta_1$--$\delta_4$ are a function of $(x,y,z,t)$ as well as the network's weights and biases, represented by the vector $\boldsymbol\uptheta$. In practice, Eq.~\eqref{equ:physics loss} is approximated by Monte Carlo sampling.\par

The final loss that we consider is a boundary condition. One major advantage of PINNs when reconstructing a flow is that boundary conditions are not always needed. However, adding a boundary condition can improve the accuracy of reconstructions when the condition is well known and has a significant impact on the phenomena of interest. Boundaries can be especially important in the context of PTV measurements of wall bounded flows because the assumption of ideal tracers may break down near the wall. We consider a surface boundary condition,
\begin{equation}
    \mathcal{L}_\mathrm{bound} = \frac{1}{\left|\mathcal{A} \times \mathcal{T}\right|}  \int_\mathcal{T} \iint_\mathcal{\mathcal{A}} \left\lVert \left[u, v, w\right]^\mathrm{T} \right\rVert_2^2 \mathrm{d}\mathcal{A} \,\mathrm{d}t,
    \label{equ:boundary loss}
\end{equation}
where the region $\mathcal{A}$ corresponds to a no-slip wall. This boundary condition is used for the synthetic and experimental boundary layer cases in Sect.~\ref{sec:DIH-PTV}.\par

An overall objective loss is formed using the physics loss, a suitable data loss, and a no-slip wall, where appropriate. In this article, we assess the conventional, displacement-based loss in Eq.~\eqref{equ:data loss:displacement} and our approximations to the SPAV loss defined by Eq.~\eqref{equ:data loss:stochastic}. The resultant loss is
\begin{equation}
    \mathcal{L}_\mathrm{total} = \mathcal{L}_\mathrm{data} + \gamma_\mathrm{phys} \,\mathcal{L}_\mathrm{phys} +
    \gamma_\mathrm{bound} \,\mathcal{L}_\mathrm{bound},
    \label{equ:total loss}
\end{equation}
where $\gamma_\mathrm{phys}$ and $\gamma_\mathrm{bound}$ are regularization parameters that must be carefully selected. The total loss is minimized by tuning the PINN's parameters, $\boldsymbol\uptheta$, with a backpropagation algorithm. Fields outputted by a trained PINN represent a balance between the PTV measurements and flow physics.\par

We stress that PINNs are not essential to our framework. SPAV can be used in conjunction with a physics loss to optimize any differentiable, parametric representation of a flow. Additional DA techniques are reviewed in \ref{app:DA}. Nonetheless, PINNs are a convenient tool for velocimetry because they provide a parsimonious model of the flow and can be implemented at a low computational cost, as we demonstrate in Sect.~\ref{sec:DIH-PTV}.\par

\section{Benchmarking SPAV data loss approximations}
\label{sec:validation}
The stochastic data loss defined in Eq.~\eqref{equ:data loss:stochastic} is a complex, nonlinear function of arbitrary localization and measurement PDFs as well as the flow field parameters, $\boldsymbol\uptheta$. We introduce three approximations to this loss, using Monte Carlo sampling, MVN PDFs, and an ellipsoidal fluid element, hereafter denoted ``MC'', ``MVN'', and ``FE'', in that order. In principle, the Monte Carlo technique will converge to the true SPAV loss with a large number of samples; the MVN model presumes Gaussian measurement and localization PDFs, which could reduce the computational cost of our technique; and in our FE approach, we accelerate this approximation by only advecting a few particles, positioned along the principal axes of the measurement PDF. Crucially, the number of samples needed for the MC and MVN methods and the veracity of the assumptions undergirding MVN and FE were not known a priori. Therefore, we assessed the accuracy and cost of these losses before using them for velocimetry.\par

We numerically test our SPAV approximations using a direct numerical simulation (DNS) from Raissi et al. \cite{Raissi2019}. The flow is a 2D cylinder wake flow with a Reynolds number of $100$. A subset of the flow with a non-dimensional size of $7 \times 5$ and 201 timesteps is extracted for testing. We randomly seed a thousand virtual particles at the initial frame. The particles are modeled as ideal tracers, governed by Eq.~\eqref{equ:particle advection:tracer}, which we implement using a second-order Runge--Kutta scheme and periodic boundaries. To mimic localization uncertainties that are characteristic of DIH-PTV, the observed particle positions are corrupted with additive Gaussian errors of zero mean. We assume a sensor positioned normal to the flow (aligned with the $y$-axis) and adopt the experimentally-determined uncertainties from Mallery et al. \cite{Mallery2019}, corresponding to non-dimensional standard deviations of $\sigma_{\mathrm{x}} = 2\times10^{-3}$ and $\sigma_{\mathrm{y}} = 5\times10^{-2}$ for our domain. These data are arranged into particle tracks: each track continues until the particle exits the domain, at which point a new track is formed at the opposite boundary.\par

Next, we train a PINN on the DNS data to obtain an idealized model of the wake flow. In other words, this PINN is not used to reconstruct the velocity or pressure fields but rather to assess our SPAV approximations. The deep, fully-connected network maps $(x,y,t)$ inputs to $(u,v,p)$ outputs with 15 layers having 250 neurons, each. Our data loss directly compares velocity outputs to the DNS solution, and we set $\gamma_\mathrm{phys}$ to unity in our objective loss. Training is conducted with the Adam optimizer using data and physics batch sizes of 10,000 points apiece; the optimizer is run at a fast learning rate of $10^{-3}$ for 5000 epochs, followed by a slow rate of $10^{-4}$ for another 5000 epochs, yielding a highly-accurate representation of the flow. Velocity fields outputted by this PINN exhibit a maximum error of $0.3\%$ relative to the DNS fields. We use the resultant architecture to evaluate our SPAV approximations with the noisy track data described above, meaning that $\boldsymbol\uptheta$ is fixed at a quasi-optimal point. Particle advection by the PINN is also implemented with a second-order Runge--Kutta method.\par

\begin{figure}[ht]
    \centering
    \includegraphics[height=8cm]{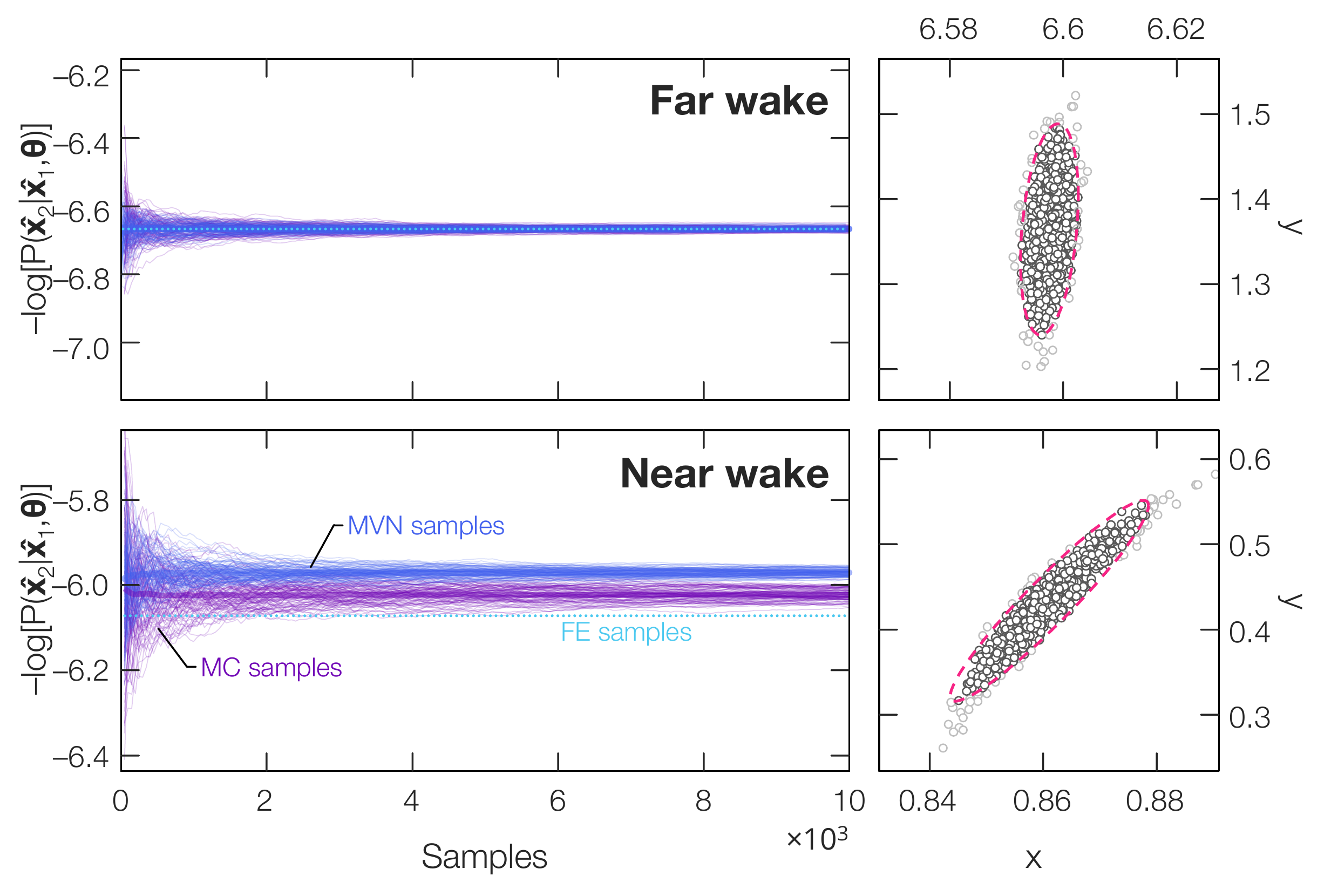}
    \vspace*{2mm}
    \caption{Testing the SPAV approximations: (left) log losses from all three variants and (right) advected particle distributions. The Gaussian distribution of particles becomes skewed in the near wake due to strong velocity gradients.}
    \label{fig:samples}
\end{figure}

Figure~\ref{fig:samples} shows a comparison of two single-particle MC, MVN and FE log losses at a single instance. These particles were selected from the near and far wake regions of the flow to illustrate how local flow characteristics could affect our approximations. For the MC and MVN losses, we conducted 100 tests with 10,000 random draws from the initial localization PDF in each case. The thin lines in Fig.~\ref{fig:samples} convey the results of one test, where the $x$-axis indicates the number of draws included in the log loss calculation. By contrast, the FE approximation requires only four particles in this 2D demonstration (six for 3D PTV), resulting in a constant estimate of the log loss. The right-hand side of Fig.~\ref{fig:samples} shows the advected distribution of particles in both cases. Naturally, the MC and MVN log loss estimates converge to a stable value with an increasing number of samples. However, it should be noted that convergence of the MVN distribution is barely faster than the MC method, which is an acute disadvantage since the MVN method relies upon strong assumptions about the measurement and localization PDFs. We also note the important trade-off between the number of samples, $M$, used to estimate the log loss for a single particle and the number of particles, $N$, used in a training batch, which are collectively limited by memory. This constraint is important because the loss calculations are less accurate for a low number of samples and stochastic gradient descent becomes less effective as the batch size decreases. For the tests reported in Sect.~\ref{sec:DIH-PTV}, we use $M = 1000$ draws and a batch size of $N = 100$ particles for both MC and MVN. This ratio was found to provide a good compromise between the accuracy of our SPAV approximations and stability of optimization.\par

Another detail that can be observed in Fig.~\ref{fig:samples} is the flow dependence of our approximations. The MC, MVN, and FE calculations agree in the far wake of the cylinder but differ in the near wake. Moreover, convergence (or lack thereof) corresponds to the spatial distribution of advected samples, which is clearly non-Gaussian in the near wake. The cloud of particles advected in this region is skewed and distorted by strong velocity gradients in the flow (n.b., these gradients are meager compared to those of even weakly turbulent flows). The banana-shaped near wake distribution of particles is poorly modeled by a Gaussian PDF, which introduces an error into the MVN estimate. The fluid element is adversely affected by outliers and, as a result, the FE log loss is the least accurate. However, these errors are not necessarily large, with average and maximum errors of 0.8\% and 3.3\% of the range of MC log losses for MVN and 1.1\% and 10.4\% for FE. Therefore, we assess the utility of these approximations in Sect.~\ref{sec:DIH-PTV}.\par

\begin{figure}[ht]
    \centering
    \subcaptionbox{\label{fig:normality:spatial distribution}}{
        \includegraphics[height=5.5cm]{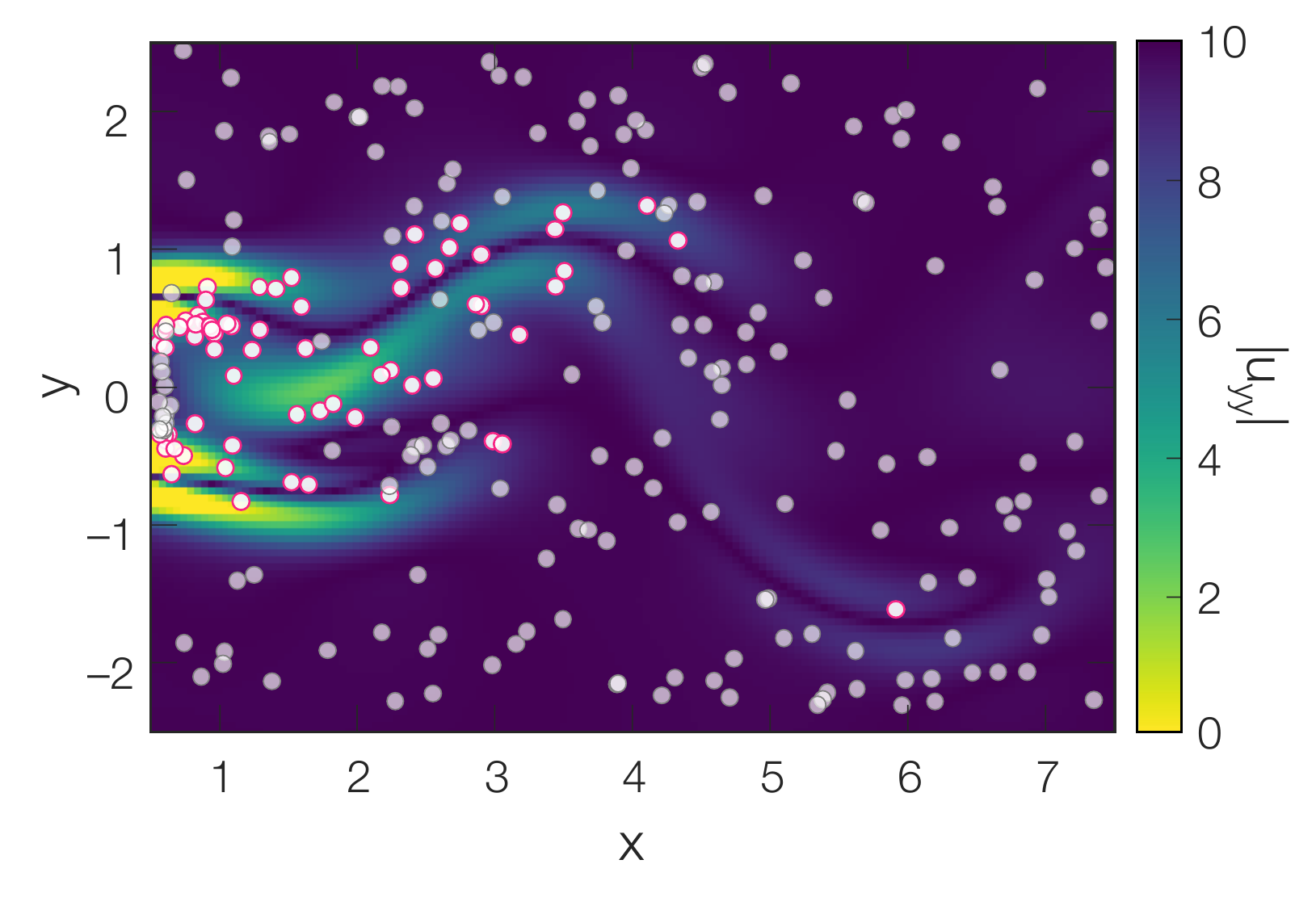}}\quad
    \subcaptionbox{\label{fig:normality:Shapiro-Wilke}}{
        \includegraphics[height=5.5cm]{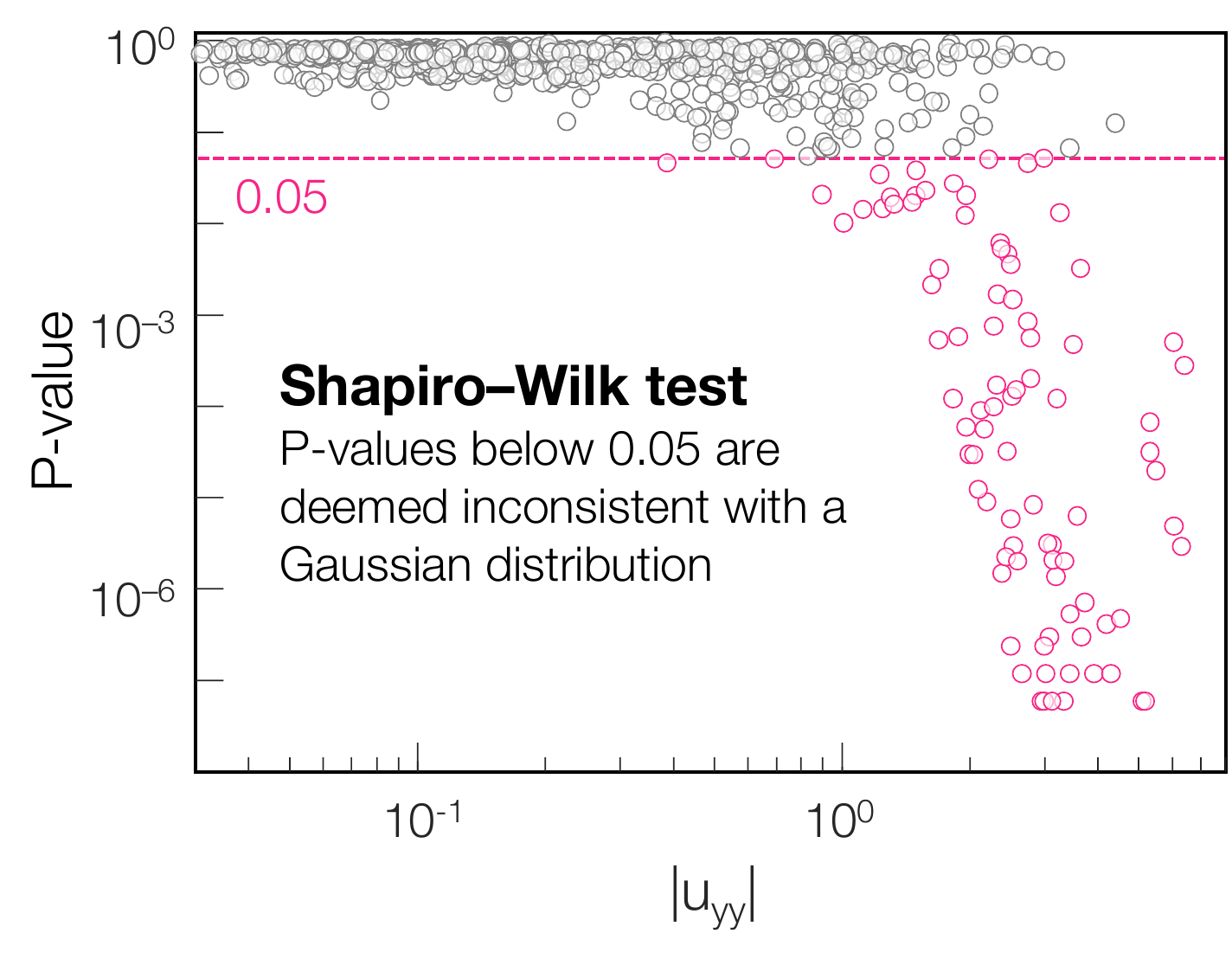}}
    \caption{Origins of non-Gaussianity: (a) advected particle positions overlaid on $|u_{yy}|$ and (b) Shapiro--Wilke $P$-values vs. $|u_{yy}|$. Strong gradients lead to non-Gaussian advected distributions.}
    \label{fig:normality}
\end{figure}

To quantify the normality of the distribution of advected samples, we used the multivariate Shapiro--Wilk test \cite{Shapiro1965}. In this test, one assumes that a distribution is normal and calculates the corresponding $P$-value, i.e., the probability of observing the target distribution assuming that it is indeed normally distributed. It is common practice to conclude that a distribution is not normal if the $P$-value falls below some critical threshold, typically 0.05. We adopt this convention for illustrative purposes but acknowledge that there is nothing fundamental about the threshold. Figure~\ref{fig:normality:spatial distribution} shows a random subset of particles at a given instance; the red particles have a $P$-value below 0.05 while the gray particles have a $P$-value greater than or equal to 0.05. Figure~\ref{fig:normality:Shapiro-Wilke} depicts $P$-values for all the particles at the same instance versus $|u_{yy}|$, i.e., the magnitude of the second partial derivative of $u$ in the spanwise direction, which is the principle axis of measurement uncertainty. Regions of the flow that exhibit velocity curvature yield a non-Gaussian distribution of advected particles, indicating that the MVN and FE techniques may perform poorly for turbulent flows if the measurement interval is long. Nevertheless, the next section demonstrates that both MVN and FE outperform the conventional data loss in accuracy, and the low cost of the FE formulation facilitates faster learning than the more accurate MC technique.\par

\section{Demonstrating SPAV for 4D DIH-PTV}
\label{sec:DIH-PTV}
While SPAV is applicable to any PTV modality, we demonstrate its performance using DIH-PTV: a low-cost technique for high-resolution velocimetry that is used to measure microfluidics \cite{Mallery2019}, near wall flow \cite{Katz2010}, and burning particles in flames \cite{Huang2021}, to name a few examples. In digital in-line holography, coherent light scatters off a target object, resulting in an interference pattern that is recorded by an imaging sensor (usually a CMOS array for PTV applications). This information can be used to localize tracer particles in a flow and follow them to form Lagrangian tracks. Unfortunately, the DoF of a typical DIH system leads to significant, anisotropic errors in the estimated particle positions. The distribution of errors is elongated normal to the camera sensor, which can bias or otherwise obfuscate the tracks, thereby corrupting estimates of the velocity and pressure fields. To compensate for this effect, DIH-PTV practitioners have introduced a host of ad-hoc methods to regularize the tracks and improve the accuracy flow fields. These methods tend to smooth-out the finer details of a flow and are themselves a source of error. SPAV, by contrast, interprets the unmodified particle tracks via models of measurement uncertainty and the underlying flow physics. This synthesis yields a marked improvement in the accuracy of 4D velocity and pressure fields in PTV.\par

This section details a comprehensive evaluation of SPAV, including its MC, MVN and FE variants, through the use of three synthetic and two experimental DIH-PTV scenarios. Further, a conventional data loss is tested in each case to provide a baseline for comparison.\par

\subsection{Measurement scenarios and implementation details}
\label{sec:DIH-PTV:details}

\subsubsection{Synthetic cases}
\label{sec:DIH-PTV:details:synthetic}
Three DNS data sets are used for synthetic testing. The first flow is a laminar 3D cylinder wake from Raissi et al. \cite{Raissi2019}; the other two flows are subvolumes of the forced isotropic turbulence and transitional boundary layer data sets from the Johns Hopkins Turbulence Database \cite{Perlman2007}. Hereafter, we refer to these cases as \textit{cylinder}, \textit{isoturb}, and \textit{TBL}, respectively. For the isoturb target, we select the central $64^3$-cell subvolume of the DNS domain; the TBL features the bottom layer of transitional turbulence, around a dimensionless distance of 430 down the plate, from $y^+ = 0$ to 126. The cylinder, isoturb, and TBL tests are sustained for 201, 51, and 51 DNS frames, respectively.\par

\begin{table}[ht]
    \renewcommand{\arraystretch}{1.25}
    \caption{Key Parameters and Dimensions of DNS Flows for Synthetic DIH-PTV Tests}
    \centering
    \begin{tabular}{c c c c c c c c}
        \hline\hline
        \multirow{2}{*}{Case} &
        \multirow{2}{*}{$Re$} &
        \multicolumn{2}{c}{Domain size} &
        \multicolumn{2}{c}{Timestep} &
        \multirow{2}{*}{Velocity field} &
        \multirow{2}{*}{No. particles} \\
        & & \multicolumn{1}{c}{DNS} & \multicolumn{1}{c}{Dim., mm} & \multicolumn{1}{c}{DNS} & \multicolumn{1}{c} {Dim., s} \\
        \hline
        Cylinder & $100$ & $7 \times 5 \times 5$ & $7 \times 5 \times 5$ & $0.08$ & $1/1250$ & $71 \times 51 \times 51$ & $1000$ \\
        isoturb & 433 & $\pi/8 \times \pi/8 \times \pi/8$ & $8 \times 8 \times 8$ & $0.002$ & $1/6520$ & $64 \times 64 \times 64$ & $4000$\\
        TBL & 430 & $6.70 \times 6.70 \times 1.46$ & $8 \times 8 \times 1.74$ & $0.25$ & $1/2240$ & $68 \times 68 \times 80$ & $2000$ \\

        \hline\hline
    \end{tabular}
    \label{tab:syn test}
\end{table}

Synthetic DIH-PTV data are generated in two steps, starting with the creation of particle fields. First, to mimic a real experiment, the DNS domains are dimensionalized. The fluid is assumed to be water with a kinematic viscosity of 1~mm$^2$/s. Virtual particles of diameter 8~$\upmu$m are uniformly seeded into the flow and ideally advected to determine their positions over time. Advection is once again conducted with a second-order Runge--Kutta scheme, subject to periodic boundary conditions. Table~\ref{tab:syn test} summarizes the key dimensionless (``DNS'') and dimensional (``Dim.'') parameters of the synthetic tests. Note that the Reynolds numbers are defined by the cylinder diameter (cylinder), Taylor microscale (isoturb), and boundary layer thickness and friction velocity (TBL).\par

The second step is to simulate holograms for our chosen measurement setups. Following the common convention in DIH, we align the $z$-axis of our measurement volume with the camera's optical axis, viz., the \textit{longitudinal} direction. As a result, the largest component of localization uncertainty lies in the $z$-direction. The hologram plane is coincident with the $x$- and $y$-axes and is positioned 7~mm away from the center of the domain. Flows are illuminated by a 633~nm laser that is positioned in-line with the camera. We use 8~mm sensors with two different pixel pitches. For the cylinder and isoturb cases, we specify an $800\times800$~px sensor with 10~$\upmu$m pixels; for the TBL case, we specify a $1600\times1600$~px sensor with 5~$\upmu$m pixels. The sensor resolution is increased in the latter scenario to better measure the particles' wall-normal motion. In other words, when an $800\times800$~px sensor is used to measure this flow at a practical frame rate, localization errors dominate genuine flow structures.\par

\begin{figure}[ht]
    \centering
    \subcaptionbox{\label{fig:isoturb:synholo}}{\includegraphics[height=4.75cm]{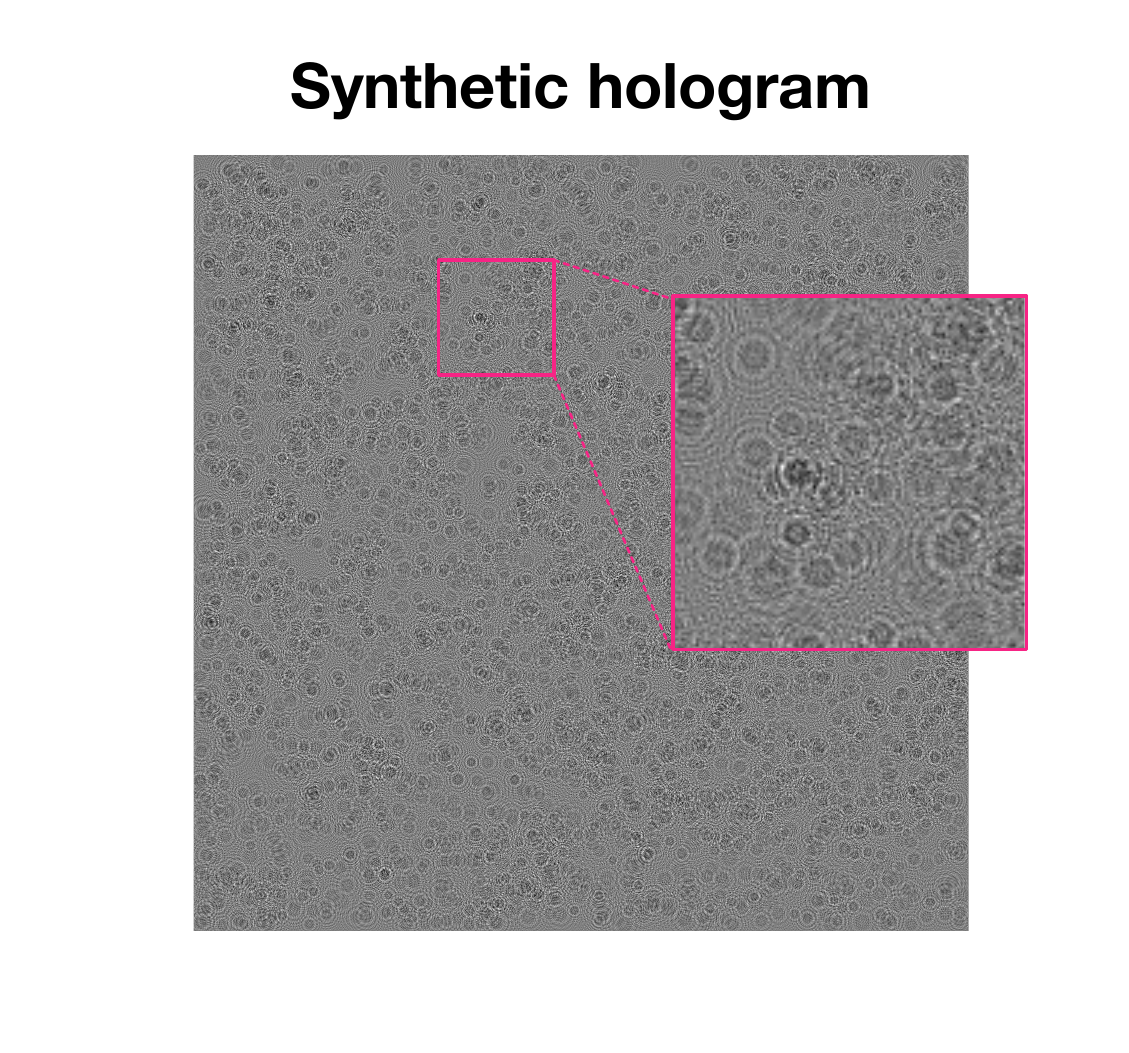}}
    \subcaptionbox{\label{fig:isoturb:locpdf}}{\includegraphics[height=4.75cm]{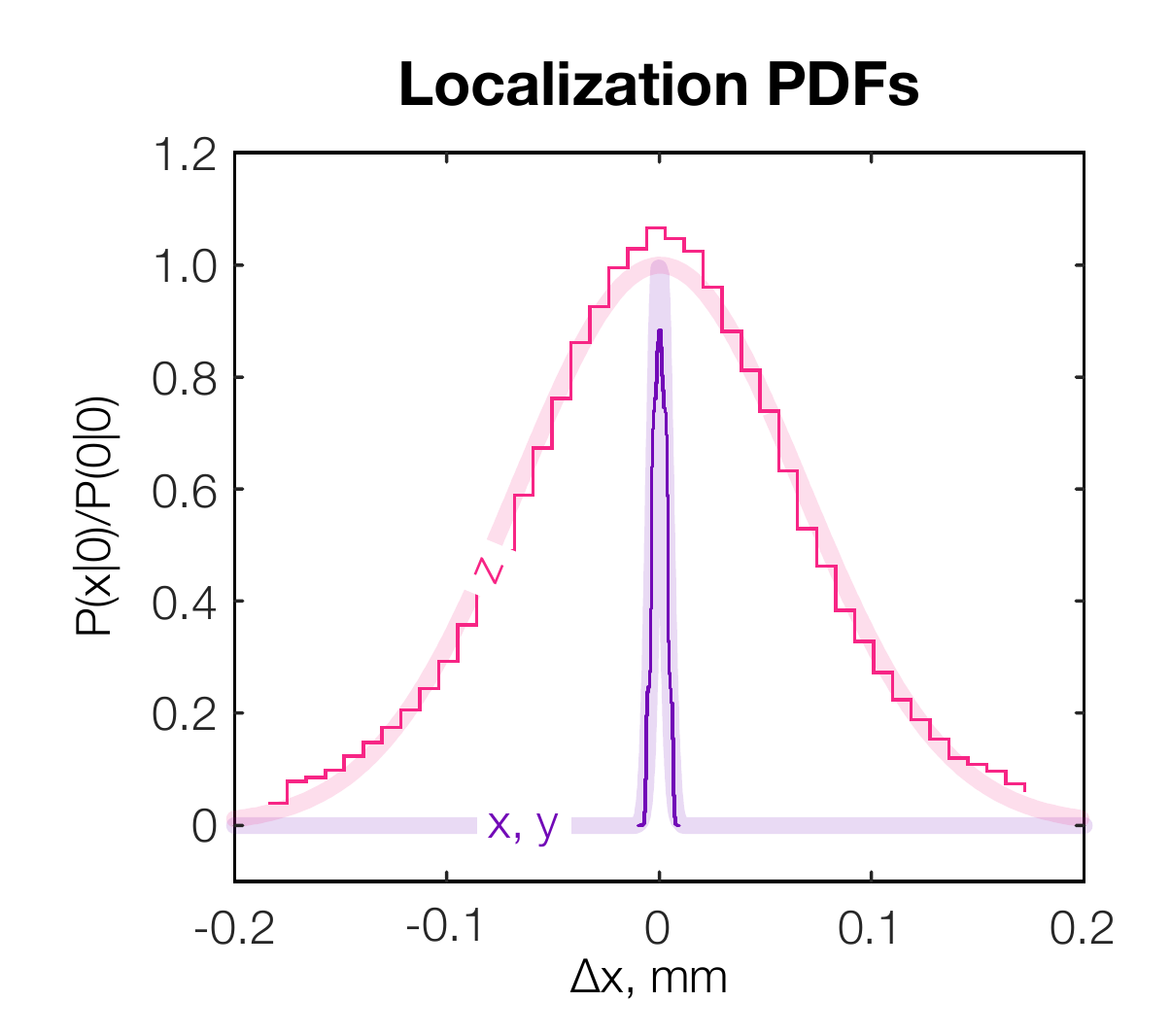}}
    \subcaptionbox{\label{fig:isoturb:tracks}}{\includegraphics[height=4.75cm]{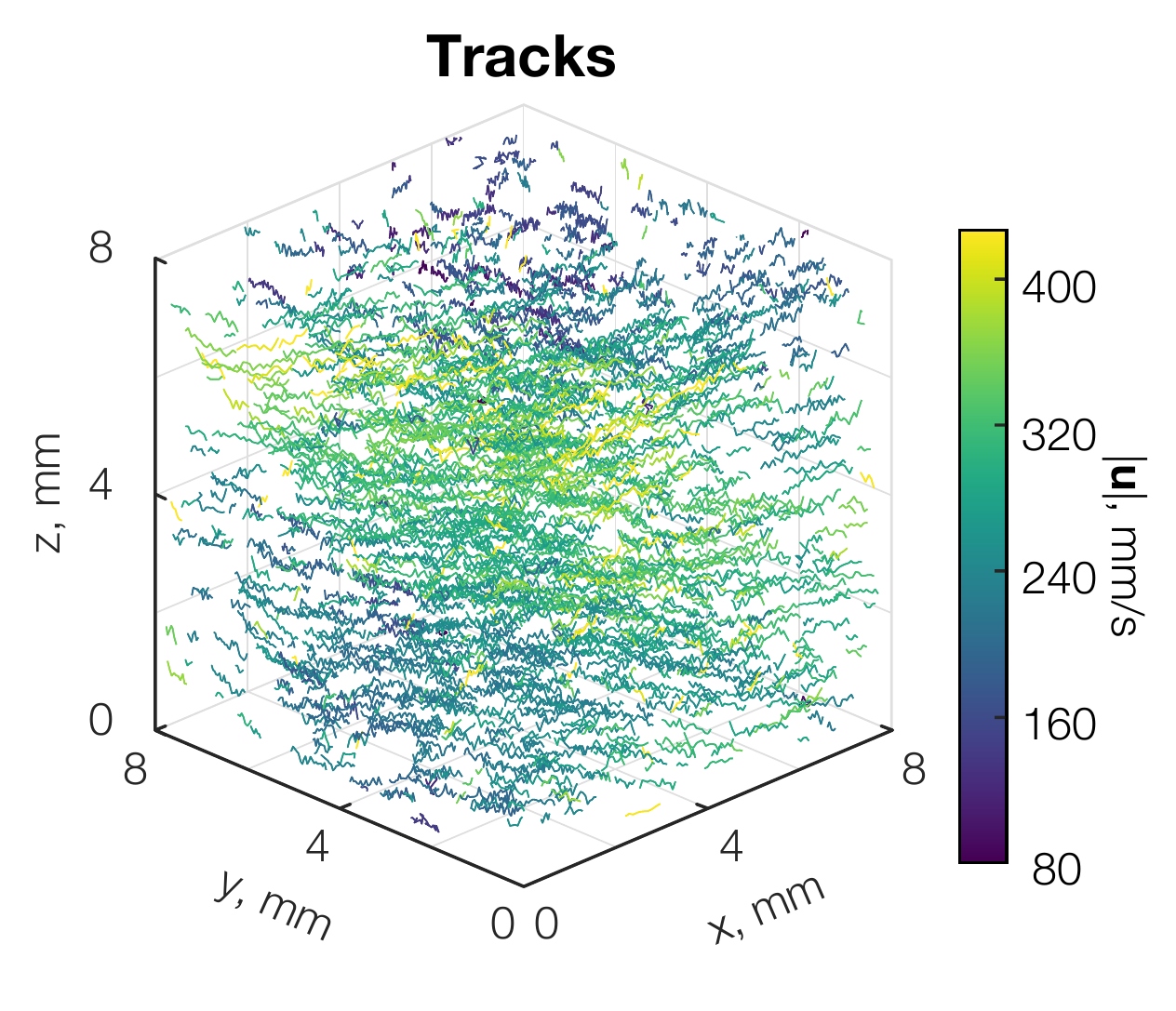}}
    \vspace*{2mm}
    \caption{Sample data from the synthetic forced isotropic turbulence case: (a) a synthetic hologram with an enlarged inset, (b) localization PDFs (the $x$- and $y$-direction PDFs overlap almost completely), and (c) DIH-based particle tracks, colored by their mean velocity magnitude. For clarity, the tracks are visibly corrupted by localization errors. Only 15\% of the tracks are visualized.}
    \label{fig:isoturb}
\end{figure}

We simulate holograms using Gao's method \cite{Gao2014}, in which the intensity field is enlarged by a factor of four with zero-padding to avoid aliasing in the Fourier domain. Furthermore, each sensor is supersampled at four times its native resolution to mitigate discretization artifacts. An algorithm for generating multi-particle holograms is provided in \ref{app:DIH-PTV overview}. Figure~\ref{fig:isoturb:synholo} shows a sample synthetic hologram from the isoturb case. Due to the high seeding density, many particle fringes overlap one another, increasing the difficulty of localization.\par

\subsubsection{Experimental cases}
\label{sec:DIH-PTV:details:experimental}
Two experimental flows are considered: laminar micro-channel and turbulent channel flow. These experiments are briefly recapitulated below to facilitate a discussion of our SPAV results. Full details of the laminar test can be found in Toloui and Hong \cite{Toloui2015} and the turbulent test is documented in Toloui et al. \cite{Toloui2017}.\par 

In the micro-channel experiment, water is pumped through a micro-channel using a motorized syringe. The flow is seeded with 2~$\upmu$m silver-coated fused silica particles to a concentration of roughly 3000~particles/mm$^{3}$. The micro-channel has 0.15~mm glass walls and a 1~mm$^{2}$ square cross-section. Flow is maintained at a Reynolds number near 10 based on the channel's hydraulic diameter. The DIH system features a $2048\times 1088$~px CMOS camera, 633~nm HeNe laser, collimated lens, and a spatial filter to shape the laser's output into a plane wave. The camera is equipped with a long working distance, infinity-corrected $5\times$ objective lens to record a magnified hologram at the outer wall of the channel. During measurement, the camera is operated at a frame rate of 338~Hz with a lateral resolution of 1.1~$\upmu$m/pixel.\par

The second experiment is conducted within a refractive index-matched ($n = 1.41$) turbulent channel flow facility at the University of Minnesota, which has a 1.2~m long smooth-wall acrylic channel with a 2500~mm$^2$ square cross-section. The channel is operated at a Reynolds number of 22,770, based on its height, and the flow is seeded with silver-coated hollow glass tracers of diameter 8--12~$\upmu$m.\footnote{$Re_\uptau$ for the turbulent channel flow experiment is estimated to be 575.} The DIH setup is similar to that in the micro-channel test, except that a $1472\times1448$~px high-speed camera, coupled with a Nikon lens (105~mm and $f$/2.8), is used to record holograms with a resolution of 10~$\upmu$m/pixel at 3080~Hz. The camera is focused on a plane around 3~mm away from the inner channel surface. In the experiments, collimated laser light is shone through the entire the channel, resulting in a large probe volume of $14.7 \times 14.4 \times 50.0$~mm$^3$.\par

\subsubsection{Particle extraction and tracking}
\label{sec:DIH-PTV:details:particle}
We employ the iterative particle extraction technique of Toloui and Hong \cite{Toloui2017}, briefly summarized in \ref{app:DIH-PTV overview}, which is tailor-made for DIH-PTV experiments that have a high particle count. Per the angular spectrum method, our reconstructions maintain the same lateral (i.e., $x$-$y$) resolution as the holograms \cite{Latychevskaia2015}. Longitudinally, we refocus the synthetic holograms onto 500, 800, and 200 planes for the cylinder, isoturb, and TBL cases, respectively, using a constant spacing of 10~$\upmu$m. To quantify the localization PDF for SPAV, i.e., $P\mathopen{}\left(\mathbf{x}|\hat{\mathbf{x}}\right)$ in Eq.~\eqref{equ:likelihood}, we compare DIH-based particle positions to their ground truth counterparts via nearest neighbor matching \cite{Toloui2015}. A sample localization PDF from the isoturb case is plotted in Fig.~\ref{fig:isoturb:locpdf}. The PDFs are well modeled as Gaussian, supporting the use of MVN PDFs in SPAV. For the isoturb setup, the standard deviation in the $z$-direction of 0.067~mm is appreciably larger than the $x$- and $y$-direction standard deviations of 0.003~mm.\par

\begin{figure}[t]
    \centering
    \subcaptionbox{\label{fig:mc meas:holo}}{\includegraphics[height=4.75cm]{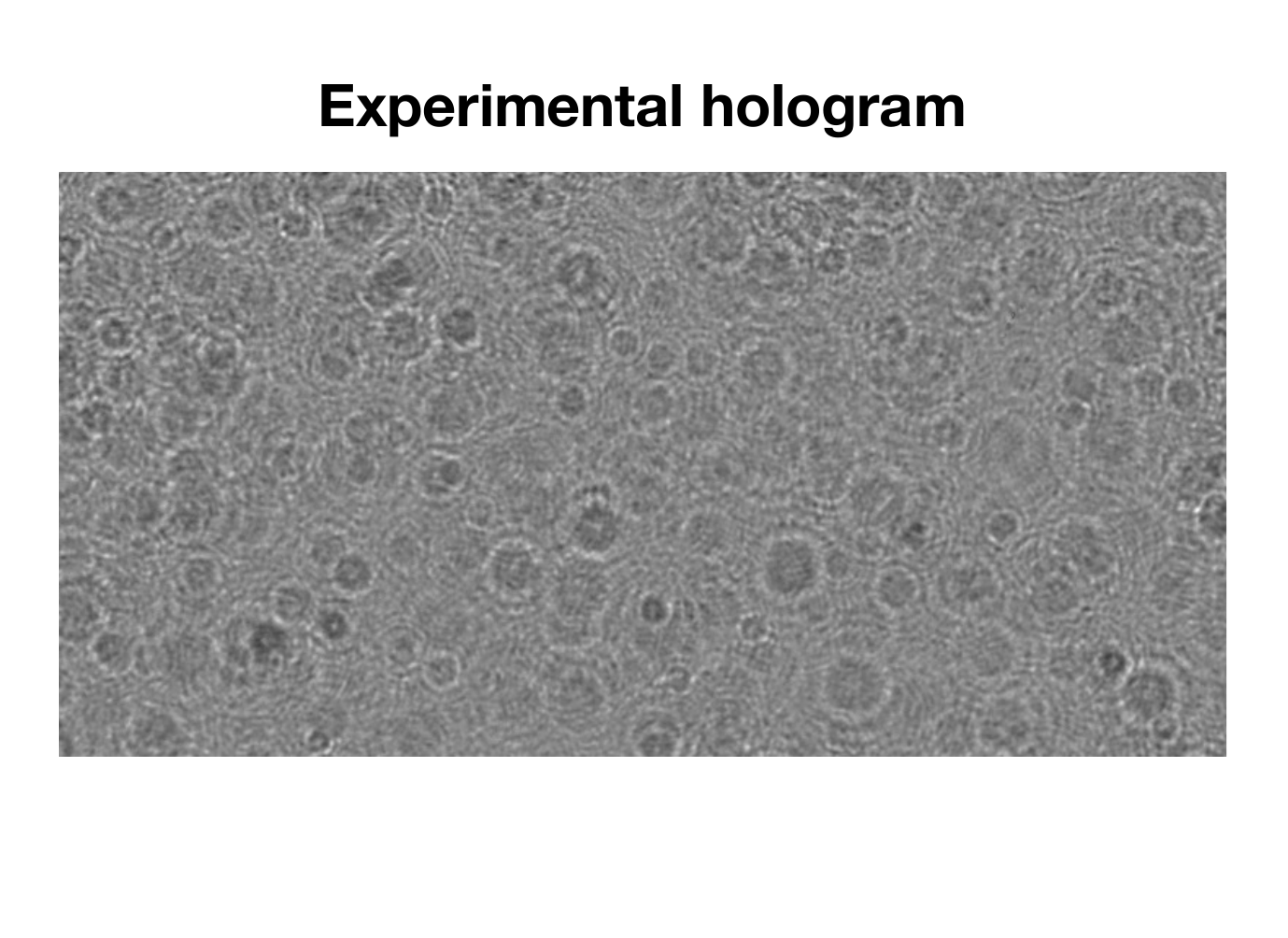}}\quad
    \subcaptionbox{\label{fig:mc meas:tracks}}{\includegraphics[height=4.75cm]{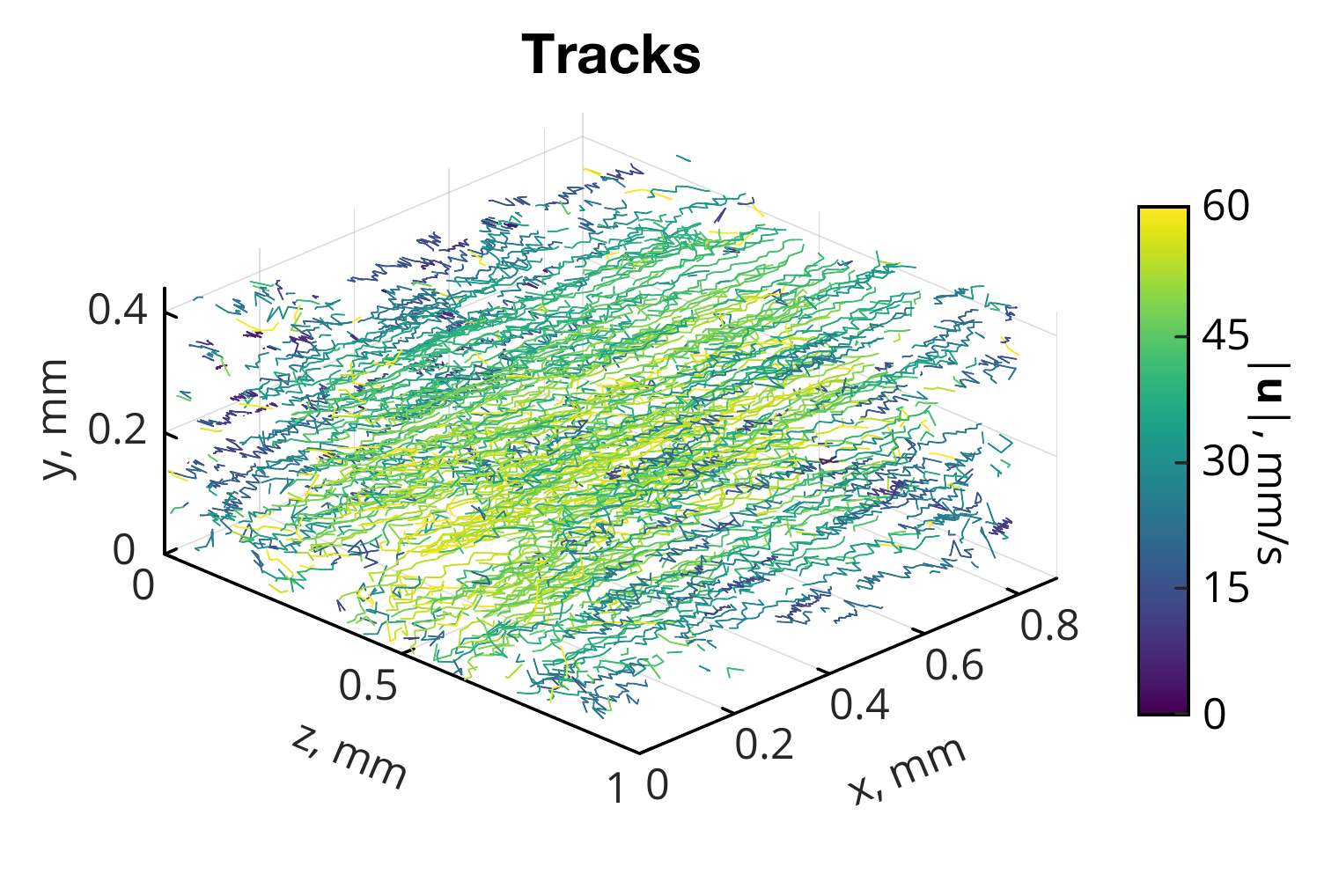}}
    \caption{Sample data from the experimental micro-channel flow case: (a) an experimental hologram and (b) DIH-reconstructed particle tracks, colored by their mean velocity magnitude. For clarity, only one-fifth of the tracks are visualized.}
    \label{fig:mc meas}
\end{figure}

Given a sequence of particle clouds, we use the Crocker--Grier algorithm to obtain Langrangian tracks. Tracks with fewer than four particles are discarded to avoid spurious matches, since ghost particles rarely persist across multiple frames. Reconstructed particle tracks from the isoturb case are plotted in Fig.~\ref{fig:isoturb:tracks}. These tracks feature strong fluctuations along the $z$-direction, which is consistent with the aysmmetric errors that are emblematic of DIH. Flow fields are reconstructed by directly feeding noisy tracks to a SPAV-enabled PINN or by passing Lagrangian velocity estimates (i.e., from Eq.~\eqref{equ:velocity estimate}) to a PINN with a conventional data loss.\par

Background subtraction must be applied to the experimental holograms to enhance their signal-to-noise ratio (SNR). We also crop the micro-channel holograms, leaving the central $800 \times 400$~px region, to avoid reflections from the glass walls. Twenty consecutive frames of this flow are numerically refocused onto 1000 longitudinal scans, separated by 1~$\upmu$m intervals. The turbulent channel case spans 40 frames; at each frame, the entire hologram is refocused onto 1024 longitudinal scans with 56.5~$\upmu$m spacing. The wall detection procedure from \cite{Toloui2017} is implemented to eliminate ghost particles outside the channel. A sample background-subtracted hologram from the micro-channel experiment is shown in Fig.~\ref{fig:mc meas:holo}. Although similar to the synthetic hologram in Fig.~\ref{fig:isoturb:synholo}, the experimental one exhibits a lower dynamic range and hence a lower SNR. Particle tracks extracted from the micro-channel holograms are plotted in Fig.~\ref{fig:mc meas:tracks}. These tracks are visibly affected by longitudinal errors in the particle positions. Nevertheless, the bulk laminar pipe flow profile is readily apparent, with fast flow along the center that slows towards the walls.\par

Experimental localization uncertainties must be be estimated to specify the SPAV data loss in each test. We thus conduct simulations that mimic the DIH system, particle concentration, and particle sizes from our tests. To account for noise, we double the magnitude of uncertainties determined by this procedure since previous works have shown that the actual DoF in a DIH experiment is approximately twice that of the corresponding synthetic test \cite{Toloui2015, Sheng2006}. As an alternative, localization uncertainties can be experimentally characterized via a controlled sample with embedded tracer particles, fixed in place \cite{Gao2013, Gao2014}.\par

\subsubsection{Network architecture and training}
\label{sec:DIH-PTV:details:network}
All PINNs used in this work are implemented in TensorFlow~1.15. Networks employed for the laminar cylinder wake and micro-channel flows are set up with ten hidden layers that have 50 neurons per output variable; the networks used for turbulent flows contain fifteen layers with 75 neurons per output variable. Swish activation functions are selected in accordance with \cite{Molnar2022a}. This architecture is empirically chosen to ensure that the PINN is expressive enough to represent the turbulent flows considered in this work. Weights are randomly initialized with a standard normal distribution and biases are set to zero at the start. Weightings of the loss term components in $\mathcal{L}_\mathrm{total}$ are optimized through a simple parameter sweep. Several sophisticated schemes have been developed to automate the loss weights, e.g., \cite{Wang2021, Wang2022d, McClenny2020}. However, we find that adaptive techniques yield marginal benefits in the presence of realistic noise \cite{Molnar2022a, Molnar2022b}.\par

Training is performed using the Adam optimizer at a fast learning rate of $1 \times 10^{-3}$ followed by a slow rate of $1 \times 10^{-4}$. We use a particle batch size of $1 \times 10^{4}$ for PINNs with a conventional data loss (``conventional PINNs'' for shorthand), which are trained for 5000 epochs at the fast rate and 2000 epochs at the slow rate. Batch sizes for PINNs with a SPAV loss (``SPAV PINNs'', ``MC PINNs'', etc.) are limited by available GPU memory. We use a batch size of 100 particles for the MC and MVN PINNs and 2000 particles for the FE PINN, since that variant is much cheaper to implement. Networks equipped with a SPAV loss are pre-trained with the \textit{conventional data loss} for 5000 epochs at the fast learning rate to speed-up convergence. Next, the SPAV losses are employed for training at the slow rate. Due to their smaller batch size, the MC and MVN PINNs are only trained for 50 epochs at the slow rate, while the FE PINNs is trained for 1000 epochs. All of these schedules are sufficient to ensure convergence. Training is conducted on an NVIDIA GeForce RTX 3090 GPU. The average computation time is around 10 hours for a conventional PINN, 20 hours for MC and MVN PINNs, and 14 hours for an FE PINN.\par

\subsubsection{Errors}
\label{sec:DIH-PTV:details:errors}
Reconstruction errors are computed in terms of the fluctuating component of a Reynolds decomposition, $\phi = \overline{\phi} + \phi^\prime$, where $\phi$ is the variable of interest, $\overline{\phi}$ is the ensemble average of $\phi$ at $(x,y,z,t)$, and $\phi^\prime$ is the fluctuating component of interest. Errors are averaged over the measurement domain,
\begin{equation}
    \langle\phi\rangle = \frac{1}{\left|\mathcal{V}\right|} \iiint_\mathcal{V} \phi \,\mathrm{d}\mathcal{V}.
    \label{equ:domain average}
\end{equation}
Normalized root-mean-square errors (NRMSEs) of the fluctuating component at a given instance are calculated as follows:
\begin{equation}
    e_\phi = \left[\frac{\left\langle \left(\phi^\prime - \phi_\mathrm{exact}^\prime\right)^2 \right\rangle}{\left\langle {\phi_\mathrm{exact}^{\prime\,^2}} \right\rangle}\right]^{1/2}.
    \label{equ:NRMSE}
\end{equation}
Time-averaged NRMSEs are denoted $\overline{e}_\phi$.\par

\subsection{Synthetic results}
\label{sec:DIH-PTV:synthetic}

\subsubsection{Laminar cylinder wake}
\label{sec:DIH-PTV:synthetic:3D cylinder}
\begin{figure}[ht]
    \centering
    \includegraphics[width=15cm]{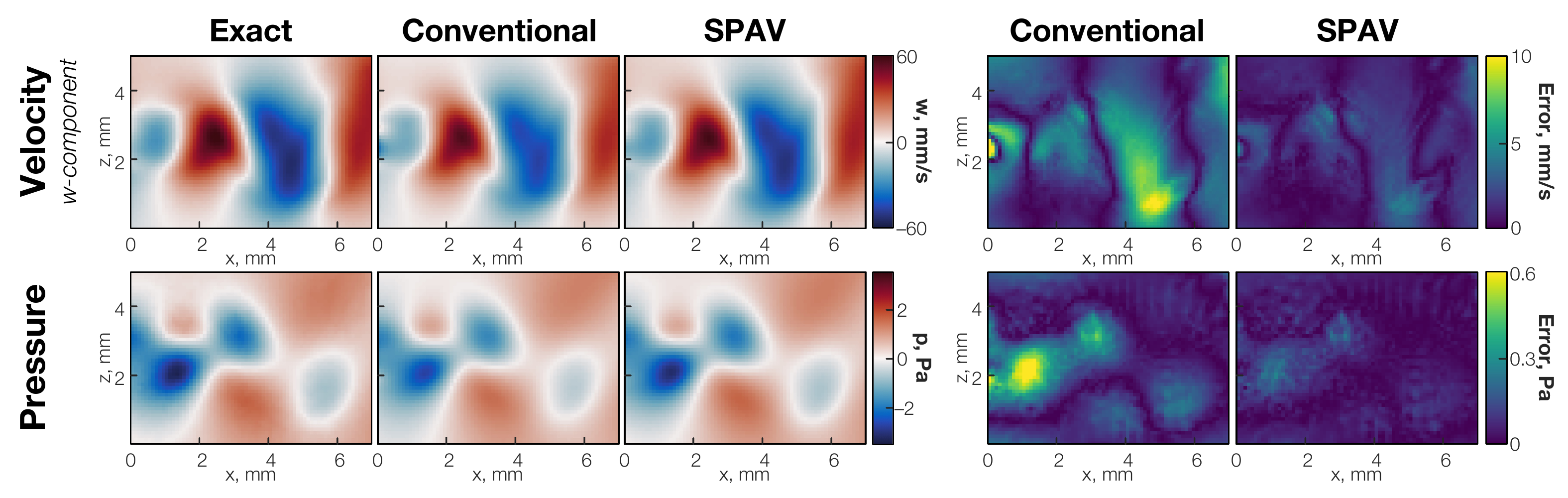}
    \vspace*{2mm}
    \caption{Exact and reconstructed cylinder wake flow fields (left) and absolute point-wise errors (right). The $u$- and $v$-velocity fields are omitted since the reconstructions are of similar quality.}
    \label{fig:Cyliner-uvwpcontour}
\end{figure}

We first present the results of our cylinder wake flow test. Figure~\ref{fig:Cyliner-uvwpcontour} compares the exact DNS data to reconstructed flow fields from conventional and MC SPAV PINNs. Velocity and pressure fields are extracted at the central snapshot and rendered on the $x$-$z$ midplane of the probe volume, perpendicular to the cylinder (itself aligned with the $y$-axis). Qualitatively, both the SPAV and conventional PINNs can recover the velocity and pressure fields to high accuracy, as evinced by the clear von K{\'a}rm{\'a}n vortex street and elevated pressures within the vortex cores. To better illustrate SPAV's performance, point-wise absolute errors are plotted on the right side of Fig.~\ref{fig:Cyliner-uvwpcontour}. Use of the conventional data loss leads to pronounced errors in the velocity fields, especially the $w$-component due to the low longitudinal accuracy of DIH. Moreover, these velocity errors limit the accuracy of the pressure field. By contrast, the SPAV results exhibit far lower errors throughout the domain. This illustrates the benefit of incorporating localization and measurement uncertainties into the reconstruction algorithm.\par

\begin{figure}[ht]
    \centering
    \includegraphics[width=15cm]{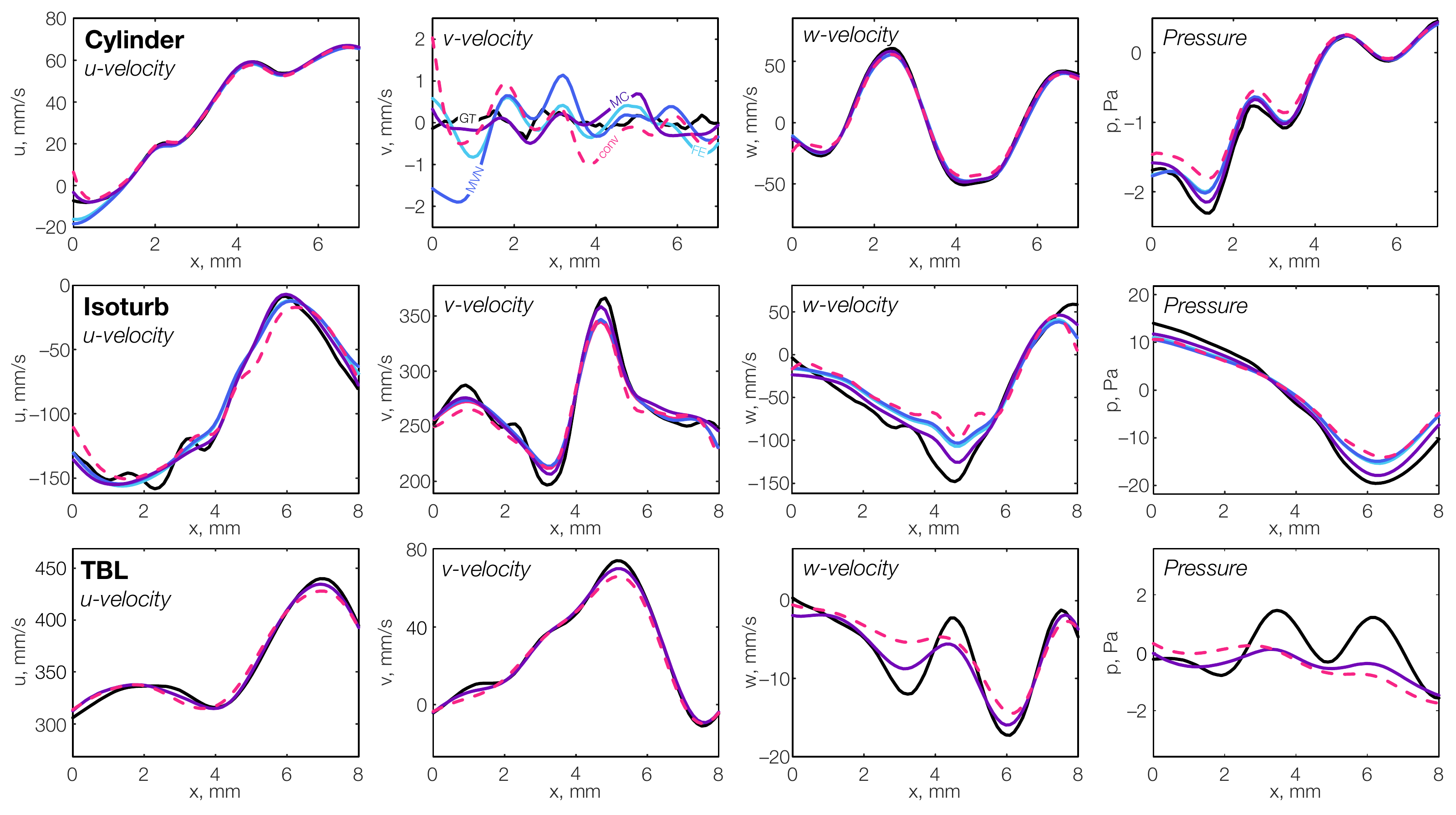}
    \vspace*{2mm}
    \caption{Instantaneous cut plots of velocity and pressure along the central $x$-axis of the domain from the conventional PINN and all three SPAV variants: (top) cylinder case, (middle) isoturb case, and (bottom) turbulent boundary layer case. Color coding: 
    ``\protect\GTline''\space for ground truth, ``\protect\Convline''\space  for conventional, ``\protect\MCline''\space for MC, ``\protect\MVNline''\space for MVN, and ``\protect\FEline''\space for FE.}
    \label{fig:VP profiles along central x axis}
\end{figure}

\begin{figure}[ht]
    \centering
    \includegraphics[width=15cm]{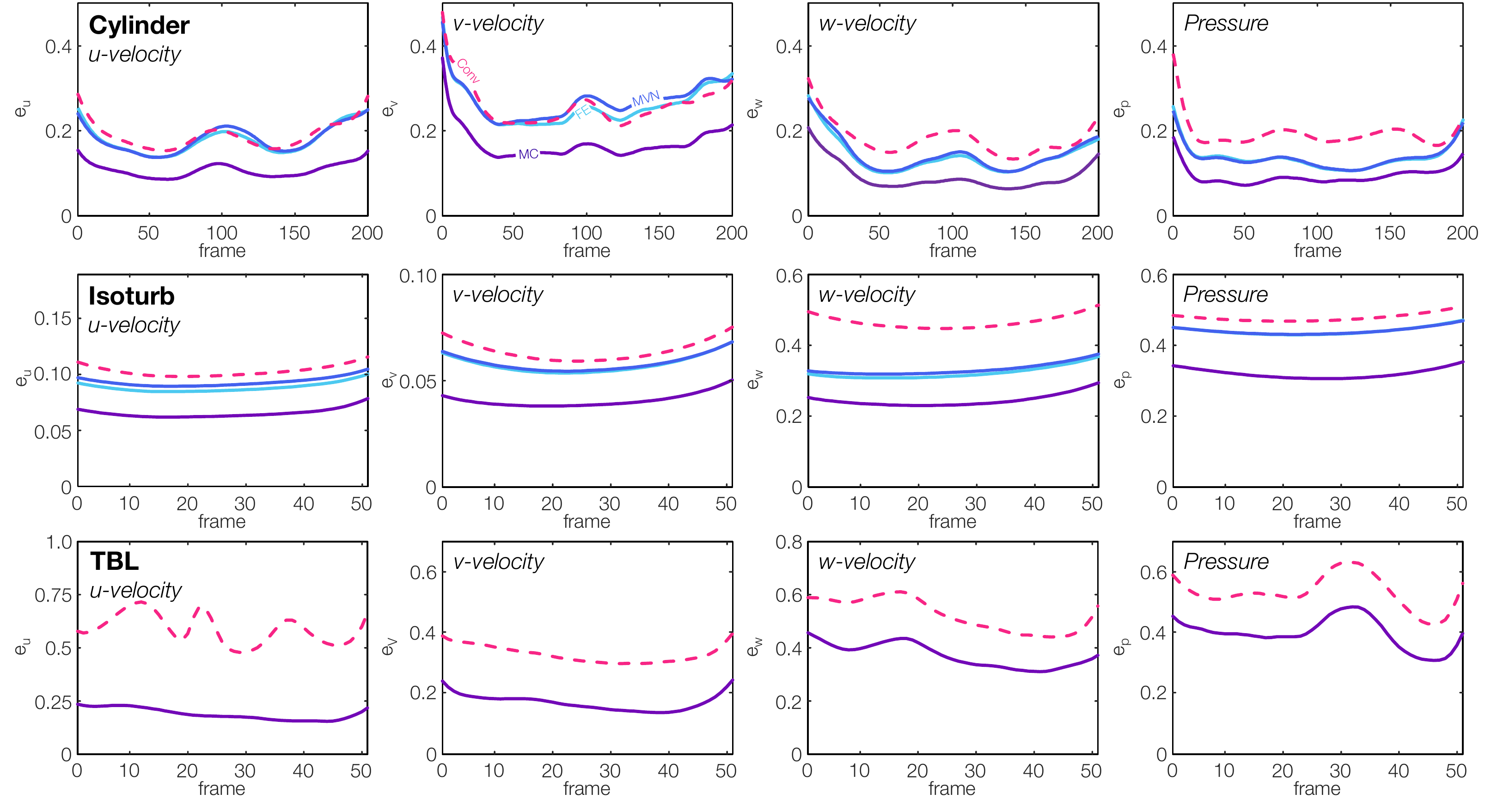}
    \vspace*{2mm}
    \caption{Comparison of velocity and pressure NRMSEs across all frames for the conventional PINN and all three SPAV variants. Color coding: 
    ``\protect\GTline''\space for ground truth, ``\protect\Convline''\space  for conventional, ``\protect\MCline''\space for MC, ``\protect\MVNline''\space for MVN, and ``\protect\FEline''\space for FE.}
    \label{fig:VP error traces}
\end{figure}

The top row of Fig.~\ref{fig:VP profiles along central x axis} depicts instantaneous velocity and pressure traces along the central $x$-axis ($y = z = 2.5$~mm) of the central snapshot for all three SPAV variants as well as the conventional data loss. We observe that SPAV reconstructions are consistently closer to the ground truth (``GT'') DNS data than the conventional estimates. Of the SPAV variants, MC yields the most accurate fields. Several large deviations are observed in the MVN and FE reconstructions in the near-wake region of the flow because these losses assume that the advected PDF is Gaussian, which is invalid in the presence of strong velocity gradients, as demonstrated in Sect.~\ref{sec:validation}. To better quantify relative performance, we calculate NRMSEs for the reconstructed fields at each frame, which are plotted in the top row of Fig.~\ref{fig:VP error traces}. Use of the MC data loss reduces errors across all fields by roughly 50\%, exceeding 5 percentage points in some cases. The conventional, MVN, and FE losses exhibit comparable accuracy for the $u$- and $v$-components of velocity (again, due to the assumption of Gaussian PDFs).\par

\subsubsection{Forced isotropic turbulence}
\label{sec:DIH-PTV:synthetic:3D isoturb}
Next is the forced isotropic turbulence, or \textit{isoturb}, case. The DNS includes a force to compensate for energy dissipation in the flow \cite{Rosales2005}. We tested isoturb reconstructions with $\mathbf{f} = \mathbf{0}$; however, while this produced accurate estimates of velocity, the quality of unforced pressure reconstructions was poor. Therefore, we include a linear force in the physics loss for all our isoturb tests,
\begin{equation}
    \mathbf{f} = \frac{\varepsilon}{3 \,\rho \,u^{2}_\mathrm{rms}} \mathbf{u},
    \label{equ:isotropic force}
\end{equation}
where $\rho$ is the fluid density, $\varepsilon$ is the mean energy dissipation rate, and $u_\mathrm{rms}$ is the root-mean-square velocity. Note that, although this procedure is valid for the isoturb simulation, it is not suitable for most experimental flows.\par

\begin{figure}[ht]
    \centering
    \includegraphics[width=15cm]{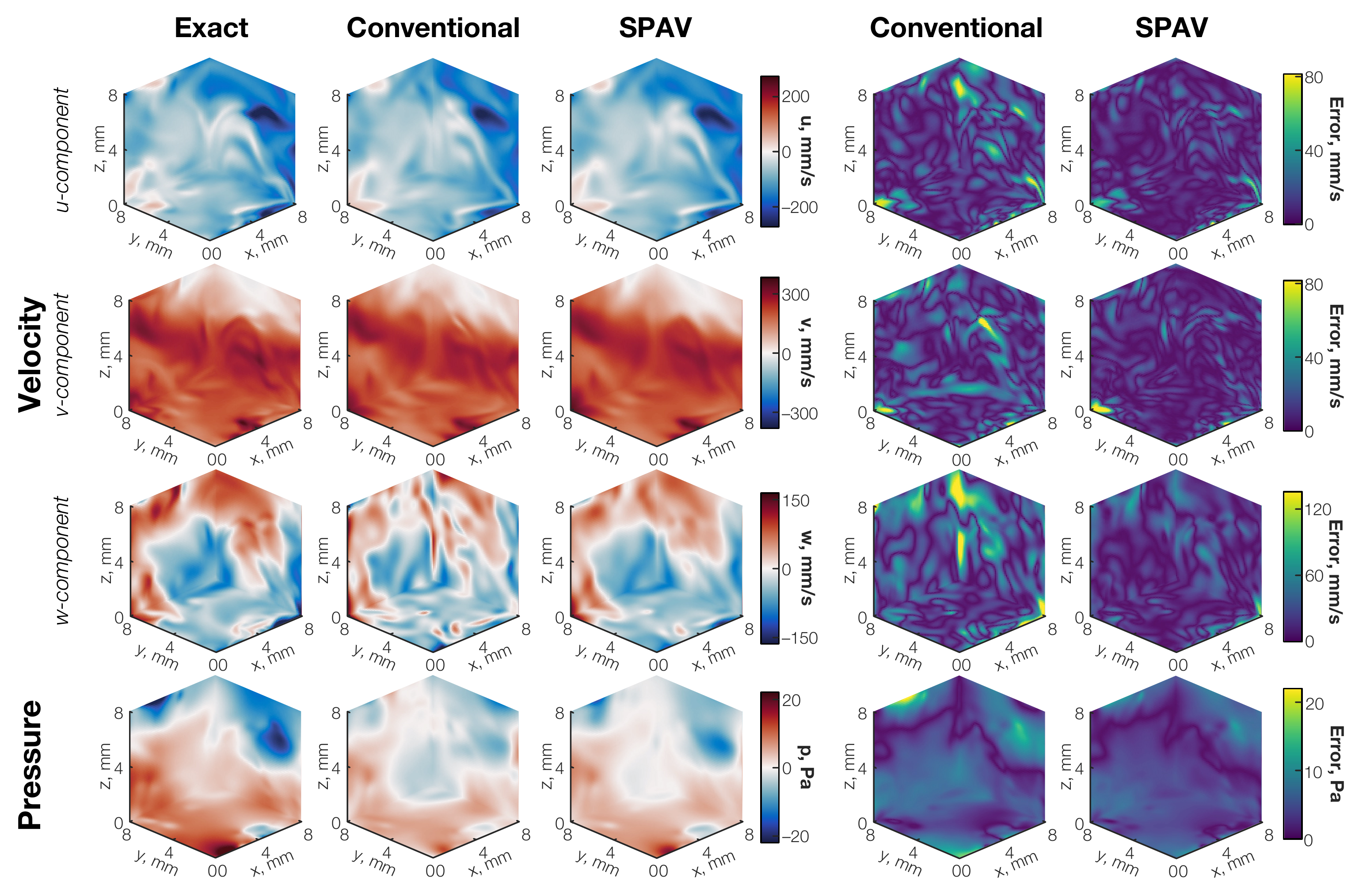}
    \vspace*{2mm}
    \caption{Exact and reconstructed cylinder wake flow fields (left) and absolute point-wise errors (right).}
    \label{fig:Isoturb-uvwpcontour}
\end{figure}

\begin{figure}[ht]
    \centering
    \includegraphics[width=15cm]{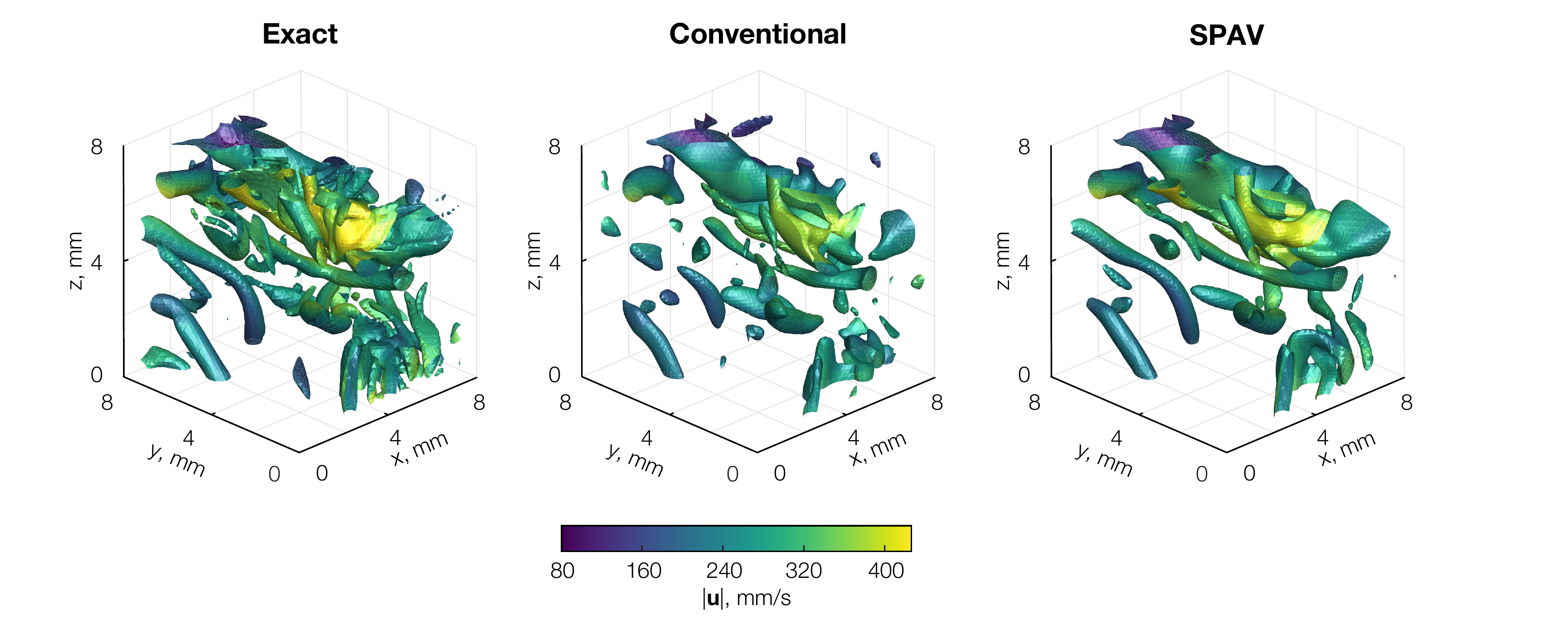}
    \vspace*{2mm}
    \caption{Coherent structures in forced isotropic turbulence from conventional and SPAV PINNs. Structures are visualized using isosurfaces of the $Q$-criterion ($Q = 8500$~s$^{-2}$) colored by the velocity magnitude.}
    \label{fig:Isoturb-Q}
\end{figure}

Figure~\ref{fig:Isoturb-uvwpcontour} shows cut plots of velocity and pressure from the DNS as well as the conventional and MC SPAV PINNs. Cuts are taken at the bottom ($z = 0$~mm), rear ($y = 8$~mm), and right ($x = 8$~mm) face of the cubic domain. Reconstructions from the conventional and SPAV PINNs are qualitatively similar to the DNS fields. Still, there are discernible differences in (1) the $w$-component of velocity, where conventional estimates are severely distorted by PTV localization errors, and (2) the pressure field, where the conventionally-trained PINN underestimates the magnitude of fluctuations. Absolute errors are plotted on the right side of Fig.~\ref{fig:Isoturb-uvwpcontour}. Once again, SPAV significantly reduces errors in the velocity and pressure fields. This reduction in error corresponds to more accurate flow derivatives and vortex detection. As an example, Fig.~\ref{fig:Isoturb-Q} compares $Q$-criterion isosurfaces computed on the ground truth velocity field and our DIH-PTV reconstructions. Velocity fields produced by a SPAV PINN contain denser coherent structures that are visibly more accurate than the structures from conventionally-obtained velocity fields.\par

Similar to the cylinder flow case, quantitative assessments are conducted by comparing velocity and pressure distributions along the central $x$-axis ($y = z = 4$~mm) in Fig.~\ref{fig:VP profiles along central x axis}. The MC data loss yields the best reconstructions due to its accurate approximation of the advected PDF, followed by the MVN, FE, and conventional PINNs in that order. NRMSEs of the estimated velocity and pressure fields across all frames are presented in Fig.~\ref{fig:VP error traces}, where the MC technique is shown to halve the reconstruction errors compared to a conventional PINN. An extended sensitivity analysis of the isoturb case that features several distributions of measurement uncertainty is provided in \ref{app:sensitivity study}.\par

\subsubsection{Transitional boundary layer}
\label{sec:DIH-PTV:synthetic:3D TBL}
The last and most challenging synthetic test is the transitional boundary layer case, i.e., \textit{TBL}. Here, we measure the bottom layer of flow and our imaging axis is aligned with the wall normal direction ($z$-axis). Similar optical setups have been used in real experiments to resolve near-wall flow from backscatter digital holograms \cite{Kumar2018b}.\par

\begin{figure}[ht]
    \centering
    \includegraphics[width=15cm]{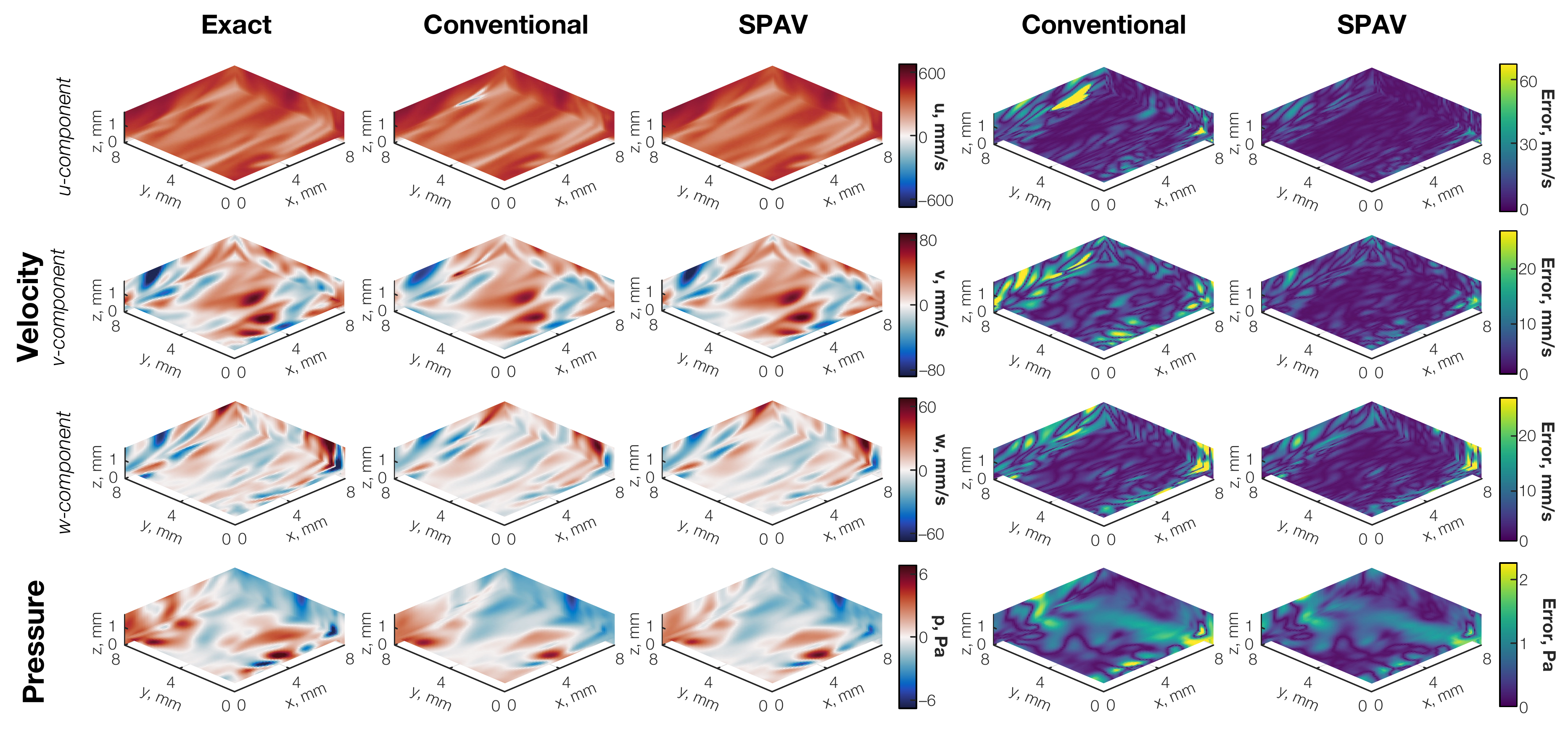}
    \vspace*{2mm}
    \caption{Exact and reconstructed boundary layer flow fields (left) and absolute point-wise errors (right).}
    \label{fig:TBL-uvwpcontour}
\end{figure}

Although the Reynolds number is similar in the isoturb and TBL scenarios, the boundary layer features three unique challenges. First, resolving fine flow structures in a thin volume requires high longitudinal accuracy. As such, we use a high-resolution sensor ($1600 \times 1600$~px) to reduce longitudinal uncertainties, although this comes at a significant computational cost. Second, the thin region limits the allowable particle concentration, beyond which the Crocker--Grier algorithm encounters a combinatoric explosion in candidate particle matches \cite{Crocker1996}. However, while tracking becomes easier with fewer particles, velocity and pressure field reconstructions are adversely affected by a low seeding density. We found the optimal number of particles for our TBL test to be around 2000 through trial and error. This challenge highlights the potential benefits of an advanced tracking algorithm for DIH-PTV that can accommodate dense particle fields. Development of such an algorithm is beyond the scope of this work, however. The third challenge pertains to the strong velocity gradients at the wall, leading to a large number of non-Gaussian PDFs that are inconsistent with the MVN and FE losses. As a result, we only assess the MC variant of SPAV for this case.\par

\begin{figure}[ht]
    \centering
    \includegraphics[width=15cm]{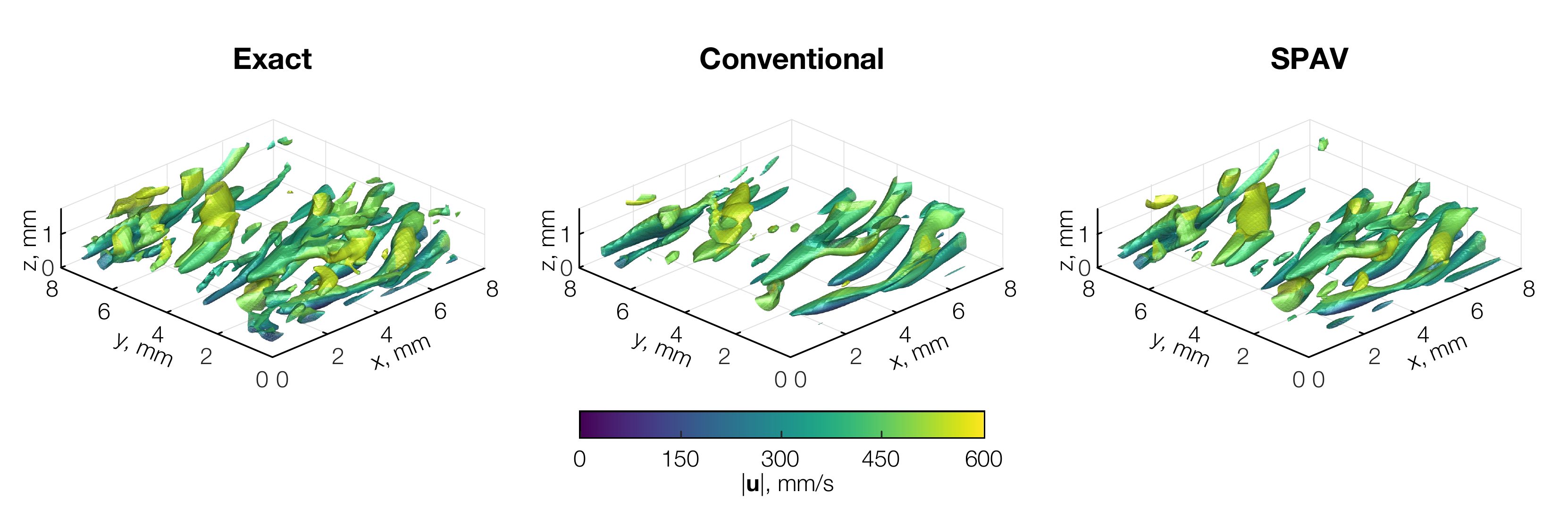}
    \vspace*{2mm}
    \caption{Coherent structures in a transitional boundary layer from conventional and SPAV PINNs. Structures are visualized using isosurfaces of the $Q$-criterion ($Q = 9450$~s$^{-2}$) colored by the velocity magnitude.}
    \label{fig:TBL-Q}
\end{figure}

\begin{figure}[ht]
    \centering
    \includegraphics[width=13cm]{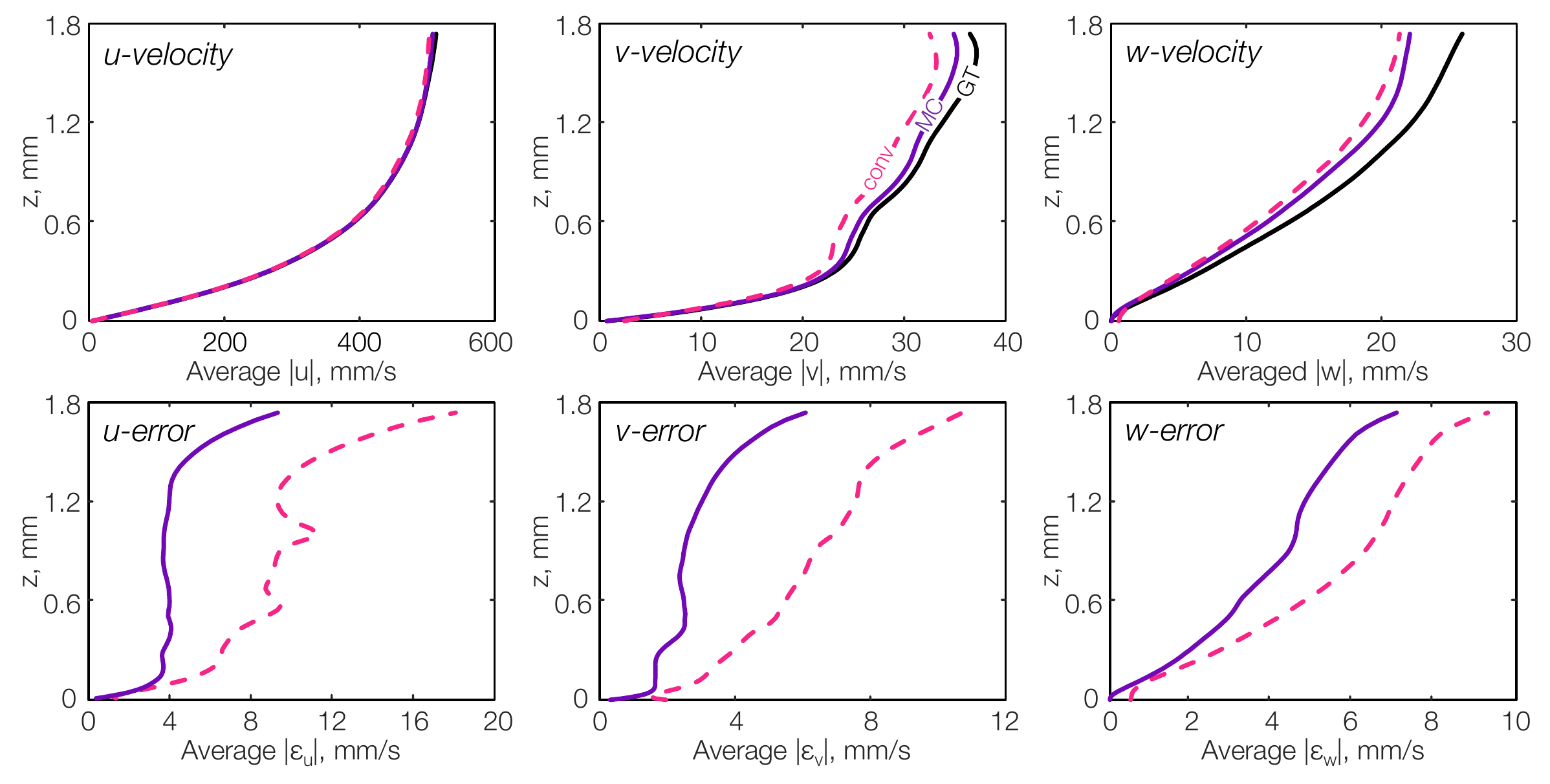}
    \vspace*{2mm}
    \caption{Mean velocity distributions (top) and errors (bottom) along the wall normal direction in the transitional boundary layer flow from the conventional and SPAV PINNs.}
    \label{fig:TBL-V profile along height}
\end{figure}

Figure~\ref{fig:TBL-uvwpcontour} compares cut plots of the transitional boundary layer. The figure shows three orthogonal planes: a quarter-height ($z = 0.43$~mm), rear ($y = 8$~mm), and left ($x = 8$~mm) plane. Velocity and pressure fields and errors are shown on the left- and right-side of the figure. The SPAV and conventional PINNs generate qualitatively similar flow fields. Slight differences are apparent in the $u$- and $v$-components of velocity as well as the pressure field. Absolute errors help to illustrate the benefit of a SPAV loss for this flow, especially along the side of the domain. Further, Fig.~\ref{fig:TBL-Q} shows ground truth and reconstructed vortical structures based on the $Q$-criterion. As in the isoturb case, our SPAV method recovers a richer set of more accurate coherent structures than can be accessed with a conventional data loss.\par

To quantify the ability of conventional and SPAV PINNs to resolve the boundary layer, we plot the mean profiles of velocity in the wall normal direction, averaged over $x$, $y$, and $t$, along with the corresponding errors. Results can be seen in Fig.~\ref{fig:TBL-V profile along height}. SPAV profiles neatly follow the DNS data, exhibiting lower errors than the conventional profiles throughout the boundary layer. This is especially true of the $w$-component profiles: the conventional PINN violates the no-slip condition at the wall, while our SPAV PINN adheres to this condition. Note that both PINNs include a no-slip penalty, per Eq.~\ref{equ:boundary loss}, but the conventional data loss is far more susceptible to longitudinal errors from a DIH system, leading to non-physical reconstructions.\par

As in the cylinder and isoturb results, we compare distributions of velocity and pressure in the TBL flow along the central $x$-axis ($y = 4$~mm, $z = 0.86$~mm) in Fig.~\ref{fig:VP profiles along central x axis} as well as NRMSEs across all the frames in Fig.~\ref{fig:VP error traces}. Only MC results are shown for SPAV in the TBL case, these results closely resemble the ground truth DNS fields and exhibit errors up to 50\% lower than those from the conventional PINN. Taken together, the cylinder, isoturb, and TBL results provide strong quantitative and qualitative evidence that stochastic particle advection can enhance the accuracy of PTV in the presence of anisotrpic measurement uncertainties.\par

\subsection{Experimental results}
\label{sec:DIH-PTV:experiments}
Following our synthetic tests, we demonstrate SPAV on two experimental data sets that feature a canonical flow: laminar micro-channel \cite{Toloui2015} and turbulent channel \cite{Toloui2017} flow. Results from our simulated studies may be used to understand the ability and limitations of the conventional and SPAV data losses in these tests.\par

\subsubsection{Laminar micro-channel flow}
\label{sec:DIH-PTV:experiments:micro-channel}
We first present the results of our micro-channel test. Due to poor particle reconstructions close to the front and back walls of the micro-channel, we limit our reconstructions to the central 0.9~mm region of the channel, from $z = 0.05$~mm to 0.95~mm. Since this flow is fully developed within the measured region of the micro-channel, it corresponds to Poiseuille flow and should exhibit a parabolic streamwise velocity profile with no transverse motion.\par

\begin{figure}[ht]
    \centering
    \subcaptionbox{\label{fig:mc:v profile}}{\includegraphics[height=4.2cm]{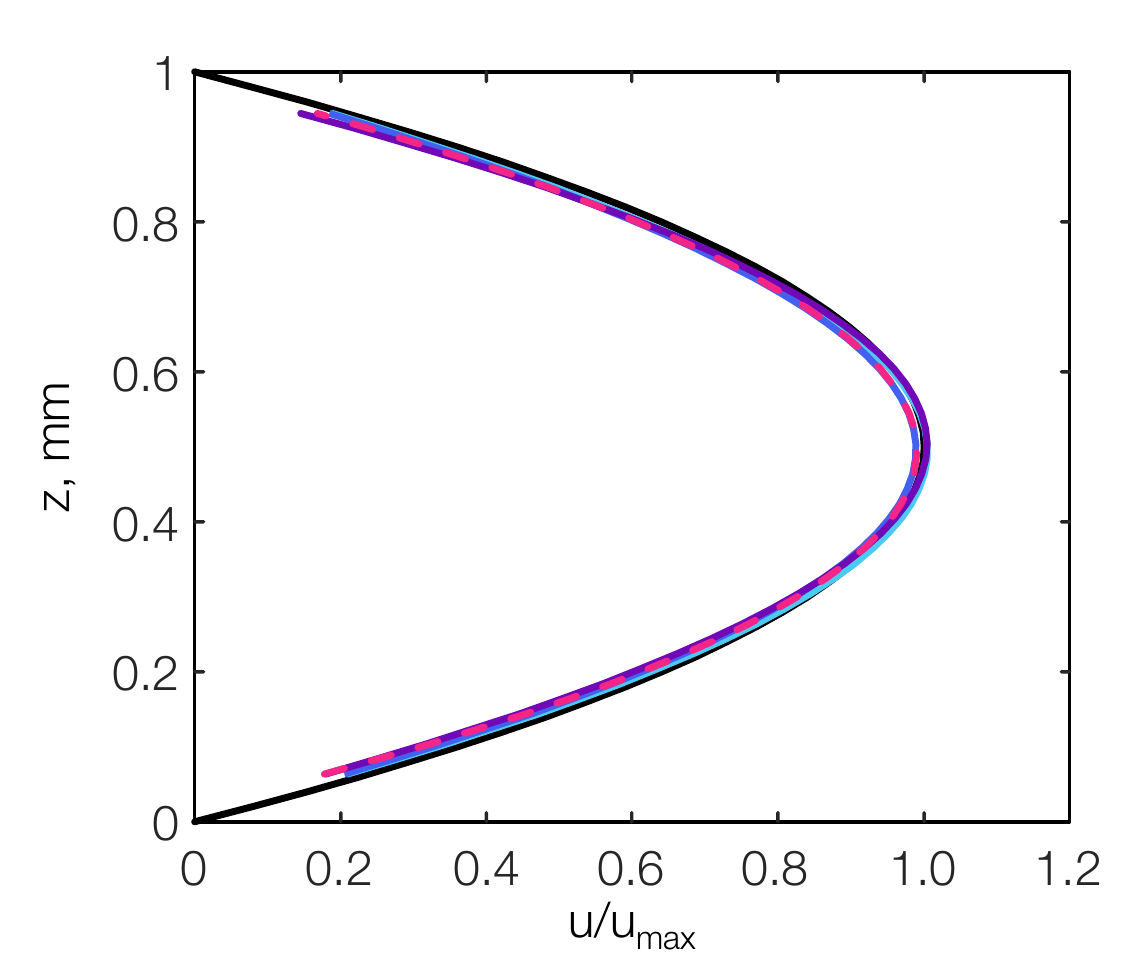}}
    \subcaptionbox{\label{fig:mc:v error profile}}
    {\includegraphics[height=4.2cm]{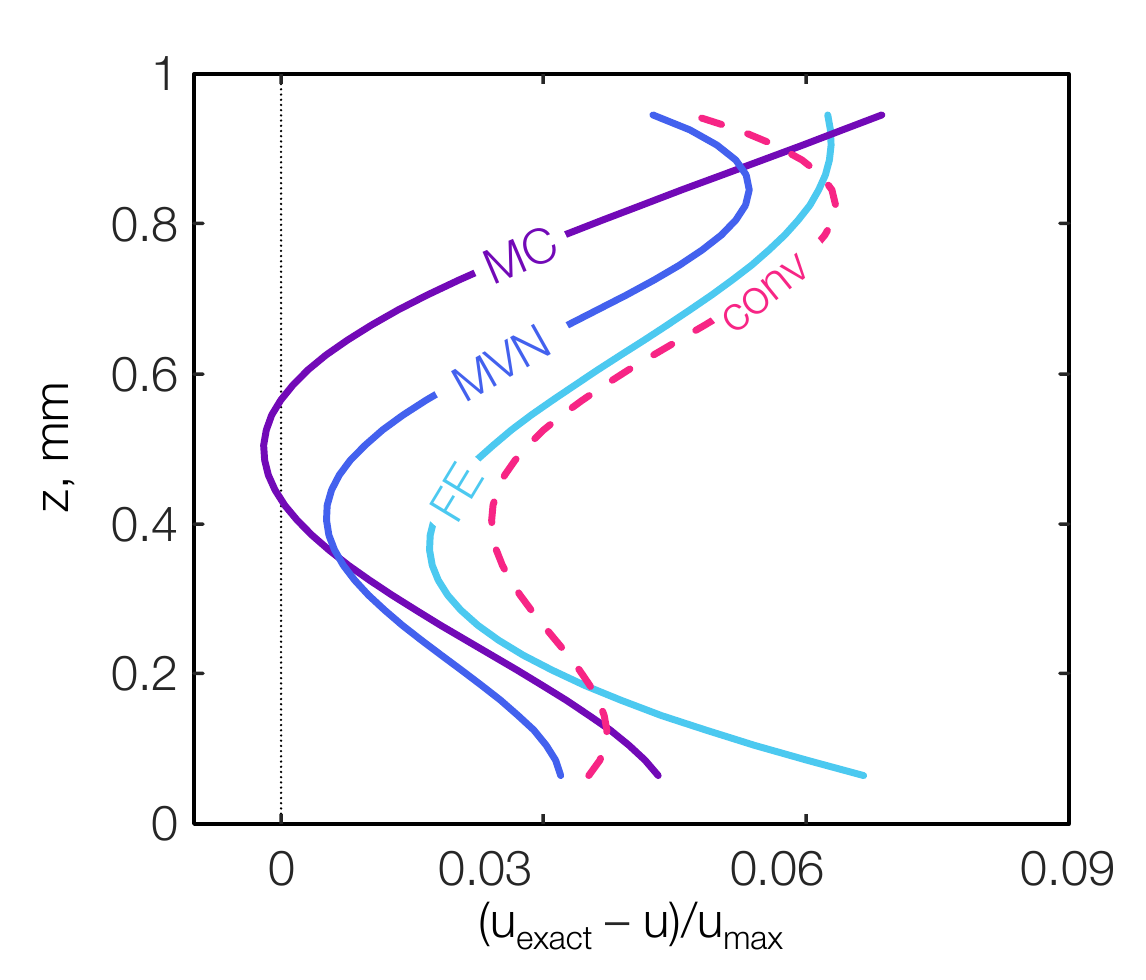}}
    \subcaptionbox{\label{fig:mc:p profile}}{\includegraphics[height=4.2cm]{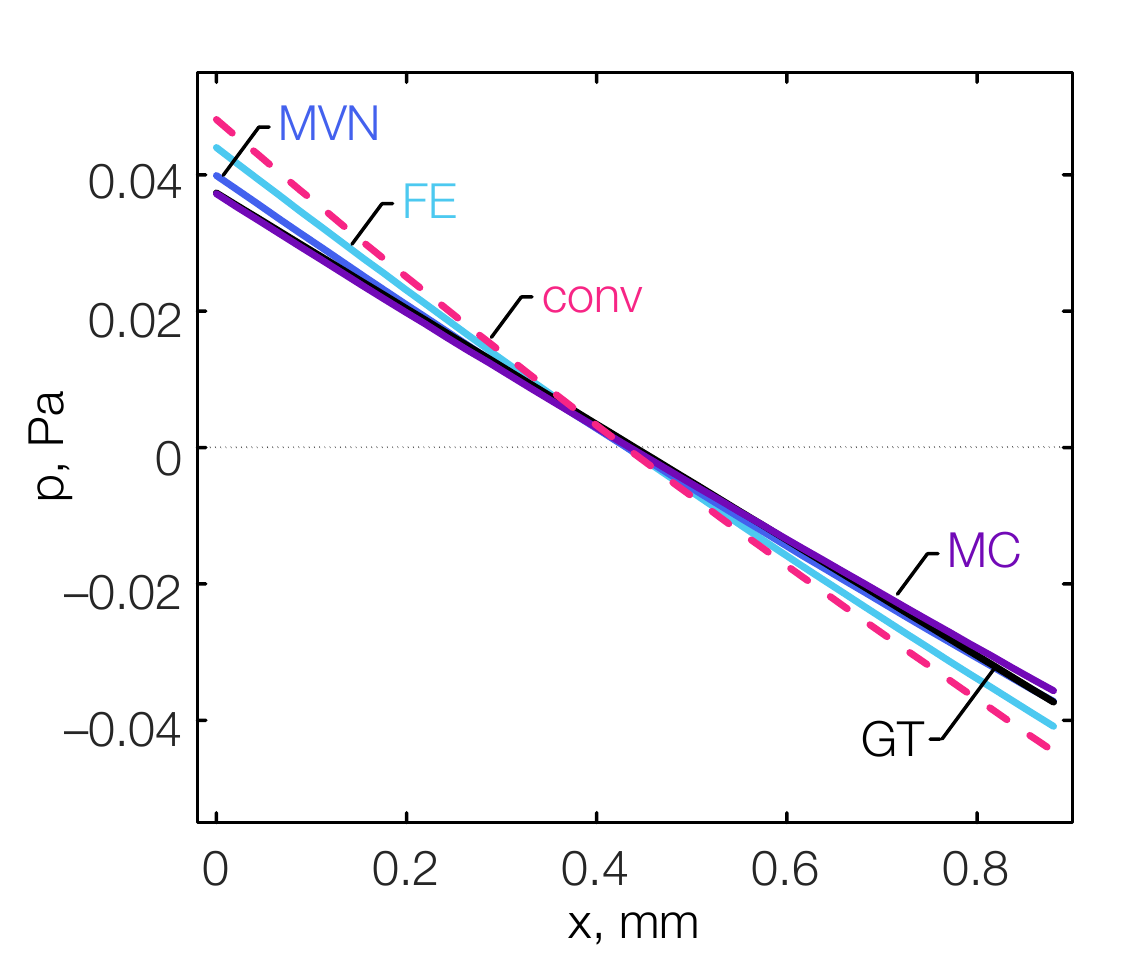}}
    \vspace*{2mm}
    \caption{Velocity and pressure distributions from a DIH-PTV micro-channel flow experiment: (a) instantaneous streamwise velocity profiles obtained by different methods vs. the analytical Poiseuille profile, (b) streamwise velocity error profiles, and (c) streamwise pressure drops estimated by different methods versus the analytical solution. Color coding: 
    ``\protect\GTline''\space for ground truth, ``\protect\Convline''\space  for conventional, ``\protect\MCline''\space for MC, ``\protect\MVNline''\space for MVN, and ``\protect\FEline''\space for FE.}
    \label{fig:mc}
\end{figure}

Instantaneous streamwise velocity profiles obtained by different methods are compared with the analytical solution on Fig.~\ref{fig:mc:v profile}. These profiles are extracted along the central spanwise axis ($x = 4.4$~mm, $y = 2.2$~mm) in the central frame and normalized by the maximum streamwise velocity, $u_\mathrm{max} = 5.3$~mm/s. This value is the ensemble average of the central streamwise velocity from the MC PINN. Estimates of $u_\mathrm{max}$ from the other PINNs are within 2\% of this value. The reconstructed velocity profiles in Fig.~\ref{fig:mc:v profile} closely match the analytical profile. Velocity errors are shown in Fig.~\ref{fig:mc:v error profile}, which indicates a maximum discrepancy of 6\%, similar to that reported in the original work of Toloui and Hong \cite{Toloui2015}. Slight differences between these results are attributed to localization uncertainties and non-ideal experimental conditions, e.g., the cross-section shape, unsteady pumping, etc. Not shown in this figure are the transverse components of velocity. Mean values of $[|v|, |w|]/u_\mathrm{max}$ -- which should equal zero -- are $[0.019, 0.018]$, $[0.018, 0.019]$, $[0.017, 0.020]$, and $[0.010, 0.009]$ for the conventional, FE, MVN, and MC PINNs, respectively.\par

Further quantitative analysis may be performed for the pressure field. The streamwise pressure gradient in fully-developed pipe flow is an analytical function of $u_\mathrm{max}$,
\begin{equation}
    \frac{\mathrm{d}p}{\mathrm{d}x} = -\frac{\mu\,u_{\mathrm{max}}}{D_\mathrm{H}^{2}},
    \label{equ:pressure drop}
\end{equation}
where $\mu$ is the dynamic viscosity and $D_\mathrm{H}$ is the hydraulic diameter of the channel. Figure~\ref{fig:mc:p profile} compares the analytical pressure field (up to a constant offset) to instantaneous estimates from the conventional and SPAV PINNs. All profiles are centered, and 3D pressure field estimates from the PINNs are averaged across spanwise ($y$-$z$) planes to obtain a streamwise pressure drop profile.  The conventional data loss yields the least accurate result, followed by the FE, MVN, and MC SPAV losses in that order. These PINNs estimate a pressure gradient of 0.107~Pa/mm, 0.097~Pa/mm, 0.088~Pa/mm, and 0.084~Pa/mm, compared to the analytical value of 0.085~Pa/mm. Note that the conventional PINN overestimates the pressure drop by 26\% for this flow. This overestimation is largely produced by the inaccurate (i.e., non-zero) estimates of spanwise velocity. By contrast, the PINN trained with an MC data loss faithfully captures the analytical gradient, with only a small error of roughly 1\%. This demonstrates the value of SPAV for obtaining a correct estimate of velocity from experimental PTV tracks.\par

\subsubsection{Turbulent channel flow}
\label{sec:DIH-PTV:experiments:turbulent channel}
\begin{figure}[ht]
    \centering
    \subcaptionbox{\label{fig:tc:tracks}}{\includegraphics[height=7cm]{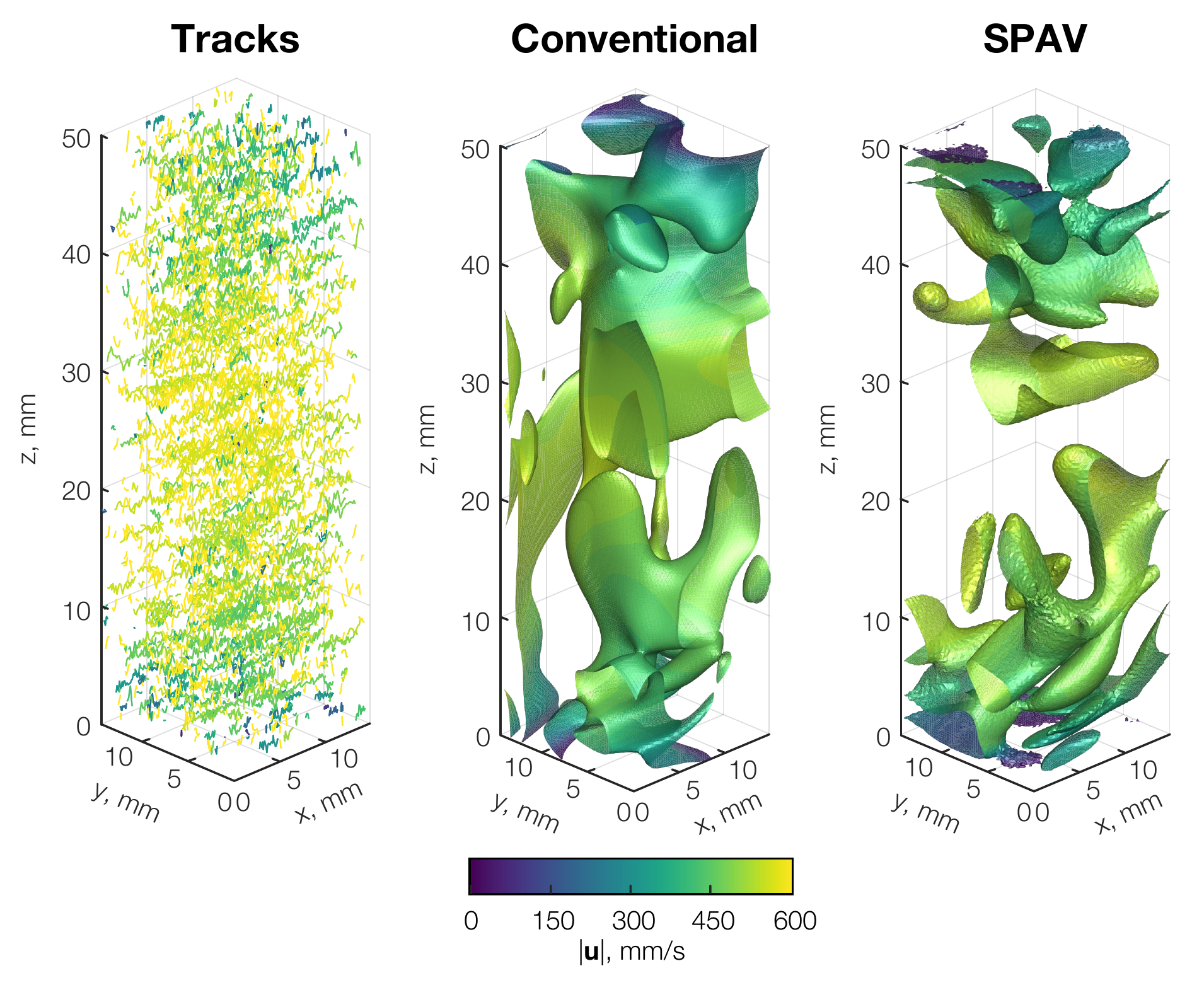}}\quad
    \subcaptionbox{\label{fig:tc:wall func}}{\includegraphics[height=7cm]{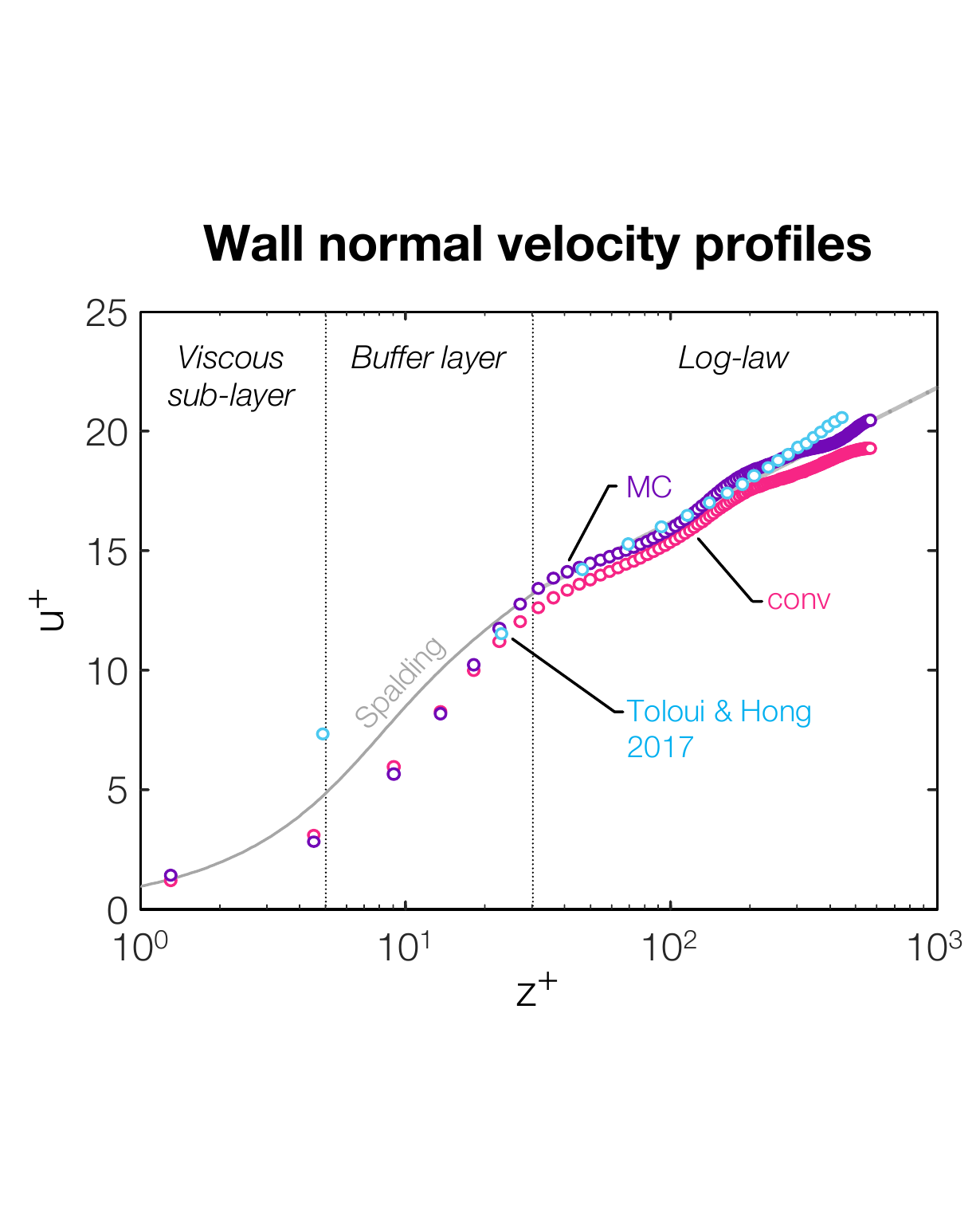}}
    \vspace*{2mm}
    \caption{Results of a DIH-PTV turbulent channel flow experiment: (a) particle tracks and coherent structures and (b) mean streamwise velocity profiles. One-tenth of the whole tracks are visualized for clarity. Vortical structures (left) are rendered with $Q$-criterion isosurfaces ($Q = 6.6$~s$^{-2}$) obtained with a conventional PINN (middle) and SPAV PINN (right). Surface color represents the flow velocity magnitudes. Spalding's wall function and Toloui's data \cite{Toloui2017} are plotted with the PINN results for reference.}
    \label{fig:tc}
\end{figure}

Lastly, we use our SPAV framework to reconstruct a turbulent channel flow from experimental holograms. A major challenge here is the large longitudinal dimension of the domain (50~mm), resulting in DIH reconstructions with low longitudinal resolution (50~$\upmu$m), which is limited by available CPU memory. Coarse longitudinal resolution exacerbates particle localization errors, leading to poor quality particle tracks. Figure~\ref{fig:tc:tracks} shows a subset of the tracks from this test, which contain strong longitudinal perturbations and even erroneous matches across consecutive frames. Such errors pose serious problems for PTV reconstruction algorithms.\par

We estimate velocity and pressure fields in this turbulent channel flow using PINNs with a conventional PINN and SPAV data loss. Only the MC variant is implemented for this case due to the strong velocity gradients. To visualize vortical structures in the channel, isosurfaces of the $Q$-criterion are also plotted in Fig.~\ref{fig:tc:tracks}. Coherent structures from the SPAV reconstruction are elongated along the streamwise wall-normal direction, growing towards the outer layer with a inclination angle of 30 to 60 degrees. These structures are consistent with past numerical and experimental studies of zero-pressure gradient smooth-wall turbulent boundary layers \cite{Ashrafian2004, Lee2011, volino2009, Wu2010}. By contrast, structures produced by the conventional PINN are totally corrupted by the large localization errors, showing high connectivity across the whole channel. These results lend credence to the velocity fields obtained using our SPAV loss.\par

Next, we examine the velocities fields by invoking the law of the wall, which describes the quasi-universal mean velocity profile of turbulent wall-bounded flows. Figure~\ref{fig:tc:wall func} compares the mean streamwise velocity profile, averaged across 40 frames (corresponding to a time span of 13~ms), from the MC and conventional PINNs, and Spalding's wall function is plotted for reference. We calculate the friction velocity to be $u_\uptau = 23$~mm/s using a logarithmic fit to the mean streamwise velocity profiles from $z = 0$~mm to 6~mm \cite{Toloui2017}; the corresponding wall unit is roughly 43.5~$\upmu$m. Results from the original work by Toloui et al. \cite{Toloui2017} are also provided for comparison. It should be noted that the original DIH-PTV workflow in \cite{Toloui2017} is different from our approach in two key aspects. First, Toloui et al. utilized a modified multi-pass Crocker--Grier algorithm with track smoothing to improve the density and accuracy of tracks, whereas we use the vanilla Crocker--Grier technique. Second, the original work employed a two-step approach to velocity reconstruction: (1) a Taylor series-based interpolation of the unstructured PTV velocity fields onto 3D grid points followed by (2) a continuity-based velocity refinement. Conversely, this work uses PINNs to reconstruct the flow from noisy tracks with either a SPAV or conventional data loss.\par

We draw two major conclusions from Fig.~\ref{fig:tc:wall func}. First, despite the use of error-laden tracks from a generic tracking algorithm, both the conventional and SPAV PINNs approximately resolve the near-wall flow profile within the viscous sub-layer. The earlier procedure from Toloui and Hong \cite{Toloui2017} barely resolved the buffer layer, despite using more accurate tracks. This speaks to the power of PINN-based DA for reconstructing flows, owning to a PINN's continuous, functional representation of the target flow. Second, the conventional PINN slightly underestimates the mean velocity in the outer layer. This may be attributable to over-smoothing of the flow to combat substantial errors in the Lagrangian tracks. Therefore, taken together with the laminar experiment, this test demonstrates the ability of our SPAV framework to improve velocity and pressure estimates in PTV despite significant localization and tracking errors.\par

\section{Conclusions and outlook}
\label{sec:conclusions}
This work presents a technique for data assimilation, termed \textit{stochastic particle advection velocimetry}, to improve the accuracy of Eulerian fields extracted from PTV measurements. Our approach incorporates an explicit particle advection model into a statistical data loss term, which accounts for arbitrary localization and tracking uncertainties. The data loss is combined with the governing physics to reconstruct velocity and pressure fields from Lagrangian particle tracks. Three numerical approximations to the SPAV loss are designed for practical use: Monte Carlo, multivariate normal, and fluid element SPAV, each having a unique level of accuracy and computational efficiency. SPAV is deployed using a physics-informed neural network, which simultaneously minimizes a set of measurement, physics, and boundary losses. We demonstrate the technique using simulated and experimental holograms from digital in-line holography PTV tests. However, SPAV is a general framework that can be applied to all forms of PTV and may be incorporated into other DA solvers, e.g., state observer methods, Kalman filters, adjoint--variational codes, etc. Several important conclusions can be drawn from this work.
\begin{enumerate}[topsep=.5ex, itemsep=-.5ex, partopsep=.5ex, parsep=.5ex]
    \item Given noisy Lagrangian particle tracks, a SPAV data loss can improve the accuracy of velocity and pressure field estimates compared to a conventional velocity loss. In our synthetic tests, SPAV reduced velocity and pressure errors by an average of 50\% and 30\%, respectively. Furthermore, in our laminar flow experiment, the pressure drop estimated by SPAV was within 1\% of the analytical value, as compared to the conventional PINN, which was off by 26\%.
    
    \item By using the fewest approximations, the MC variant of SPAV consistently produces the most accurate reconstructions, irrespective of the flow characteristics. Meanwhile, the MVN and FE techniques only provide marginal improvements over the conventional data loss for turbulent flows, in which strong velocity gradients generate non-Gaussian advected particle PDFs.
    
    \item SPAV also enhances the accuracy of flow derivatives compared to a conventional workflow. In our simulated and experimental tests, alike, SPAV-based coherent structures (viz., isosurfaces of the $Q$-criterion field) were more detailed and accurate than those obtained with a standard velocity loss.
\end{enumerate}
Ultimately, SPAV provides a pathway to extract more accurate high-order derivatives, statistics, and structures of turbulent flows from experimental PTV data. Quantities of interest may include Reynolds stresses, dissipation rate fields, and coherent structures, which can potentially provide new insights into turbulence.\par

We have identified several avenues for future research to stabilize, validate, and enhance the performance of SPAV.
\begin{enumerate}[topsep=.5ex, itemsep=-.5ex, partopsep=.5ex, parsep=.5ex]
    \item The speed and domain size of PINN-based SPAV can be increased through GPU parallelization and domain decomposition, for instance, using an extended PINN \cite{Jagtap2021}. The enhanced expressivity of an extended network could help practitioners to resolve the finer scales of an intense turbulent flow.
    
    \item Inertial effects could be incorporated into the PAV or SPAV framework to account for non-ideal PTV measurements in extreme scenarios, such as cases with non-negligible thermophoretic or drag forces.
    
    \item Lastly, the SPAV framework can be leveraged to improve the accuracy of other PTV techniques, including conventional multi-camera, plenoptic, and synthetic aperture PTV. Further, SPAV can be combined with a high-fidelity data assimilation technique, like an adjoint--variational method, as reviewed in \ref{app:DA}.
\end{enumerate}\par

\appendix
\renewcommand{\thesection}{Appendix \Alph{section}}

\section{Review of data assimilation for PIV and PTV}
\label{app:DA}
This appendix briefly reviews data assimilation techniques that have been applied to PIV and PTV. DA confers several benefits. First, physical constraints are enforced -- albeit weakly, in some cases -- which serves to enhance the spatio-temporal resolution of estimates and suppress noise. Second, latent (i.e., not-directly-measured) state variables can be inferred from the measurements, for instance, estimating pressure from particle image data and thereby providing richer insight into the flow. Third, when using a high-fidelity solver, the flow may be faithfully simulated beyond the measurement horizon, called forecasting, e.g., to resolve key statistics. This review groups DA techniques for particle-based velocimetry into four categories: observer methods, adjoint methods, physics-informed interpolation, and physics-informed machine learning. The former two categories are often regarded as ``measurement-enhanced simulation'' while the later two are better thought of as ``physics-enhanced post-processing''.\par

To the best of our knowledge, every demonstration of DA for PIV and PTV to date has featured a na{\"i}ve comparison of the measured and predicted velocity fields. Embedding an explicit advection model like SPAV into a DA algorithm has the potential to improve the accuracy of all the techniques reported below.\par

\subsection{Observer methods}
\label{app:DA:observer}
Observer methods employ a control theoretic framework and treat the flow as a dynamical system whose states -- namely, velocity, pressure, and other fields -- are estimated with a numerical model. This model is tuned by the ``observer'' to match experimental measurements. Observer DA begins with an approximate initial flow state, which is evolved using a CFD simulation to predict future states. Real measurements of the flow are compared to synthetic data, generated from the predicted states, and discrepancies between the measured and modeled data are used to construct a feedback signal. Subsequent flow state predictions incorporate this feedback to minimize the measurement residuals with the goal of reproducing true states of the flow. Two leading observer methods that have been used for PIV and PTV DA are the state observer method \cite{Nisugi2004} and Kalman filter (KF) method \cite{Takehara2000}, designed for deterministic and stochastic systems, respectively.\par

State observers directly estimate flow states with a CFD algorithm that has been modified to incorporate feedback. Design of the feedback mechanism is a key aspect of these algorithms. A common approach is to specify a force in the momentum equations, formulated in terms of residuals between the CFD-based velocity fields and PIV data, to reduce the residuals at future timesteps \cite{Yamagata2008, Saredi2021}. Given an appropriate feedback mechanism, this procedure yields flow fields that gradually approach the true state, at which point the feedback signal should vanish. In an early example, Hayase and Hayashi \cite{Hayase1997} used a state observer to reconstruct synthetic turbulent flow in a square duct. Residuals between the predicted and ``measured'' velocity fields were used to drive the pressure boundary conditions with a proportional controller. Later, Imagawa and Hayase \cite{Imagawa2010} implemented a state observer to reconstruct synthetic turbulent flows in a duct. Their feedback comprised a force in the momentum equations that was proportional to the velocity residuals. More recently, Saredi et al. \cite{Saredi2021} incorporated a state observer into a Reynolds-averaged Navier--Stokes (RANS) model using time-averaged tomographic PIV data; they successfully reconstructed the turbulent flow fields around a wall-mounted bluff body. State observers are relatively cheap and simple to implement and hence attractive for aerodynamic design and optimization problems \cite{Hayase2015}. However, the deterministic framework only works well with high-fidelity measurements. Moreover, design and implementation of the feedback mechanism is largely heuristic, lacking a strong theoretical foundation.\par

In KF DA, the flow is conceived as a stochastic system. Flow fields are treated as random variables that are characterized by PDFs, which encode one's state of knowledge about the fields. The basic framework is similar to a deterministic observer, except that the CFD model is used to predict both the state estimate and its covariance (uncertainty), which are incorporated into the updated state via a Kalman filter. The simplest KF techniques are limited to linear models and Gaussian PDFs, but most flow problems are high-dimensional, nonlinear, and may exhibit non-Gaussian uncertainties, especially in the context of turbulence \cite{Colburn2011}. Several KF variants have been developed to tackle these problems, such as the extended KF and ensemble KF (EKF and EnKF). For instance, Suzuki \cite{Suzuki2012} designed a reduced-order EKF algorithm to assimilate 2D PTV measurements of a planar jet with a DNS. The authors devised a cheap, linear approximation to the covariance matrix, which only holds for weakly nonlinear systems. By contrast, EnKFs can approximate arbitrary distributions by propagating an ensemble of perturbed states through the full nonlinear dynamical model. Hence, nonlinear effects on the system statistics are properly resolved. Several works have demonstrated EnKF for flow DA. Deng et al. \cite{Deng2018} utilized EnKF with a RANS model to recover the flow field of a 2D turbulent jet from time-averaged PIV data. The EnKF-optimized RANS parameters led to a more accurate simulation of the real flow. Mons et al. \cite{Mons2016} employed a modified EnKF, called an ``ensemble Kalman smoother'', to reconstruct unsteady 2D cylinder wakes from sparse synthetic velocity data with a DNS model. The authors compared this approach to variational DA, showing similar performance at a much lower computational cost and model complexity. Nevertheless, most demonstrations of KF DA for 2D velocimetry to date have been conducted with an EnKF. For 3D turbulent flows, a large number of ensemble members, i.e., CFD runs, is required to reliably approximate the system statistics, posing a significant computational cost.\par

\subsection{Adjoint methods}
\label{app:DA:adjoint}
Adjoint--variational DA algorithms seek the solution to a constrained optimization problem. In the context of fluid flow, the objective loss is a scalar that may compare a wide range of measured and modeled data, e.g., Lagrangian tracks from a PTV system, pressure traces from a series of taps, wall shear stress estimates from images of a stress sensitive coating, and so on. The flow is parameterized in terms of a control vector, such as the initial state of all the fields and/or time-resolved boundary conditions, which usually has a large number of dimensions \cite{Wang2022b}. The control vector is employed in a \textit{forward} CFD simulation, which ensures that the resulting flow fields satisfy the governing equations. Simulated flow fields are used to estimate synthetic data, which are compared to real measurements via the loss. This approach provides greater flexibility than observer methods, which generally require a direct relationship between measured signals and computed flow fields.\par

In adjoint optimization, one computes the sensitivity of the objective loss with respect to the control vector. To do this, another set of equations, called the \textit{adjoint equations}, are derived using Lagrange multipliers to convert the original constrained problem into an unconstrained one \cite{Wang2022b}. Solving the adjoint equations is akin to a \textit{reverse} simulation, and the adjoint solution can be used to calculate the gradients of interest. The control vector is then updated by steepest descent, i.e., stepping down the objective surface along the gradient. This forward--adjoint--gradient descent cycle is repeated until an optimum is reached. Unlike observer methods, each step of an adjoint--variational DA algorithm produces flow fields for the full time horizon of an experiment, which can improve accuracy and facilitate forecasting: simulating the flow beyond the measurement interval \cite{Chandramouli2020}.\par

Although originally developed for meteorological applications like weather prediction \cite{Le1986}, adjoint--variational solvers have been extensively applied to fluid dynamics problems in engineering. Early works include the reconstruction of 2D cylinder wakes from PIV measurements, based on either a DNS \cite{Gronskis2013} or RANS simulation \cite{Foures2014}. However, extending this approach to fully 4D variational DA (4DVar) comes at an exorbitant cost due to the repeated forward and backward DNSs \cite{Chandramouli2020, He2020, He2022}. To address the high cost of 4DVar, Chandramouli et al. \cite{Chandramouli2020} reduced the simulation's spatial resolution by introducing a dynamic error term associated with the unresolved scales (essentially a subgrid closure). Chandramouli's algorithm was validated with synthetic 3D PIV measurements of a turbulent cylinder wake flow, showing promise for real experiments. Another strategy to reduce the cost of 4DVar involves lowering its temporal resolution, as proposed by He et al. \cite{He2020, He2022}. A time domain decomposition is performed and 4DVar is applied sequentially across the temporal subdomains. The authors solve the forward equations at all timesteps, as in conventional CFD, but they also conduct an adjoint loop and optimization in subdomains where measurement data is present. In this case, the control vector is a forcing term added to the momentum equations. He et al. validated their sequential 4DVar technique by estimating velocity and pressure fields in a circular jet with experimental tomographic PIV measurements. Their demonstration constitutes one of the only 4DVar tests with experimental data. However, their reconstructions were only piece-wise continuous, and the flow estimates exhibited nonphysical jumps across the temporal subdomains.\par

Despite continuing efforts to develop 4DVar, its application to highly turbulent flow remains a challenge because the adjoint model fails to estimate useful gradients when the system is too chaotic \cite{Lea2000}. Advanced methods like least squares shadowing \cite{Wang2014} have been developed and show potential to tackle this problem.\par

\subsection{Physics-informed interpolation}
\label{app:DA:interpolation}
Physics-informed interpolation is used to fill-in coarse PIV data or sparse PTV tracks. In these algorithms, interpolation is subject to soft constraints like a divergence penalty \cite{deSilva2013, Wang2016} or loss derived from the incompressible Navier--Stokes equations \cite{Schneiders2016, Schanz2018}. Often, a continuous representation of the flow is implemented using splines or radial basis functions, whose coefficients are obtained from an optimization procedure. Different from the observer or adjoint--variational methods, interpolation techniques do not rely on a CFD model to predict flow states. Instead, these states are projected onto the chosen basis from measured data and optimized by penalizing large deviations from the data and governing equations. Consequently, interpolation-based DA tends to be cheaper and faster than observer, adjoint, and machine learning methods. It should also be noted that there are many techniques for estimating pressure from PIV or PTV data that do not attempt to improve the reconstructed velocity field (see \cite{Zhang2020} for a recent example). However, this section is focused on algorithms that utilize physics to enhance Eulerian estimates of \textit{velocity}.\par

Two popular interpolation schemes that are actively used and developed by the PTV community are FlowFit \cite{Gesemann2016, Schanz2016} and VIC+ \cite{Schneiders2014, Schneiders2016}. The original formulation of FlowFit takes time-resolved PTV tracks as input and outputs more accurate, space-filling velocity and pressure fields. These fields are represented by a 3D B-spline basis. The objective loss comprises (i) the disparity between measured and interpolated velocity and acceleration data -- the latter of which can be inferred from the momentum equations as well as the particle tracks -- and (ii) the aggregate divergence of the velocity and Eulerian acceleration fields. VIC+ uses the same inputs but only outputs a velocity field, which is supported by radial basis functions. In VIC+, the vorticity and acceleration fields are deduced from the vorticity transport equation. Both methods culminate in a nonlinear optimization problem that can be minimized by established algorithms, e.g., the limited-memory Broyden--Fletcher--Goldfarb--Shanno algorithm. Note that, although VIC+ employs a continuous basis to represent the field variables, derivatives thereof are calculated by projecting the fields onto a Cartesian grid and applying finite differences. By contrast, partial derivatives of a PINN and FlowFit are exact, analytical derivatives of the neural network and B-spline basis functions, respectively. Furthermore, existing formulations of FlowFit and VIC+ assume that the flow is incompressible.\par

Recent developments of FlowFit and VIC+ help to promote temporal smoothness. Previous versions only utilized snapshot measurements and lacked temporal consistency. FlowFit now incorporates artificial Lagrangian tracers (virtual particles), which are seeded into the domain and advected between timesteps to facilitate temporal coupling \cite{Ehlers2019, Ehlers2020}. These artificial tracers constrain the material derivative in undersampled regions of the flow, thereby suppressing nonphysical acceleration events. Key modifications to VIC+ include the ``time-segment-assimilation'' method, which estimates velocity data from full particle tracks. The flow is evolved forward and backward via the incompressible, inviscid vorticity transport equation, and a total cost function -- comparing measured and modeled velocity fields over the entire time horizon -- is minimized to recover velocity data at the central frame \cite{Scarano2022}.\par

It should be noted that the advection of artificial tracers in the modified FlowFit algorithm is distinct from PAV. In FlowFit, tracers are advected by the instantaneous velocity field estimate at one timestep to inform the following estimate, and this is repeated for subsequent timesteps. The result is a series of 3D inverse problems. By contrast, in PAV, we perform particle advection and flow optimization simultaneously across the entire time horizon, which amounts to a 4D inverse problem.\par

\subsection{Physics-informed machine learning}
\label{app:DA:ML}
There has been a rapid uptake in the use of machine learning to interpret measurements and assimilate them with a physical model. Many researchers train deep neural networks to denoise or improve the resolution of sparse PIV or PTV data, e.g., \cite{Fukami2019, kim2021, Rabault2017, Lee2017}. This is usually done with a supervised model in a purely data-driven manner, which requires a large, high-fidelity, and accurately labelled training set. As a corollary, these techniques do not always perform well on novel data sets. It is thus necessary to incorporate physics into the DA process to develop a generalizable machine learning method for PTV.\par

Physics-informed machine learning was recently established as a robust, physics-based approach to functional regression \cite{Raissi2019, Cai2022}. The technique, described in detail in Sect.~\ref{sec:method:PINN}, employs a deep neural network to approximate flow fields in functional form. The function, defined by the network's weights, biases, activation functions, and architecture, maps spatio-temporal coordinates to the flow fields of interest. Derivatives of the PINN are used to evaluate the governing equations as well as predict measurements of the flow. Residuals from the equations are added up in a physics loss and synthetic data from the PINN are compared to real measurements in a data loss; these losses may be augmented with boundary condition losses where applicable. Lastly, an aggregate loss is minimized by backpropagation to yield a function that approximately satisfies physical laws and replicates experimental measurements.\par

While PINNs can be used to solve well-posed forward problems, they are especially useful in the context of inverse analysis \cite{Cai2022}.\footnote{Even when PINNs are used to solve a forward problem, the technique essentially amounts to an inverse solver.} Numerous papers have shown the potential of PINNs to regularize reconstructions and recover latent states of a flow \cite{Molnar2022a, Molnar2022b}. Some noteworthy examples of PINN-based velocimetry have come from Han and coworkers \cite{Han2021}, who tested flow around a car's side mirror using LaVision's 4D Lagrangian robotic PTV system; Di Leoni et al. \cite{DiLeoni2022}, who reconstructed the shear layer behind a backward-facing step in a water tunnel using sparse STB-based particle tracks; Wang et al. \cite{Wang2022a}, who employed a PINN to improve tomographic PIV measurements of the 3D wake behind a hemisphere; and Soto et al. \cite{Soto2022}, who enhanced the temporal resolution of a 2D PIV test using a PINN to fuse synthetic PIV snapshots with fast point probe data. These tests and others \cite{vonSaldern2022} demonstrate the ability of PINNs to perform flow field DA with multi-resolution, multi-modal data.\par

Based on this preliminary success, researchers have been actively developing methods to enhance PINNs for PIV and PTV. PINNs tend to produce overly-smooth velocity fields, eradicating the fine structures present in a turbulent flow \cite{Wang2022a}. This is due to both noise in the data and the network's limited expressivity. In this paper, we demonstrate how SPAV can compensate for noise and improve the accuracy of reconstructions. There are several strategies to increase expressivity, including the use of tailored architectures, e.g., with ResNet \cite{Bu2021} or SIREN \cite{Pan2022} layers, and domain decomposition with extended PINNs \cite{Jagtap2021, Alhubail2022}. Other researchers have improved the performance of PINNs by embedding boundary conditions like periodicity into the network's architecture \cite{Du2021}, using adaptive activation functions \cite{Jagtap2020}, or dynamically weighting the loss function components \cite{Wang2021}.\par

\section{Overview of DIH-PTV}
\label{app:DIH-PTV overview}
Digital-in-line holography is a 3D imaging technique that is widely used to characterize microscopic objects. Time-resolved DIH measurements of a particle-laden flow can be utilized for PTV, termed DIH-PTV. A standard plane-wave DIH-PTV setup consists of a laser, beam forming optics, and a camera. These elements are usually set up in-line for PTV, with the camera and laser positioned on either side of the target flow. Light from the laser scatters off tracer particles and the reference and scattered waves interfere, forming a hologram on the camera's sensor. This information can be numerically refocused to reconstruct a 3D optical field from which particle positions can be extracted. Following these particles in time yields Lagrangian tracks that are suitable for Eulerian reconstruction. This appendix contains a brief review of DIH theory, an algorithm for hologram simulation, reconstruction of the optical field, and some particle extraction methods.\par

\subsection{Hologram formation}
\label{app:DIH-PTV overview:formation}
\begin{figure}[ht]
    \centering
    \includegraphics[height=5cm]{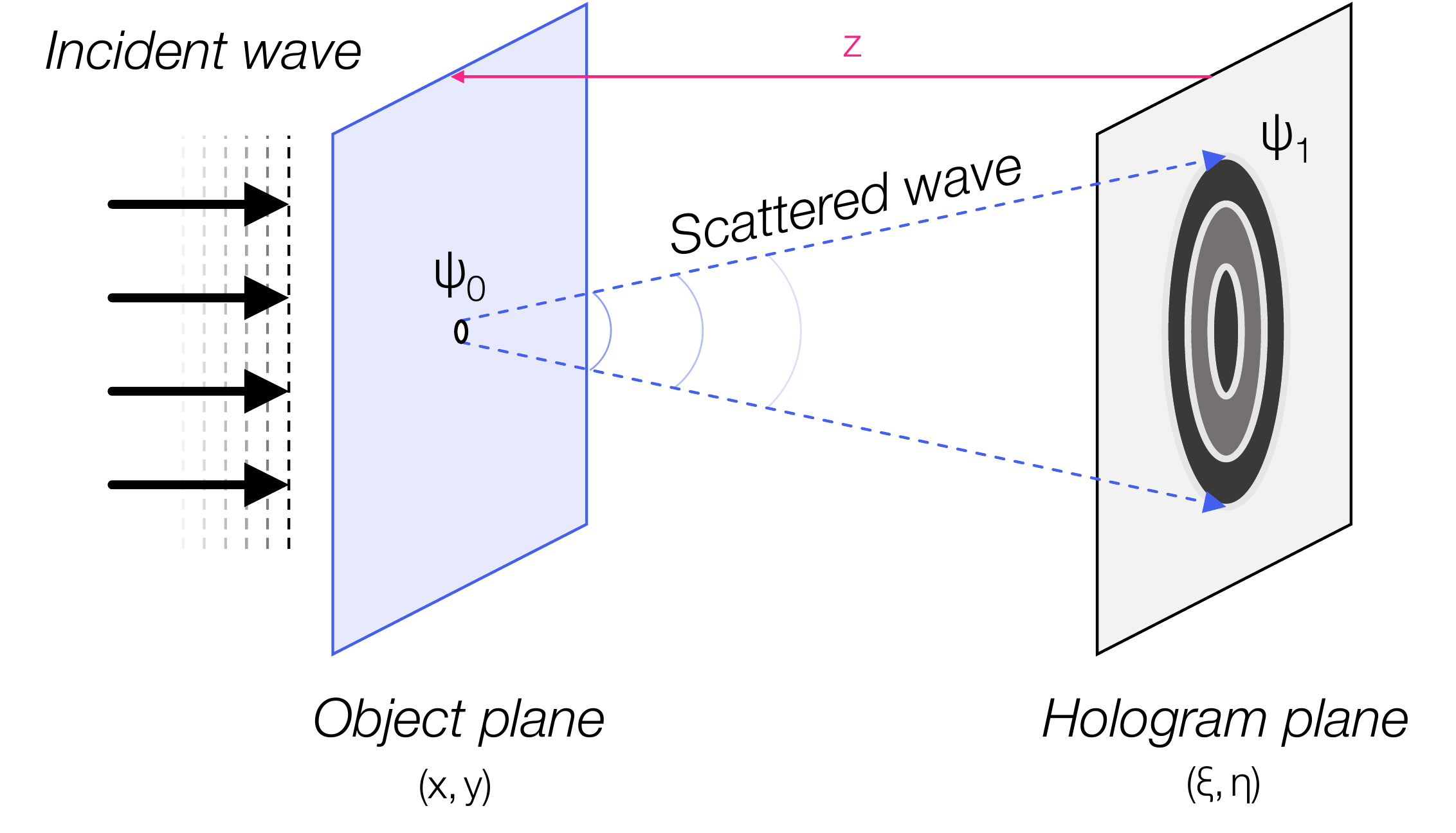}
    \caption{Graphical depiction of in-line hologram formation for plane wave DIH.}
    \label{fig:DIH schematic}
\end{figure}

In DIH-PTV, a plane or spherical wave of light is shone through a particle field towards a camera, which records the resultant interference patterns. These patterns carry information about the phase and magnitude of the light field which can be leveraged to reconstruct it. Figure~\ref{fig:DIH schematic} depicts a plane wave scattering off a quasi-point particle at $\Psi_0$ and then forming a hologram on the imaging sensor. The resultant field can be treated as a linear combination of the undisturbed ``reference wave'', $R$, and an ``object wave'', $O$, produced by scattering off the disturbance at $\Psi_0$. Both fields code the complex amplitudes of the electrical field of a light wave.\footnote{Note that separating the electric field into $R$ and $O$ is a mathematical treatment. In reality, only the total field is physically meaningful.}\par

Assuming a homogeneous, isotropic medium (e.g., free space or a lens with a constant refractive index), scalar diffraction theory can be used to describe the propagation of the electric field of a light wave \cite{Goodman2005}. The reference and object waves interfere with each other, and the complex wave amplitude in the hologram plane can be approximated by the Fresnel--Kirchoff diffraction formula, assuming the far-field limit (meaning that the distance from the particle to the sensor is orders of magnitude larger than the particle size),
\begin{equation}
    E\mathopen{}\left(\xi,\eta\right) = -\frac{\mathrm{i}}{\lambda} \iint \left[O(x, y) + R(x, y)\right] \times \frac{\exp\mathopen{}\left(\mathrm{i} \,k \,|\mathbf{r}| \right)} {|\mathbf{r}|} \, \mathrm{d}x \,\mathrm{d}y.
    \label{equ:forward propagation: FK formula}
\end{equation}
In this expression, $E$ is the electric field incident upon the hologram; $\xi$ and $\eta$ are sensor coordinates that are aligned with $x$ and $y$, respectively; $\mathrm{i} = \sqrt{-1}$; $\lambda$ is the wavelength of light and $k = 2\pi/\lambda$ is the wavenumber; and $\mathbf{r}$ is a vector from the object at $\Psi_0$ to a point in the hologram plane, e.g., $\Psi_1$ in Fig.~\ref{fig:DIH schematic}. The magnitude of $\mathbf{r}$ is
\begin{equation}
    |\mathbf{r}| = \sqrt{(x - \xi)^2 + (y - \eta)^2 + z},
    \label{equ:distance}
\end{equation}
where $z$ is the distance from the hologram to the object plane. According to the interference theory, the recorded hologram can be mathematically expressed as
\begin{equation}
    H\mathopen{}\left(\xi,\eta\right)
    = \left|R(\xi,\eta) + O(\xi,\eta)\right|^2 = |R|^2+|O|^2+R^*O+R O^*,
    \label{equ:detector intensity}
\end{equation}
where $(\cdot)^*$ denotes the complex conjugate. Here, $|R|^2$ is the constant background created by the reference wave, alone, $|O|^2$ is the intensity of the object wave, which is orders of magnitude smaller than that of the reference wave, and the terms $R^*O$ and $RO^*$ correspond to interference (fringes) in the hologram.\par

\subsection{Synthetic hologram generation}
\label{app:DIH-PTV overview:simulation}
Based on the hologram formulation theory outlined above, holograms can be simulated by choosing a suitable object model and diffraction point-spread function (PSF), which describes the propagation of a light wave from the object plane to a sensor. We start by simulating the hologram produced by a solitary object and extend this model to multi-object scenarios, including particle fields. Consider the $i$th spherical particle of radius $r_{i}$ and location $(x_{i}^\prime, y_{i}^\prime)$ in the object plane. The effect of this particle on the electric field can be modeled with an amplitude transmittance function (a.k.a. a mask function) \cite{Gao2014},
\begin{subequations}
    \begin{empheq}[left={T_{i}(x, y) = \empheqlbrace\,}]{align}
        & 0, \,\sqrt{(x-x_{i}^\prime)^2 + (y-y_{i}^\prime)^2} \leq r_{i} \\
        & 1, \,\sqrt{(x-x_{i}^\prime)^2 + (y-y_{i}^\prime)^2} > r_{i}
    \end{empheq}
    \label{equ:object transmittance}%
\end{subequations}
In this case, the distance between the particle and hologram plane is $z_{i}^\prime$. After interacting with the particle, the complex wave amplitude exiting the object plane, $E_{\mathrm{exit},i}$, may be expressed as 
\begin{equation}
    E_{\mathrm{exit},i}(x,y) = E_{\mathrm{inc},i}(x,y) \,T_{i}(x, y),
    \label{equ:exit wave}
\end{equation}
where $E_{\mathrm{inc},i}$ is the wave incident on the particle. The exiting wave, $E_{\mathrm{exit},i}$, contains both the reference and object waves, $E_{\mathrm{exit},i} = R_{i} + O_{i}$, and propagates towards the sensor plane. Using the angular spectrum method, one can approximate this propagation process with a convolutional integral, transforming Eq.~\eqref{equ:forward propagation: FK formula} into
\begin{equation}
    E\mathopen{}\left(\xi,\eta\right)=\iint E_{\mathrm{exit},i}(x,y) \,g\mathopen{}\left(x-\xi, y-\eta, z_{i}^\prime\right) \mathrm{d}x \,\mathrm{d}y = E_{\mathrm{exit},i}(x,y) \otimes g\mathopen{}\left(x, y, z_{i}^\prime\right),
    \label{equ:forward propagation: angular spectrum}
\end{equation}
where $\otimes$ is the convolution operator and $g$ is the diffraction PSF. This function is typically modeled by a Rayleigh--Sommerfield kernel,
\begin{equation}
    g(x,y,z) = \frac{1}{\mathrm{i}\lambda}\frac{\exp\mathopen{}\left[\mathrm{i} \,k \,\sqrt{x^2 + y^2 + z^2}\right]}{\sqrt{x^2 + y^2 + z^2}}.
    \label{equ:forward propagation: RS kernel}
\end{equation}
For efficient computation, the convolution in Eq.~\eqref{equ:forward propagation: angular spectrum} is usually converted to a multiplication in the Fourier domain, often using a fast Fourier transform,
\begin{equation}
    E\mathopen{}\left(\xi,\eta\right) = \mathcal{F}^{-1}\mathopen{}\left\{\mathcal{F}\mathopen{}\left[E_{\mathrm{exit},i}(x,y)\right] \times  \mathrm{exp}\mathopen{}\left[\mathrm{i} \,k \,z_{i} \sqrt{1 - \left(\lambda \,\omega_\mathrm{x}\right)^{2} - (\lambda \,\omega_\mathrm{y})^{2}} \right] \right\},
    \label{equ:forward propagation: FFT}
\end{equation}
where $\mathcal{F}(\cdot)$ is the Fourier transform, with inverse $\mathcal{F}^{-1}(\cdot)$, and $\omega_{\mathrm{x}}$ and $\omega_{\mathrm{y}}$ are spatial frequencies in the Fourier domain. Note that other diffraction kernels like the Fresnel kernel are available \cite{Latychevskaia2015}. However, the Rayleigh--Sommerfield kernel used in the angular spectrum method entails minimal assumptions and is thus widely applicable and widely adopted. A detailed explanation of diffraction kernels for DIH can be found in \cite{Kim2010}.\par

Thus far, Eqs.~\eqref{equ:object transmittance}--\eqref{equ:forward propagation: FFT} describe the simulation of a hologram for a single, solitary particle. PTV features dense particle fields, in which the light field incident on a particle may be affected by forward scattering from upstream particles and may thus affect downstream particles, in turn. We incorporate these interactions in our simulation by following the method of Gao \cite{Gao2014}; conversely, the magnitude of backwards scattering is orders of magnitude weaker and may be neglected \cite{Wriedt2012}.\par

For a multi-particle hologram, the particles are first sorted by distance to the sensor in descending order, that is, $z_{i+1} < z_{i}$. The complex wave amplitude exiting the $(i+1)$th particle plane can be computed using the wave exiting the $i$th particle plane such that
\begin{equation}
    E_{\mathrm{exit},i+1}(x,y) = \left[\smash{\underbrace{E_{\mathrm{exit},i}(x,y) \otimes g\mathopen{}\left(x, y, z_{i}^\prime - z_{i+1}^\prime\right)}_{E_\mathrm{inc,i+1}}}
    \vphantom{0_0\left.0_0^\prime\right.} \right] T_{i+1}(x,y). \vphantom{\underbrace{0_0}_{0_0}}
    \label{equ:forward propagation: multi-scattering}
\end{equation}
In planar DIH, the wave approaching the first particle is assumed to be a uniform plane wave: $E_{\mathrm{inc},1} = 1$. Finally, the multi-particle hologram is expressed as
\begin{equation}
    H\mathopen{}\left(\xi, \eta\right) = \left|E_{\mathrm{exit},N}(x,y) \otimes g\mathopen{}\left(x, y, z_{N}^\prime\right)\right|^{2},
    \label{equ:forward propagation: multi-scattering hologram}
\end{equation}
where $N$ is the number of particles and $E_{\mathrm{exit},N}$ is inductively computed via Eq.~\eqref{equ:forward propagation: multi-scattering}.

\subsection{Reconstruction}
\label{app:DIH-PTV overview:reconstruction}
``Reconstruction'' in DIH involves computing the object wave from the recorded holograms, which is generally performed in two steps. The first step is to ``numerically illuminate'' the hologram with the reference wave,
\begin{equation}
    R\mathopen{}\left(\xi, \eta\right) H\mathopen{}\left(\xi,\eta\right) = \underbrace{\left(|R|^2+|O|^2\right) R}_\text{(i)} + \underbrace{O|R|^{2}}_\text{(ii)} + \underbrace{O^*R^{2}}_\text{(iii)}\,,
    \label{equ:object intensity}
\end{equation}
where the multiplication of $R$ and $H$ represents the illumination of the hologram. In Eq.~\eqref{equ:object intensity}, (i) is a constant background, (ii) constitutes the real object image to be extracted, and (iii) is the twin image, which generates a distorted wavefront leading to an undesirable virtual image. In the second step, the light wave exiting the hologram is backpropagated to the object plane, assuming forward diffraction by the Fresnel--Kirchoff kernel,
\begin{equation}
    E\mathopen{}\left(x,y,z\right) = \frac{\mathrm{i}}{\lambda} \iint R\mathopen{}\left(\xi, \eta\right) H\mathopen{}\left(\xi,\eta\right) \times \frac{\exp\mathopen{}\left(-\mathrm{i} \,k \,|\mathbf{r}| \right)} {|\mathbf{r}|} \,\mathrm{d}\xi \,\mathrm{d}\eta.
    \label{equ:back propagation: FK formula}
\end{equation}
Similar to the forward hologram simulation, Eq.~\eqref{equ:back propagation: FK formula} is computed by the angular spectrum method,
\begin{equation}
  E\mathopen{}\left(x,y,z\right) = \left[R\mathopen{}\left(\xi, \eta\right) H\mathopen{}\left(\xi, \eta\right)\right] \otimes g\mathopen{}\left(\xi, \eta, -z\right),
  \label{equ:back propagation: RS kernel}   
\end{equation}
where the negative sign before $z$ represents the direction of backpropagation. Taking the reference wave to be unity, the convolution in Eq~\eqref{equ:back propagation: RS kernel} is computed in the Fourier domain,
\begin{equation}
  E\mathopen{}\left(x,y,z\right) = \mathcal{F}^{-1}\mathopen{}\left\{\mathcal{F}\mathopen{}\left[H\mathopen{}(\xi, \eta)\right] \times \exp\mathopen{}\left[-\mathrm{i} \,k \,z \sqrt{1 - \left(\lambda \,\omega_\mathrm{x}\right)^2 - \left(\lambda \,\omega_\mathrm{y}\right)^2} \right]\right\}.
  \label{equ:back propagation: FFT}   
\end{equation}
This process is repeated for a series of $x$-$y$ planes, often called ``scans'', which are stacked along the $z$-axis to form the reconstructed optical field. Note that the reconstruction process in Eqs.~\eqref{equ:back propagation: RS kernel} and \eqref{equ:back propagation: FFT} is based on the sensor intensity. Some other works describe an equivalent process that uses a \textit{contrast} hologram that is centered and normalized by the background image \cite{Berg2022, Latychevskaia2015}. The background can be obtained from a calibration image with no particles.\par

\subsection{Particle extraction}
\label{app:DIH-PTV overview:particle extraction}
To obtain particle tracks in DIH-PTV, individual particles must be extracted from the reconstructed 3D optical field. This usually involves applying a threshold to isolate filaments associated with an individual particle and then computing the centroid of each filament. There are three main challenges associated with extraction. Firstly, due to the extended DoF of DIH, extracted particle filaments are highly elongated in the longitudinal direction and it is difficult to precisely pinpoint the waist of filament. Secondly, 3D particle segmentation requires sophisticated image processing (e.g., denoising, SNR enhancement, and morphological operations) and heuristic parameter selection (e.g., filter sizes, intensity thresholds). In other words, DIH usually entails a lot of arbitrary human intervention. Thirdly, cross-interference between overlapping particle fringes decreases the SNR of holograms and reduces the number of detectable particles from the particle field, which worsens with the seeding density. This effect limits the minimum spacing of Lagrangian tracks, which poses difficulties when reconstructing a turbulent flow.\par

To address these challenges, we employ the DIH particle extraction method introduced by Toloui and Hong \cite{Toloui2015}. Their algorithm is built upon the backpropagation method described above, which is augmented with three improvements: one to tackle each challenge. We briefly sketch this method below; interested readers are referred to \cite{Toloui2015} for more details. The first improvement aims to enhance the longitudinal accuracy of the 3D optical field via deconvolution. The deconvolved optical field is given by
\begin{equation}
  I_\mathrm{deconv} = \mathcal{F}^{-1}\mathopen{}\left[
  \frac{\mathcal{F}\mathopen{}\left(I_\mathrm{PSF}\right)^* \mathcal{F}\mathopen{}\left(I_\mathrm{P}\right)}
  {\mathcal{F}\mathopen{}\left(I_\mathrm{PSF}\right)^* \mathcal{F}\mathopen{}\left(I_\mathrm{PSF}\right) + \beta} \right],
  \label{equ:extraction: deconvolution}   
\end{equation}
where $I_\mathrm{PSF}$ represents the intensity distribution of the optical system's PSF, $I_\mathrm{P}$ is the reconstructed 3D particle field in Eq.~\eqref{equ:back propagation: FFT}, $I_\mathrm{P} = |E|^2$, and $\beta$ is a small constant that prevents division by zero. The PSF can be determined by reconstructing the synthetic hologram of a single particle located at the center of the measurement volume. This 3D deconvolution models the DIH-reconstructed intensity field as a convolution of the true particle field with the system's PSF. Reversing (\textit{deconvolving}) this process can thus improve longitudinal resolution.\par

The second improvement involves a series of well-defined 3D image processing steps to avoid human intervention: 3D local SNR enhancement, automatic thresholding, and 3D segmentation. These procedures overcome the inhomogeneous SNR distribution across the deconvolved particle fields, such that a single intensity threshold can be automatically chosen for the entire domain to segment 3D particles from the background.\par

Lastly, the third improvement introduces an iterative inverse particle extraction (IIPE) method to increase detectable particle concentration. IIPE is able to remove detectable particles from the hologram, one by one, and reveal dim fringes that were previously hidden. Holograms of individual particles are removed by filling the target particle's in-focus cross-section in the reconstruction plane with a mean background value. This is followed by a forward process to generate the updated hologram, as described above in \ref{app:DIH-PTV overview:simulation}, which does away with the fringe.\par

\section{Measurement uncertainty sensitivity study}
\label{app:sensitivity study}
The DIH-PTV tests reported in Sect.~\ref{sec:DIH-PTV} exhibit a large, anisotropic distribution of localization uncertainty. This appendix presents a sensitivity analysis in which the magnitude and degree of anisotropy are varied for the forced isotropic turbulence case. Tests performed for this analysis are similar to the tests in Sect.~\ref{sec:DIH-PTV:synthetic:3D isoturb}, except that there is no synthetic DIH procedure. Instead, the particle positions are corrupted by additive errors having a prescribed magnitude and $z$-direction skew. Other than that, the domain size, measurement interval, number of particles, advection procedure, PINN architecture, training schemes, etc. of the sensitivity study are carried over from the isoturb case, as detailed in  Sect.~\ref{sec:DIH-PTV:details}. All additive errors are drawn from a centered Gaussian distribution with a non-dimensional standard deviation of $1.25\times 10^{-4}$, $2.5\times 10^{-4}$, $5\times 10^{-4}$, $1\times 10^{-3}$, $2\times 10^{-3}$, or $4\times 10^{-3}$, corresponding to 0.03\%, 0.06\%, 0.13\%, 0.25\%, 0.5\%, or 1\% of the \textit{isoturb} domain, respectively. In all cases, $\sigma_\mathrm{x} = \sigma_\mathrm{y}$, and we only ran tests with $\sigma_\mathrm{z} \geq \sigma_\mathrm{x}, \sigma_\mathrm{y}$, assuming that depth sensing would be less accurate than dot finding, resulting in a total of 21 tests.\par

\begin{figure}[ht]
    \centering
    {\includegraphics[height=7cm]{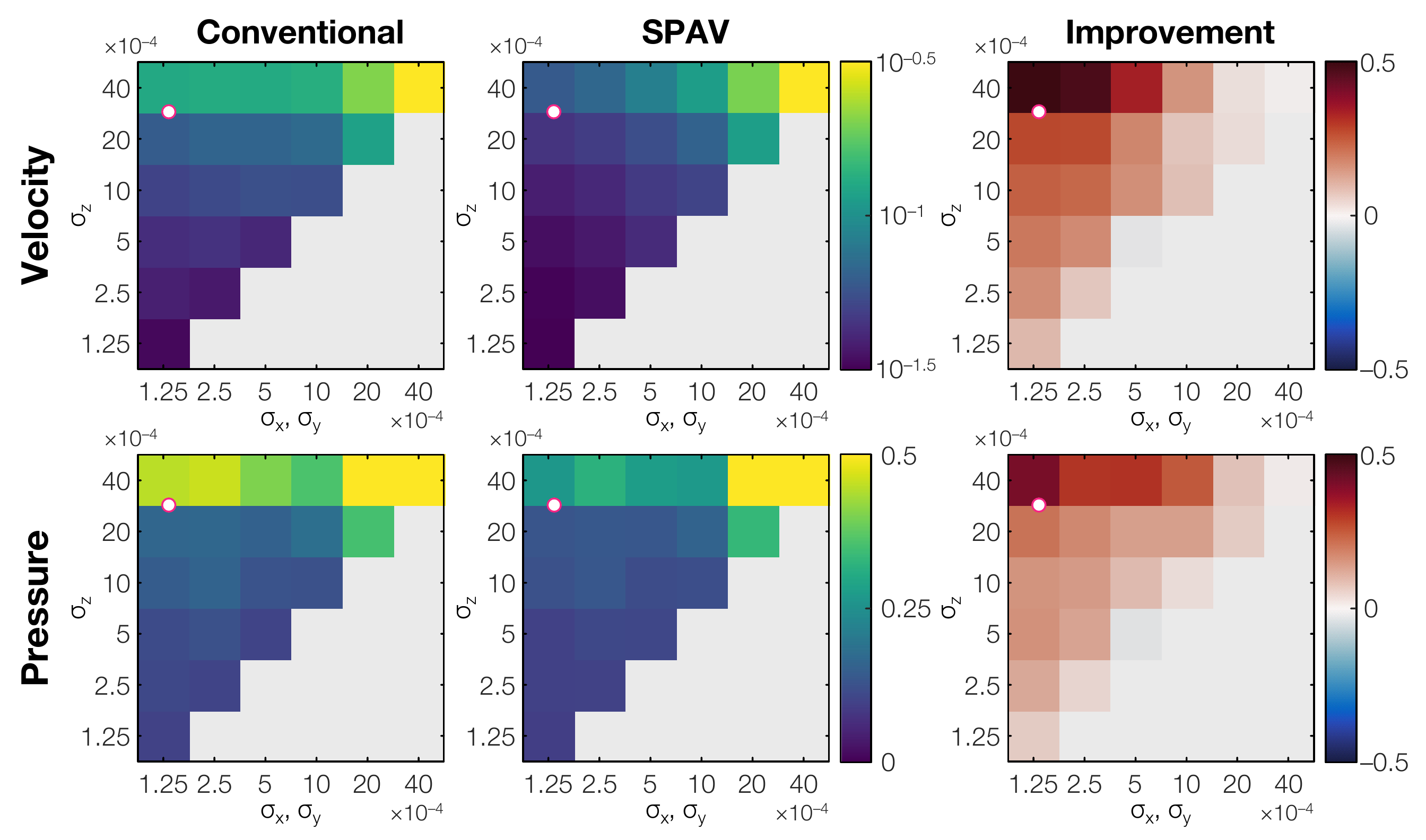}}
    \vspace*{2mm}
    \caption{Measurement uncertainty sensitivity analysis for the isoturb case: velocity (top row) and pressure (bottom row) errors from conventional (left) and SPAV (middle) PINNs as well as the relative difference (right). The fuchsia dot represents the measurement uncertainty from the DIH-PTV isoturb test in Sect.~\ref{sec:DIH-PTV:synthetic:3D isoturb}}
    \label{fig:sensitivity}
\end{figure}

Figure~\ref{fig:sensitivity} presents mean reconstruction errors for the isoturb flow, produced by the conventional and MC SPAV techniques. Naturally, reconstruction errors increase with both the magnitude and skew of localization errors. The conventional and SPAV methods exhibit broadly similar trends, but SPAV is consistently superior to the conventional method, as illustrated in the third column of Fig.~\ref{fig:sensitivity}. Here, the relative performance of these methods is visualized in terms of the reduction of error, i.e.,
\begin{equation}
    \frac{\overline{e}_\phi^\mathrm{conv} - \overline{e}_\phi^\mathrm{SPAV}}{\overline{e}_\phi^\mathrm{conv}}.\nonumber
\end{equation}
The benefit of SPAV is most pronounced for anisotropic errors, with reductions of error up to 50\% for distributions that are characteristic of a DIH-PTV system. The differences are less significant for isotropic errors, although it is possible that these results would change with further optimization of the loss component weights. SPAV's relative advantage may be explained by effectively discounting measurement components (i.e., the $x$-, $y$-, and $z$-components of $\mathbf{x}$) as a function of their uncertainty. Under this interpretation, when the measurement uncertainty is isotropic, both the conventional loss and SPAV utilize equal weights and should exhibit similar performance. However, in practice, anisotropic localization errors are prevalent in PTV, so it is reasonable to expect that SPAV will enhance the accuracy of Eulerian velocity and pressure field estimates compared to a generic data loss based on Eq.~\eqref{equ:velocity estimate}.\par

\section*{Acknowledgements}
This material is based upon work supported by the Erlangen Graduate School in Advanced Optical Technologies at the Friedrich-Alexander-Universit{\"a}t Erlangen-N{\"u}rnberg.\par

\section*{Data availability statement}
The data that support the findings of this study are available upon request from the authors.\par

\bibliographystyle{osajnl2} 

\end{document}